\documentclass{paper}
\usepackage{mathrsfs}
\usepackage{a4}
\usepackage{fancyhdr}
\usepackage{times}

\usepackage{graphicx}
\begin{document}
%
%
%
%
%
\title{A Nonlinear Instability Theory for a Wave System causing Transition in Spiral Poiseuille Flow}
\author{Venkatesa Iyengar Vasanta Ram\\Institut f\"ur Thermo- und Fluiddynamik\\Ruhr Universit\"at Bochum, 44780 Bochum, Germany}
\maketitle
%
\section*{Abstract}
This paper is on the effect of nonlinearity in the equations for propagation of disturbances on transition in the class of {\bf Spiral Poiseuille Flows}. The problem is approached from the fundamental point of view of following the growth of  initially infinitesimally small disturbances into their nonlinear stage when the effect of Reynolds stresses makes itself felt. To this end a {\bf rational scheme of an iterative solution method} for the set of {\it Generalised Nonlinear Orr-Sommerfeld, Squire and Continuity Equations} is proposed. The present proposal for {\bf Spiral Poiseuille Flows} closely follows that put forth in 1971 by Stuart and Stewartson in their seminal papers on the influence of nonlinear effects during transition in the {\bf bench-mark} flows, which are the flow in the gap between concentric circular cylinders  (Taylor instability), and in the plane walled channel flow \cite{stuart1,stewartsonstuart}.

The basic feature of the proposed method is the introduction of an {\bf Amplitude Parameter} and of a {\bf slow/long scale variable} to account for the effect of {\bf growing disturbances} within the framework of a {\bf rational iteration scheme}. It is shown that the effect of amplified disturbances is captured, as in the {\bf bench-mark flows}, by a {\bf Ginzburg-Landau} type differential equation for an {\bf Amplitude Function} in terms of suitably defined {\bf slow/long- scale variables}. However, the coefficients in this equation are  numbers that depend upon the flow parameters of the {\bf Spiral Poiseuille Flow}, which are the Reynolds Number, the Swirl Number, and the geometric parameter of transverse curvature of the flow geometry.
\section{Introduction}
In the highly ramified research area of transition in fluid flows, see eg.  \cite{prandtltietjens, albumvandyke,
 japsocmechengrs, tritton, koschmieder}, a branch that has repeatedly drawn attention through books, a variety of other publications like text books, monographs and reviews devoted to the subject, is transition in {\it spiral flows}. This is a class of flows recognisable through the streamline pattern they exhibit, that is characterised  by the topological feature of being made up of helically wound streamlines. The wide interest transition in {\it spiral flows} holds is perhaps explicable through its inherent scientific interest on its own, coupled with its relevance for gaining insight into the behaviour of a wide variety of flows, in nature as multifarious as in   technological applications.

 A consequence of the helical winding of streamlines is that, even
{\it spiral Poiseuille flows}, which form only a sub-class of {\it spiral flows}, are susceptible to transition through the simultaneous action of two disturbance propagation mechanisms, widely  regarded as basically distinct,  see eg. \cite{taylorscientificpapers, stuart, shen, batchelor, cclin, chandrasekhar, schlichtinggersten}.
These are: 1) the {\it Tollmien-Schlichting mechanism}, and 2) the {\it Taylor mechanism}. The former of these two is a mechanism that induces transition already in flows
with straight and parallel streamlines, whereas the latter is crucially dependent for its effectiveness on the streamlines being curved. A noteworthy feature of {\it spiral flows}
undergoing transition is that even the {\it spiral Poiseuille flow}, which is a {\it spiral flow} of relative simplicity,  exhibits in its transition behaviour complexities that are
traceable to the outstanding differences between salient flow patterns and gross properties
of the flow when transition is induced by disturbances propagating according to laws {\bf governed primarily by one and only one of the two mechanisms}, {\it i.e.} of either
{\it Tollmien-Schlichting} or {\it Taylor}. The differences are manifest in an equally pronounced fashion in experiment as well as in theory, making it natural to enquire into the nature of these differences in
{\it spiral flows} when these undergo transition under the influence of a combined action of both the two disturbance propagation mechanisms. The present work is part of an ongoing
investigation into the differences in transition behaviour of {\it spiral flows} when the {\bf decisive governing transition mechanism} may shift from one to the other. To this end
we first establish a theoretical framework suited for a detailed comparative study of the differences arising in {\it spiral Poiseuille flows} that is firmly grounded
in first principles for disturbance propagation. As a preliminary step in this study we address the question of propagation of {\bf infinitesimally small disturbances} in
{\it fully developed spiral Poiseuille flows}, and then extend the same to account for effects of {\bf nonlinearity} due to {\bf increase of disturbance amplitude}. The method we follow to reach this objective is one proposed and illustrated
by Stewartson and Stuart in their seminal papers, \cite{stuart1, stewartsonstuart}.

The common starting point for transition studies hitherto is based on the conjecture that all details of fluid motion related to transition are explicable through the 
{\bf Navier-Stokes Equations} that are known from standard literature such as e.g. \cite{lamb, batchelor}. The justification offered for this conjecture is that all the 
scales of fluid motion hitherto observed to be involved in transition phenomenon, whether in its initial stages or in its later course, are far removed from both sub-microscopic and 
astronomical scales, see e.g. \cite{feynman, krsreenivasan}. This conjecture has found almost universal acceptance as the basis for explanation of all stages of 
transition, although the origin of the plethora of scales and details of the fluid motion thereon have not yet been entirely unravelled. Many questions that arise in the
context of flows in transition have indeed been posed and answered, and phenomena explained on the basis of the conjecture just referred to. But, there are many questions 
still remaining open, amongst which are those related to transition initiated by physically distinct disturbance propagation mechanisms acting simultaneously! The course
transition then takes, particularly under the influence of nonlinearity in the equations governing disturbance propagation, continues to occupy a prominent position among 
subjects of enquiry into fluid flow phenomena. The present work is intended as but a step towards narrowing down the existing gap in understanding of the details in this 
parameter range of transition. A route taken to this end has been through examination of the course of transition in selected archetypal flow examples that can undergo 
transition through simultaneous action of several {\bf distinct} transition inducing disturbance propagation mechanisms, taking explicitly into account the effects of 
nonlinearity of the equations for disturbance propagation. 

Our choice of the class of {\bf Spiral Poiseuille Flows} for this purpose was motivated by prospects the author considers this class of flows offers, for gaining physical insight into the origin of the multitude of flow patterns it exhibits during transition, see eg, \cite{brockmannetal} and references therein.
An advantage offered by the {\bf Spiral Poiseuille Flow} for such a study is that in both the limits of
the swirl being either very small or very large, transition is initiated by one and only one disturbance propagation mechanism all by itself. These limiting cases 
have both been subjects of thorough  studies on their own, which cover effects of nonlinearity as well. The insight gained through these studies makes them suitable to be bench-marks against which flow behaviour during transition in the {\bf Spiral Poiseuille Flow} can be set. The comparison, therefore, holds promise of illuminating the shift of the transition
mechanism from one to the other in the entire parameter range bridging the limiting cases of {\bf Spiral Poiseuille Flows}, hence our choice for this flow.

The search for the origin of scales and evolution of details of fluid motion on these scales forms an integral part of transition research today, and the root cause of 
the multiplicity of scales is conjectured to lie in the {\bf patterns of instability} of the state of fluid motion towards small disturbances that are inevitably present. The
expectation is that the observed complexity of fluid motion on the different scales would be explicable as a result of the nonlinearity of the equations of fluid motion. 
The starting point for explaining the origin of the variety of scales and features of fluid motion thereon has hitherto remained the well-known {\bf Navier-Stokes Equations} 
for fluid motion set out in standard published literature in fluid dynamics, just referred to. The starting point will be kept the same in the present work too. It may be 
in order to point out in this context that a relatively recent study, \cite{vvr}, shows the {\bf linearised equations for transition inducing disturbance propagation in 
Spiral Poiseuille Flows} do contain, {\bf on suitable formulation}, the established equations of {\bf both} the limiting cases, hitherto regarded distinct.

A straightforward approach towards the goal of illuminating the bridge between the limiting cases would be to trace the spatio-temporal evolution of disturbances in the flow 
in question, such has been done in e.g. \cite{hasoonmartin}; and to follow up its outcome with an analysis for extracting the scales and features of motion over
these scales displayed by the disturbance. However, the present state of understanding does not permit drawing generally valid conclusions on the relations between the 
characteristics of disturbance propagation on the newly arising scales and their effects in arbitrary basic flows, in particular not for arbitrary intervals in time and 
space, leaving us with no choice but to undertake such a study in archetypal basic flows. This provided the motivation for the present work. The emphasis placed hereby is on
accounting for the effects of the nonlinear terms in the equations for disturbance propagation in the class of flows under investigation. The study thus attempts to go a 
step further than the classical linear theory, such as e.g. in \cite{hasoonmartin,diprimapridor, ngturner} and \cite{cotrellpearlstein}, which are limited to infinitesimally 
small-amplitude disturbances only.  
%
\subsection{Scope and outline of the present work}
The task of tracing the spatio-temporal evolution of a particular disturbance to a certain given basic flow calls for integration of well-tailored analytical and numerical 
methods with each other. Analytical approaches potentially illuminate the scales and the nature of motion on the different scales involved, thus unveiling the possibly 
hidden mechanisms during flow transition. Algorithms, developed for solving the governing equations, yield numerical values for physical quantities, from which 
further features of interest, like scaling behaviour, can -hopefully - be extracted. Concurrence of findings from the two approaches, analytical and numerical, would then 
instill confidence in the insight gained. Perturbation methods are well suited for such tasks, see e.g. \cite{murdock, kevorkiancole, nayfeh, schneider, hinch}. A perturbation approach that appears particularly suited
for transition studies of {\bf Spiral Poiseuille Flows} is the one developed, tested and demonstrated for use in the two limiting cases of large and small swirl by Stuart 
and Stewartson \cite{stuart1, stewartsonstuart}. The work of Stuart \cite{stuart1} illustrates this method for studying transition in the flow with circular streamlines in 
the gap between concentric circular cylinders. That of Stewartson and Stuart \cite{stewartsonstuart} demonstrates its applicability to the plane-walled channel flow. It
is relevant in this contex to redraw attention to each of these flow examples being archetypal basic flows with {\bf only one} transition inducing disturbance propagation 
mechanism, standing in contrast to the class of present flows of enquiry, which exhibit {\bf two distinct transition inducing disturbance propagation mechanisms acting 
simultaneously}.

The object of the present work is to derive from first principles the set of equations that governs the propagation characteristics of transition inducing disturbances in 
the class of {\bf Spiral Poiseuille Flows}. The focus of attention in this study is on the {\bf effect of nonlinear terms in these equations as the propagation mechanism 
shifts from one of a 'Tollmien-Schlichting' type to the other of a 'Taylor' type}. To this end the author has chosen to tread along a route that attempts to harmonise analytical 
and numerical approaches, following the example set by Stewartson and Stuart in their seminal papers, {\it viz.} \cite{stuart1} and \cite{stewartsonstuart}. In order to keep 
the size of this paper within limits, the scope of the present paper is restricted only to the analytical side of the task, as Stewartson and Stuart have done in theirs,
\cite{stewartsonstuart}. In order to facilitate comparison of the findings from our present work with that for the limiting cases, the route of our present paper is charted out to closely follow that laid out by Stewartson and Stuart in \cite{stewartsonstuart}. Our procedure is sketched in outline below. \\

The flow geometry, the reference quantities and the governing equations are introduced in Sec. 2. The section that follows, Sec. 3, outlines the salient steps in the approach of Stuart
and Stewartson, \cite{stewartsonstuart}, that circumvents the blind alleys encountered when established {\it classical methods for studying wave propagation} are
employed in a straightforward manner to follow the propagation characteristics of transition inducing disturbances in the class of {\bf Spiral Poiseuille Flows}. Key concepts involved in this approach
are the introduction of an {\bf amplitude parameter} and of {\bf slow/long scales}, in terms of which the problem is readdressed in order to arrive at a
solution that meets the requirements of a {\bf rational theory}. In Sec. 4 the problem is reformulated to address the question of transition in Spiral Poiseuille Flow along
lines of Stuart and Stewartson in \cite{stewartsonstuart}. The reformulated problem is implemented in Sec. 5 to arrive at a Ginzburg-Landau-type equation that accounts for 
the growth effects of the initially infinitesimally small nonlinear terms in the equations for propagation of disturbances during transition of Spiral Poiseuille flow. The reformulation enables to recognise both the common ground and the differences in the dominant mechanisms at work in cases of mild/small and strong/large swirl. They emerge naturally from the equations themselves. A brief discussion of the physical insights gained from these equations follows in  the closing section of this work, Sec. 6.

%
%
\subsection{The flow geometry and reference quantities}
 We denote the radii of the outer and the inner cylinders through $R_o$ and $R_i$ respectively, the angular velocity of the inner cylinder through $\Omega$, and the imposed
axial pressure gradient through $\frac{dP_G}{dx}$. The topologically outstanding feature of the streamline pattern of the class of flows realisable from this arrangement is, as stated earlier,
the helical winding of its streamlines. A dynamically salient feature consequential to helical winding of the streamlines is the susceptibility of the class of flows in question 
to transition through combined action of the {\bf two distinct physical mechanisms of disturbance propagation}, {\it viz.} the {\bf Tollmien-Schlichting} and the 
{\bf Taylor} mechanisms. For quantification of the strength of these two distinct mechanisms we introduce characteristic velocities of their own, {\it viz.} $U_{refxp}$ and 
$U_{ref \varphi}$ respectively. The former we define through the imposed axial pressure gradient, $ \frac{dP_G}{dx}$, with $ U_{refxp}^2 = - \frac{H^2}{2 \mu} \frac{dP_G}{dx}$, 
and the latter through $U_{ref \varphi} = \Omega R_i$. Here $H$ stands for the semi gap-width, {\it i.e.} $2H= (R_o-R_i)$, and $\mu$ and $\nu$ in that order for the dynamic and 
the kinematic viscosity respectively. For purposes of the present work we choose as the reference velocity $U_{ref}$, the vectorial sum of the two characteristic velocities 
defined  through $U_{ref} = \sqrt{ (U_{refxp}^2 + U_{ref \varphi}^2) }$. The advantages offered by this choice will be evident on inspection of the corresponding propagation 
problem of {\bf infinitesimally small-amplitude disturbances}, which excludes nonlinear terms, {\it vide} \cite{vvr}. These advantages are retained even when the initially small disturbance grows in amplitude to warrant inclusion of nonlinear effects. On defining the {\bf Swirl Parameter}, $S$, as the
ratio $ \frac{U_{ref \varphi}}{U_{refxp}} $, it then measures the relative strengths of the {\bf Tollmien-Schlichting} and the {\bf Taylor} mechanisms during transition of the
{\bf Spiral Poiseuille Flow}. $S$ may then assume values between $0$ and $ \infty $, taking the value $0$ when the flow undergoes transition only {\it via} the 
{\bf Tollmien-Schlichting} mechanism, and $ \infty $ when it is through the {\bf Taylor} mechanism alone.

The governing equations for the current problem are the equations of motion of an incompressible fluid together with the continuity equation, which are known from standard 
texts, e.g. (\cite{lamb}, \cite{batchelor}, \cite{schlichtinggersten}). In the present work we use polar co-ordinates, $(r, \varphi, x)$, and the form of the equations in this 
co-ordinate system may also be found in the books just cited. Shedding light on the shift of the transition inducing mechanism is well served if the following set is chosen 
as reference quantities, hence our choice: for length, the semi gap-width $H$, for velocity, the {\bf vectorial sum of the two characteristic velocities}, $U_{ref} $, and 
for time $\frac{H}{U_{ref}}$. The Reynolds number, $Re$, is therefore $Re = \frac{U_{ref} H}{\nu}$.  The {\bf Swirl Parameter}, $S$, introduced earlier, is a further flow 
parameter governing transition of the flow under investigation. The geometrical parameter $ \epsilon_R = \frac{(R_o - R_i)}{(R_o + R_i)} $ also enters the 
problem of disturbance propagation, making the phenomenon of {\bf transition in Spiral Poiseuille Flow} depend upon the three parameters $Re, S$ and $\epsilon_R $. In the 
basic flow of the problem of current interest the (dimensional) axial and the azimuthal velocity components are: $U_{ref}(1+S^2)^{-\frac{1}{2}}~ U_{Gx}$ and 
$U_{ref} S (1+S^2)^{-\frac{1}{2}}~U_{G \varphi}$, with 
$U_{Gx} = \big((1-y^2) - \epsilon_R \frac{y(1-y^2)}{3} + O(\epsilon_R^2) \big)$ and $U_{G \varphi} = \big( \frac{(1-y)}{2} - \epsilon_R \frac{(1-y^2)}{2} + O(\epsilon_R^2) \big)$, 
where $U_{Gx}$ and $U_{G\varphi }$ are dimensionless functions of the wall-normal coordinate $y$ alone. 
The pressure field in the basic flow is given by  $\frac{\partial P_G}{\partial x} = constant $ and  $\frac{\partial P_G}{\partial r} = \rho \frac{U_{ref}^2}{H} \frac{U_{G \varphi}^2}{r} $. For the 
purposes of the present work it is more convenient to work with {\bf additive departures} of the velocity and pressure fields from the basic flow, so we write the velocity components and 
the pressure as sums of these quantities in the basic flow and {\bf disturbances}, denoted through the subscript $s$, as $u_x = U_{Gx} + u_{sx}, u_{\varphi} = U_{G \varphi} + u_{s \varphi}$ and $u_r = 0 + u_{sr}$.
On rewriting the equations of motion in terms of disturbances as described, it will be seen that there is no room for confusion if the subscript $s$ to denote a disturbance 
is dropped, so henceforth the subscript $s$ for the disturbance will be dropped. The equations of motion, together with the equation of continuity, then yield a set of four 
equations for the four unknowns, which are the three components of the velocity and the pressure, $(u_x, u_{\varphi}, u_r)$ and $p$. In classical works on disturbance 
propagation characteristics these equations, linearised for infinitesimally small disturbances around a given basic flow, form the starting point. 
{\bf In our present work the non-linearities are retained}. We will refer to this set of equations as {\bf equations in conventional variables}.

In the course of the present work it will be seen that the phenomenon in question is rendered more transparent if it is viewed in terms of other variables too. We refer to these as {\bf strong swirl variables} and {\bf Taylor variables}. Their definitions are gathered together in {\bf Appendix 1}:

\subsection{The governing equations}
In order to keep transparency between the present work and the extensive literature on fluid flow transition hitherto gathered, we work with a system most often used in
published literature. This is a system in which the disturbance in pressure is eliminated as an unknown leading to the well-known Orr-Sommerfeld equation. In their original 
form, the equations of motion are four in number, {\it vide} \cite{lamb, batchelor, schlichtinggersten, tritton} and are for four unknowns, $(u_x, u_{\varphi}, u_r, p)$. To 
eliminate pressure as an unknown from the four equations, there are two distinct procedures which we refer to as the {\bf Orr-Sommerfeld} and the {\bf Squire} procedures. 
For a concise description of the difference between these two procedures the reader is referred to \cite{vvr}. On elimination of pressure disturbance as an unknown through 
both these procedures we are left with three equations for the three unknowns $(u_x, u_{\varphi}, u_r)$. {\bf These equations are inclusive of the effects of nonlinearities}.  
We refer to them as the set of {\bf generalised nonlinear Orr-Sommerfeld}, {\bf generalised nonlinear Squire} and {\bf continuity} equations. They are the governing equations 
for our problem and the starting point for our present studies. The unknowns are only the three components of the velocity disturbance from the basic flow in the particular 
geometric configuration in question. The governing equations for our problem, whose elements are given in detail in {\bf Appendix 1}, may be written in a compact matrix 
notation as follows: 
\begin{equation}
 \mathscr D \mathbf u = \mathscr N.  
\label{governingequation}
\end{equation}  
In (\ref{governingequation}), $ \mathscr D $ denotes the linear part of the matrix differential operator, and $ \mathscr N $ an operator whose elements act on the non-linear
terms. Elements of $\mathscr N$ are composed of sums of derivatives of the tensor product of the velocity vector $\mathbf u$ with itself, $ \mathbf u \otimes \mathbf u$. 
Both $\mathscr D$ and $\mathscr N$ act on the unknown velocity disturbance $ \mathbf u $ which is a column vector comprising the three components of the velocity disturbance 
$(u_r, u_{\varphi}, u_x)^T $. \footnote{In some  stages of this work, to aid transparency in following the course of steps the elements of $\mathbf u$ and $\mathscr N$ are arranged in the form of diagonal matrices instead of column vectors. We will do so when we see advantages for such use, however, using the same notation for both the column vector and the diagonal matrix. We hope no confusion will arise on this account. If we see a possibility of confusion between the two usages we will indicate in the text the meaning in which it is to be understood.}
%
%
\subsubsection{The differential operators}
The linear differential operator $\mathscr D$ may itself be written in a matrix form as: 
\begin{eqnarray}
\mathscr D = \left(
\begin{array}{ccc}
 \mathscr D_{OSr}   &  \mathscr D_{OS \varphi}   &  \mathscr D_{OSx}\\
 \mathscr D_{Sqr}   &  \mathscr D_{Sq \varphi}   &  \mathscr D_{Sqx}\\
 \mathscr D_{Cor}   &  \mathscr D_{Co \varphi}   &  \mathscr D_{Cox}
\end{array}
\right), \; 
\mathbf{u} =
\left( 
\begin{array}{c}
u_{r} \\ u_{\varphi} \\ u_{x}
\end{array}
\right), 
\mathscr N = \left(
\begin{array}{c}
\mathscr N_{OS} \\ \mathscr N_{Sq} \\ \mathscr N_{Co}
\end{array}
\right),
\label{doperatorswirl}
\end{eqnarray}
where the subscripts, $OS, Sq, Co$ to $\mathscr D$ indicate the source of the operator, i.e. whether it is derived from the {\bf Orr-Sommerfeld}, {\bf Squire} or  
{\bf Continuity} equation. The defintions of the elements of operators in (\ref{doperatorswirl}) are also given in {\bf Appendix 1}: 

With the swirl parameter, $S$, defined through $S = \frac{\Omega R_i}{U_{refxp}}$, the solution of the problem posed is dependent upon the two flow parameters, 
$(Re, S)$ and the geometric parameter $\epsilon_R$.
%

\section{Effects of mild nonlinearity in the governing equations on propagation characteristics of transition-inducing disturbances in Spiral Poiseuille Flow}

\subsection{A sketch of the approach of Stewartson and Stuart}

The approach of Stewartson and Stuart may be understood as one through which the effect of Reynolds stresses arising on the onset of transition from amplification of initially small-amplitude
disturbances is accounted for through modification of the ansatz employed in the classical theory. This modification is associated with a reformulation of the problem, which 
in turn enables a theory to be worked out that falls within the framework of a {\bf rational theory} in the sense of Van Dyke \cite{vandyke}, see also \cite{murdock}. The 
outlines of the reformulated theory may be paraphrased through its salient steps which are as follows:

\begin{enumerate} 
 \item The {\bf initial step} is obtaining a solution to the problem of propagation of {\bf infinitesimally small disturbances} to a {\bf known basic flow}. In this 
classical step, an arbitrary disturbance, parametrized through its {\bf wavenumber, frequency and amplitude}, is governed by the equations of fluid motion, 
{\it vide} \cite{lamb, batchelor}, that are linearised around the basic flow of inquiry. These equations for the propagation of {\bf infinitesimally small-amplitude disturbances} 
assume the form of an eigenvalue problem that is characterised by its {\bf dispersion relation} being {\bf independent of the amplitude}. The eigenvalue and the eigenfunctions, which describe the {\bf shapes} of the disturbance, are not known {\it a priori}. They have to be obtained as a part of the solution.
\item 
The dispersion relation, obtained during solution of the eigenvalue problem, permits an initial classification of disturbances into {\bf exponentially growing} or 
{\bf decaying} disturbances, thus enabling a {\bf neutrally stable surface} to be delineated in the space of the parameters that governs the evolution of disturbances. 
However, the {\bf classical solution}, in its form at this stage, is not, despite its illuminative feature, employable as a starting point for treatment of effects caused by 
{\bf growing disturbances}, say through straightforward iteration. Reasons for this limitation of this apparently straightforward route will be evident on closer examination
of the classical solution, which is to follow.
 \item Stuart and Stewartson circumvent the blind alley encountered by modifying the ansatz in the initial step to account for the effects of growing disturbances. Their
reformulation of the problem involves the following outstanding features. These are: 

\begin{itemize}  
 \item Characterisation of the {\bf strength of the nonlinear terms}, which do not necessarily remain infinitesimally small, through an {\bf Amplitude Parameter}. It should be noted in this context that the {\bf Amplitude Parameter} does not directly appear as a parameter present in the equations themselves or in their boundary conditions. Since the solution of the equations sought is governed by the set of {\bf flow and geometric parameters} present in the equations and the prescribed boundary conditions, Stuart and Stewartson {\bf postulate} a relation between the {\bf Flow and Geometric Parameters} that are prescribed and the {\bf Amplitude Parameter} characterising the solution. We follow a similar approach in our present work too.

\item Modification of the ansatz of the classical problem through introduction of an {\bf Amplitude Function} that evaluates the effect of growing disturbances in terms of 
suitably defined {\bf slow/long scales}. Utilization of this modified ansatz hinges upon certain {\bf solvability conditions} having to be fulfilled for further iteration steps to be successful. Setting up these {\bf solvability conditions} is the crucial task in their procedure.
 \item Derivation of an {\bf equation that captures the evolution of growing disturbances} in terms of the freshly introduced {\bf slow/long scales}. This equation is the 
{\bf solvability condition} that has to be fulfilled for the {\bf modified ansatz to lead to an iteratively solvable scheme}, thus circumventing the blind alley encountered in the initial ansatz. The
outcome of the reformulated problem may be expected to capture the effects of {\bf growing disturbances} as long as the values of the parameters in the space of the 
dispersion relation, where the growth rate of the disturbance goes through a (local) maximum, is located close enough to the surface of neutral stability of the classical 
problem. 
\end{itemize}
This inherent limitation of the eigenvalue problem as posed surfaces immediately on discretizing the set of differential equations governing propagation of infinitesimally small disturbances, followed by
 casting the same into its matrix form. In the notation used by Strang in the books authored by him, {\it vide} \cite{strang1, strang2, strang3}, but with the notation for the eigenvalue changed from $\lambda$ to $\omega$, it is meaningful to distinguish between equations that may
be brought close to the form $ \mathbf A \mathbf x = \omega \mathbf x $, and such, that may be brought to the form $\mathbf A \mathbf x = \mathbf b $. The former, where $\mathbf A$ is a
known matrix, $\mathbf x$ the unknown vector, {\it viz.} the eigenfunction, and $\omega$ the unknown eigenvalue, is an eigenvalue problem that admits the trivial solution, $\mathbf x \equiv \mathbf 0$,
that is, however, not of interest. But the matrix equation $\mathbf A \mathbf x = \omega \mathbf x$ may admit {\bf nontrivial} solutions, {\it i.e.} $\mathbf x \not \equiv 0$,
if the wavenumber of the disturbance, $\lambda$, and its frequency, $\omega$ are {\bf related to each other through the dispersion relation}, and these solutions are indeed of fundamental interest!
{\bf For infinitesimally small disturbances the dispersion relation is independent of the amplitude}. Again in Strang's notation, in the latter problem, $\mathbf A \mathbf x = \mathbf b $, $ \mathbf A$
is the known matrix, $\mathbf x$ the {\bf unknown} vector, and $\mathbf b$ a {\bf given} vector. Solving the inhomegeneous problem, $\mathbf A \mathbf x = \mathbf b$, requires the matrix $\mathbf A$ to
be invertible, whereas for finding a nontrivial solution to $\mathbf A \mathbf x = \omega \mathbf x$ the matrix $\mathbf A$ has to be non-invertible! We refer to this latter solution as {\bf the classical solution} of the eigenvalue problem.

If, as will be seen to be the case in the present problem, after execution of the initial step which is an eigenvalue problem, and so belongs to the group of equations of the kind
$ \mathbf A \mathbf x - \omega \mathbf x = 0 $, the equation to be solved turns out to be of the kind $ \mathbf A \mathbf x - \omega \mathbf x = \mathbf b $
with $ \omega $ an eigenvalue and $\mathbf b$ a nontrivial specified vector, both {\bf known after execution of the initial step}, the question arises as to whether the inhomogeneous equation for a given right-hand side is {\bf solvable} at all. Strang observes, {\it vide} \cite{strang1, strang2, strang3} that this is indeed solvable if the given vector $ \mathbf b $ obeys the condition that it lies in the column space of the matrix, $ \mathbf A $. This is the {\bf solvability condition} for the inhomogeneous problem $ \mathbf A \mathbf x  - \omega \mathbf x = \mathbf b$.

In our present work the eigenvalue problem to start with is only slightly different from Strang's example. It is not of the form $ \mathbf A \mathbf x = \omega \mathbf x $ standing in contrast against $ \mathbf A \mathbf x = \mathbf b$;
 but of the form $ \mathscr L \mathbf u +  \omega \mathscr M \mathbf u = 0 $ standing in contrast against  $\mathscr L \mathbf u + \omega \mathscr M \mathbf u = \mathbf b $.
The matrices $\mathscr L $ and $\mathscr M $ follow from the differential equations of the problem in question, $\mathbf u$ the unknown column vector, and the right-hand side, $\mathbf b $, a given vector that is known from the previous iteration step.

Extending Strang's example to the question of solvability of the inhomogeneous equation $\mathscr L \mathbf u - \omega \mathbf M \mathbf u = \mathbf b$, we interpret Strang's formulation to be applicable to our present problem if the vector $\mathbf b$ lies in the column space of $\mathscr L  \mathbf u- \omega \mathscr M \mathbf u $, or, in other words, the vectors on both sides of the equation are aligned with each other.

The present work requires a formulation of the {\bf solvability condition} in analytical terms and this will be presented in Section 5.
The approach of Stewartson and Stuart is based on the conjecture that the location of maximum growth of the disturbance is sufficiently close to the surface of
neutral stability that is obtainable from the dispersion relation according to the classical theory of infinitesimally small disturbances.
\end{enumerate}

{\bf Step 1} in the above outline may be regarded as essentially solved after the classical solution has been obtained. So, attention is focussed in what follows mainly on the course of steps following {\bf Step 1}.

\section{The amplitude parameter and the slow/long scale variables}

The {\bf Amplitude parameter and the Slow/Long scale variables} in terms of which the effects of growing disturbances during transition are dealt with in \cite{stuart1, stewartsonstuart} depend  crucially upon the growth of disturbances that were
neglected in the derivation of equations for disturbance propagation in the classical theory. An analytical expression for the {\bf growth} of these terms is contained in 
the classical {\bf elementary wave ansatz} used for the solution of the equation (\ref{governingequation}) with $\mathscr N=0$, which is:
\begin{equation}
 \mathbf u = \mathbf a \exp (\imath \Theta) + \mathbf a^\ast \exp (- \imath \Theta).
\label{ansatzelementarywave}
\end{equation}
In (\ref{ansatzelementarywave}) above, $\mathbf a$ is a column vector and the {\bf phase}, $\Theta$, a linear function of only the variables, $x, \varphi$ and $t$ which will soon be declared to be the
set of {\bf fast/short scale variables} in our problem. The phase, $\Theta$, is given by the following expression: 
\begin{equation}
 \Theta = \Theta (x, \varphi, t) = \lambda_x x + n_{\varphi} \varphi - \omega t, 
\label{defnthetaproblm1}
\end{equation}
where, for following the temporal evolution of disturbances the wavenumbers $(\lambda_x, n_{\varphi})$ are real and prescribed, and the frequency $\omega$, regarded as an 
unknown, is permitted to be complex. Due to constraints of the flow geometry, only integer values are permissible for $n_{\varphi}$. 

Substitution of the ansatz according to (\ref{ansatzelementarywave}) in the governing equation (\ref{governingequation}) with the \underline{right-hand} \underline{side $\mathscr N$ set equal to
$ \mathbf 0 $} yields the set of ordinary homogeneous differential equations that defines the eigenvalue problem for propagation of infinitesimally small disturbances in the {\bf Spiral Poiseuille Flow}. This set of ordinary differential equations defining the {\bf classical eigenvalue problem for the class of Spiral Poiseuille flows } may be written in a compact matrix form as follows:
\begin{equation}
 \mathscr L {\bf a} -  \omega \mathscr M \mathscr {\bf  a} = 0,
\label{mlswirl}
\end{equation}
where $\bf a$ is a column vector and $\mathscr L$ and $\mathscr M$ are matrix operators as follows:
\begin{eqnarray}
\mathscr L = \left(
\begin{array}{ccc}
 \mathscr L_{OSr}   &  \mathscr L_{OS \varphi}   &  \mathscr L_{OSx}\\
 \mathscr L_{Sqr}   &  \mathscr L_{Sq \varphi}   &  \mathscr L_{Sqx}\\
 \mathscr L_{Cor}   &  \mathscr L_{Co \varphi}   &  \mathscr L_{Cox}
\end{array}
\right),
\mathscr M = \left(
\begin{array}{ccc}
 \mathscr M_{OSr}   &  \mathscr M_{OS \varphi}   &  \mathscr M_{OSx}\\
 \mathscr M_{Sqr}   &  \mathscr M_{Sq \varphi}   &  \mathscr M_{Sqx}\\
 \mathscr M_{Cor}   &  \mathscr M_{Co \varphi}   &  \mathscr M_{Cox}
\end{array}
\right),
\label{lmoperatorswirl}
\end{eqnarray}
The elements of the column vector of the eigenfunctions $\bf a$ are as follows:
\begin{equation}
{\bf  a} = ( A_r, A_{\varphi}, A_x)^T.
\label{eigenfunctionclassical}
\end{equation}
The expressions for the matrix differential operators $\mathscr L $ and $ \mathscr M $ are given in {\bf Appendix 2}.

Its dispersion relation may be written as follows:
\begin{equation}
 \omega = \omega (\lambda_x, n_{\varphi}; Re, S, \epsilon_R)
\label{dispersionrelclassical}
 \end{equation}

A wave packet is describable through {\bf Fourier superposition of elementary waves according to (\ref{ansatzelementarywave})}. Fourier-superposition of waves of the type 
(\ref{ansatzelementarywave}) over the entire permissible range of wavenumbers leads to the following expressions for the solution of $\mathbf u$ : 
\begin{equation}
 \mathbf u = \sum_{n_{\varphi}= 0}  ^{\infty}  \int_{\lambda_x = -\infty}^{\lambda_x = \infty} \mathbf a (y; \lambda_x, n_{\varphi}, Re, S, \epsilon_R) \exp \left(\imath \Theta(\lambda_x, n_{\varphi}) \right) d \lambda_x + c.c. 
\label{fouriersuperpswirl}
\end{equation}
In order to get an expression for $\mathbf u$ in the variables used in the original formulation of the governing equations, (\ref{governingequation}), a back-transformation 
from the {\it Fourier-space} to the space $(t, x, y, \varphi)$ has to be undertaken. However, for the purposes of the present work it is not necessary to have the back-transformed 
solution in all generality, since it suffices to have an {\bf asymptotic solution} for large values of ($x, t$), with $\frac{x}{t} $ held finite. This problem has been worked out 
extensively in published literature, in particular for water waves, see e.g. \cite{whitham, lannes, constantinescherjohnsonvillari}. Summarising Whitham \cite{whitham}, 
Chapter 11.3 , we may write: the {\it form} of the expression for an elementary wave according to the ansatz (\ref{ansatzelementarywave}, \ref{defnthetaproblm1}) also 
remains valid for large values of $x$ and $t$ when $\frac{x}{t}$ is held fixed and finite, if the amplitude $\mathbf a$, the wavenumber 
$\lambda_x, $ and the frequency $\omega$ are not held constant but are permitted to {\bf vary slowly}. The phase $\Theta$ is then given by 
$\Theta = x \lambda_x(x, t) + \varphi n_{\varphi}(x, t) - t \omega (x, t)$. For a more thorough discussion of this aspect of the solution the reader is referred to Whitham's book, \cite{whitham}, Chapter 11.3.

%
%


Since, in the course of the present work, although the ansatz for the classical eigenvalue problem according to (\ref{ansatzelementarywave}) will be employed only in a modified form, it is useful to have the set of ordinary differential equations associated with the classical ansatz according to (\ref{ansatzelementarywave}) ready in their matrix form close to $ \mathbf A \mathbf x = \omega \mathbf x $ used by Strang  \cite{strang1, strang2, strang3} for explanation of the salient difference between a matrix eigenvalue problem and the more classical inhomogenous form $ \mathbf A \mathbf x = \mathbf b $. This set of ordinary differential equations may be written in the form $ \mathscr L \mathbf a - \omega \mathscr M \mathbf a = \mathbf 0$, which is found in published literature in the area of hydrodynamic stability, e.g. \cite{schmidhenningson}. They are given in {\bf Appendix 2}. In the present work, it is more appropriate to illustrate the difference in matrix formulation between the eigenvalue problem and the inhomogeneous problem, through setting the examples $ \mathscr L \mathbf a - \omega \mathscr M \mathbf a = 0 $ and $\mathscr L \mathbf a - \omega \mathscr M \mathbf a = \mathbf b$ against each other, which is an extension in spirit of the examples used by Strang, \cite{strang1, strang2, strang3}, {\it viz.} $\mathbf A x = \omega \mathbf x$ and $\mathbf A \mathbf x = \mathbf b$.

For the purpose of gaining an insight into the analytical structure of the (nonlinear) terms in $\mathscr N$ it suffices to examine the particular case when 
$\mathbf u$ is a sum of only two waves alone. 
Setting the velocity vector $\mathbf u$, regarded as a column vector of {\bf real} quantities, as a superposition of only two waves, viz.  ${\it wave1} $ and ${\it wave2}$,
where the {\bf wave-amplitudes}, $ (\mathbf a^{(wave1)},\mathbf a^{(wave2)}) $, and the frequency in the {\bf phases}, $ (\Theta_1, \Theta_2 ) $ are permitted to be
{\bf complex}, we may write $\mathbf u$ as follows:  \begin{eqnarray}
\mathbf u = \mathbf a^{(wave1)} \exp \left\{i(\Theta_1)\right\} + \mathbf a^{(wave1) \ast} \exp \left \{-i(\Theta_1) \right\} +   \nonumber  \\
              \mathbf a^{(wave2)} \exp \left \{i(\Theta_2)\right \} + \mathbf a^{(wave2) \ast } \exp \left \{-i(\Theta_2) \right\},
\label{zweiwellenansatzuu}
\end{eqnarray} 

In (\ref{zweiwellenansatzuu}), the ($*$) following
the wave-amplitude $\mathbf a$ indicates the {\it conjugate complex} of the preceding quantity.
The analytical structure of the tensorial nonlinear term, $\mathbf u \otimes \mathbf u $, with respect to the {\bf fast/short-scale variables}, $(x, \varphi, t)$,   is obtainable through straightforward multiplication of the involved terms. This is seen to be a sum of terms of different periodicity. For the specific example, when the flucuating motion is a superposition of only two waves, $ {\it wave 1} $ and ${\it wave 2}$, of periodicity $\Theta_1$ and $\Theta_2$ respectively, the nonlinear term $ \mathscr T $  comprises terms of periodicity $0, 2\Theta_1, 2\Theta_2, (\Theta_2+ \Theta_1)$ and $(\Theta_2 - \Theta_1)$ only, so that $\mathscr T$ may formally be written to explicitly exhibit its periodic structure with respect to the {\bf fast/short-scale variables}, which is as follows:
\begin{eqnarray}
 \mathscr T = \mathscr T^{(0)} \exp \left\{ i(0)\right \} + \mathscr T^{(2 \Theta_1)} \exp \left\{ 2i (\Theta_1)\right\}  + \mathscr T^{(2 \Theta_2)} \exp \left \{ 2i (\Theta_2)\right \}  +  \nonumber   \\  \mathscr T^{(\Theta_1 + \Theta_2)} \exp \left \{ i (\Theta_1 + \Theta_2) \right \} + \mathscr T^{(\Theta_1- \Theta_2)} \exp \left\{i(\Theta_1 - \Theta_2)   \right\} + c.c.  \label{tensorproduktuotimesu}
 \end{eqnarray}
In the above expression, (\ref{tensorproduktuotimesu}), we have set the superscript in brackets () to indicate the periodicity of the term, which may be $0, 2 \Theta_1, 2 \Theta_2, ( \Theta_2+ \Theta_1) $ or $(\Theta_2 - \Theta_1)$. In the course of this work we will choose $\Theta_1$ to lie on the surface of neutral stability for which the imaginary part of $\omega, \omega_i = 0$, and denote this choice with  $\Theta_N$. The expression in (\ref{tensorproduktuotimesu}) will henceforth be wrtitten as follows:
\begin{eqnarray}
 \mathscr T = \mathscr T^{(0)} \exp \left\{ i(0)\right \} + \mathscr T^{(2 \Theta_N)} \exp \left \{ 2i (\Theta_N)\right \}  + \mathscr T^{(2 \Theta_2)} \exp \left \{ 2i (\Theta_2)\right \}  +  \nonumber   \\  \mathscr T^{(\Theta_N + \Theta_2)} \exp \left \{ i (\Theta_N + \Theta_2) \right \} + \mathscr T^{(\Theta_N- \Theta_2)} \exp \left \{i(\Theta_N - \Theta_2)   \right \} + c.c.
\label{tensorproduktunotimesu2}
 \end{eqnarray}
  
%
\subsection{The amplitude parameter of the problem}
The expression for the tensor product $\mathbf u \otimes \mathbf u $, with $\mathbf u$ given by (\ref{ansatzelementarywave}) and $\Theta$ by (\ref{defnthetaproblm1}), shows 
that the inverse of the characteristic time for the temporal growth of the nonlinear term (which is exponential) is $(\delta \omega)_i$, the imaginary part of the change in the (complex) 
frequency of the wave packet, which is approximated here through superposition of two waves only. The 
quantity $(\delta \omega)_i$ for which we introduce the notation $\epsilon_A$, remains small when the departure from the surface of neutral stability in the space of the 
dispersion relation is small and is therefore suited to be regarded as the amplitude parameter sought, and introduce the notation $\epsilon_A =(\delta \omega)_i $. 

Following Stuart and Stewartson \cite{stuart1, stewartsonstuart}, we now hypothesise that the (complex) frequency $\delta \omega$ is analytic in terms of the set of variables in the dispersion relation. Under this hypothesis
$\delta \omega$ may be written in terms of the partial derivatives of its frequency w.r.t. the set of variables in the dispersion
relation evaluated on the surface of neutral stability. In particular, we linearise the dispersion relationship around the surface of neutral stability expressing the 
departure of the (complex) frequency from its value on this surface through: 
\begin{eqnarray}
 \omega - \omega_N = \delta \omega =        \left(\frac{\partial \omega}{\partial \lambda_x}\right)_N  (\lambda_x - \lambda_{xN}) +  \left(\frac{\partial \omega}{\partial Re} \right)_N (Re - Re_N) + \left(\frac{\partial \omega}{\partial S} \right)_N (S - S_N)
\label{defndeltaomegasprpoisflow}
\end{eqnarray}
All the partial derivatives above are to be understood as complex. The asymptotic expansions for $\mathbf u$ and $\mathscr D$ that are shortly to follow (in Section 4, there
(\ref{asymptexpmathbfumathscrd})) are to be understood as expansions in terms of the amplitude parameter $\epsilon_A$ defined through the relation,
\begin{equation}
\epsilon_A = (\delta \omega)_i 
\label{defnepsilona}
\end{equation}
Since we are restricting our attention to the side of the surface of neutral stability on which disturbances are growing, $\epsilon_A$ and $(\delta \omega)_i$ are positive. 

At this stage of the work, the amplitude parameter $\epsilon_A$ as defined in (\ref{defnepsilona}) is not yet known, since the eigenvalue has to be solved for in the initial step. The question of the relation between $\epsilon_A$ and the flow parameters governing the solution is therefore still open, and it will be dealt with in Sec. 5.2.2 that is to follow.
\subsection{The slow/long-scale variables}
We now introduce, in addition to the originally introduced variables, $(y, \varphi, x, t)$, now to be understood as {\bf fast/short-scale variables}, a set of new variables,
which we refer to as {\bf slow/long-scale variables}, for studying the effects of nonlinearity in the propagation of transition inducing disturbances in Spiral Poiseuille
Flow. The {\bf slow/long scale variables} are defined through the following relations: 
\begin{equation}
 \tau = \epsilon_A  t , \xi_x = \epsilon_A^{\frac{1}{2}} \left( x - \left(\frac{\partial \omega}{\partial \lambda_x} \right)_{N,r}  t \right),
\label{defntauxixswirl}
\end{equation}

 The introduction of the additional set of variables for a proper description of the changes of the variables over differing scales requires a reformulation of the problem involving rewriting the equation of motion, $\mathscr D \mathbf u = \mathscr N$, in terms of the {\bf extended set of variables},
$(t, y, \varphi, x, \tau, \xi_x)$. This reformulation calls for replacement of the derivatives w.r.t. $(t, y, x)$ in the original formulation through an asymptotic expansion of the derivatives according to the following rule, denoted by an arrow, $ \rightarrow $,
see eg. \cite{kevorkiancole, nayfeh}  :
%
%

\begin{eqnarray}
 \frac{\partial}{\partial x} \rightarrow \frac{\partial}{\partial x} + \epsilon_A ^{\frac{1}{2}} \frac{\partial}{\partial \xi_x} +  O(\epsilon_A)  \nonumber    \\
  \frac{\partial}{\partial t} \rightarrow \frac{\partial}{\partial t} + \epsilon_A ^{\frac{1}{2}} \left(- \left(\frac{\partial \omega}{\partial \lambda_x} \right)_{N,r} \frac{\partial}{\partial \xi_x} \right)  + \epsilon_A \frac{\partial}{\partial \tau} + o(\epsilon_A )
\label{dervexpswirl}
\end{eqnarray}

The meaning of these additionally introduced variables is evident on inspection of their definitions according to (\ref{defntauxixswirl}).

At this stage, the derivative $\left(\frac{\partial \omega}{\partial \lambda_x} \right)$
is yet unknown, but it will be seen in the following section that it is determinable from the
solution of the classical eigenvalue problem. Anticipating the result in advance, we merely state here that it follows from the {\bf solvability condition to the order} $O(\epsilon_A)$.

%
\section{Reformulation of the problem}  
The starting point for reformulation of the problem is the set of governing equations of motion, (\ref{governingequation}), written in terms of the {\bf extended set of variables},
$(t, y, \varphi, x, \tau, \xi_x )$. This reformulation calls for replacement of the derivatives w.r.t. $(t, y, \varphi, x)$ in the original formulation through
the new set $(t, y, \varphi, x, \tau, \xi_{x} )$ according to the scheme (\ref{dervexpswirl}) .

\subsection{The structure of the solution sought} 

In the study of transition in {\bf Spiral Poiseuille Flows}, an outstanding feature of the solution, which is a disturbance to the known basic flow, may be described in words as its structure that is
characterised by changes over variables with widely differing {\bf scales}. The scales are {\bf fast/short} as well as {\bf slow/long}. A parameter that may be regarded as small 
distinguishes between these scales. The solution will be sought in the form of an asymptotic expansion in terms of an {\bf amplitude parameter} $\epsilon_A$ that will be defined 
through the characteristic time for the {\bf temporal growth rate of a wave packet of small-amplitude disturbances}. This  characteristic time, when evaluated according to 
the classical theory for propagation of small-amplitude disturbances in the known basic flow, may be regarded as small in the neighbourhood of the surface of neutral stability. 
Since $\epsilon_A$ depends upon the {\bf mechanism of growth} of small-amplitude disturbances which may vary from point to point on the surface of neutral stability, there is a necessity for a  more
precise definition of $\epsilon_A$, and this will follow in the next subsection. For the present, it suffices to write the unknown, which is the velocity disturbance vector
$\mathbf u$, in the form of an asymptotic expansion in terms of the {\bf amplitude parameter} $\epsilon_A$. The {\bf amplitude parameter}, $\epsilon_A$ is different from, but related to
the {\bf flow parameters} of the problem, which are $(Re, S, \epsilon_R)$. We may regard this relation as a hypothesis for the present, whose fulfilment requires closer examination; it may differ from point
to point on the surface of neutral stability for the basic flow under consideration. This relation will become more transparent during the further course of this work. Since 
the rate of change of the unknown, temporal as well as spatial, depends upon variables of widely differing scales, it will be necessary to write the matrix differential 
operator, $\mathscr D $, and the vector on the right-hand side, $\mathscr N$, also as an asymptotic expansions in terms of the {\bf amplitude parameter} $\epsilon_A$. We 
therefore write: 
  
\begin{eqnarray}
 \mathbf  u \simeq \epsilon_A^{\frac{1}{2}} \mathbf u_1 + \epsilon_A \mathbf u_2 + \epsilon_A^{\frac{3}{2}} \mathbf u_3 + O(\epsilon_A^2)    \nonumber   \\  
 \mathscr D \simeq \mathscr D_1 + \epsilon_A^{\frac{1}{2}} \mathscr D_2 + \epsilon_A \mathscr D_3 + O(\epsilon_A^{\frac{3}{2}})    \nonumber   \\  
 \mathscr N \simeq \epsilon_A^{\frac{1}{2}} \mathscr N_1 + \epsilon_A \mathscr N_2 +  \epsilon_A^{\frac{3}{2}} \mathscr N_3  + O(\epsilon_A^2)
\label{asymptexpmathbfumathscrd}
\end{eqnarray}
The reason for writing the asymptotic expansions above in powers of $\epsilon^{\frac{1}{2}}$ will be clear on setting up the iteration scheme that is shortly to follow.
In the asymptotic expansion for $\mathscr D$ in (\ref{asymptexpmathbfumathscrd}), the leading term, $\mathscr D_1$ describes changes only over the {\bf fast/short scale} variables
whereas the further terms, $\mathscr D_2$ and $\mathscr D_3$, describe changes  in terms of both {\bf fast/short} and {\bf slow/long scale variables}. From the relation between the non-linear terms in $\mathscr N$ and the velocity disturbance, $\mathbf u$ ({\it vide} {\bf Appendix 1}), it follows that the vector $\mathscr N_1 \equiv \mathbf 0$. Furthermore, it will be seen necessary to take into
account {\bf the relation between the amplitude parameter and the departure of the flow parameters from their values on the surface of neutral stability}, which will be introduced in Sec. 5.1. The elements of these matrix differential operators which now depend upon the location of the point on the surface of neutral stability are also given in
{\bf Appendix 1}. For the present, it suffices to note that the governing equation, (\ref{governingequation}), together with the asymptotic expansions for $\mathbf u$, $\mathscr D$ and $\mathscr N$ according to (\ref{asymptexpmathbfumathscrd}) may formally be written as follows:
\begin{eqnarray}
 \mathscr D \mathbf u  \simeq \epsilon_A^{\frac{1}{2}} (\mathscr D_1 \mathbf u_1) +
 \epsilon_A (\mathscr D_1 \mathbf u_2 + \mathscr D_2 \mathbf u_1) +
 \epsilon_A^{\frac{3}{2}} \left (\mathscr D_1 \mathbf u_3 + \mathscr D_2 \mathbf u_2 + \mathscr D_3 \mathbf u_1 \right).
 \label{governingequationadditional}
\end{eqnarray}
Subsequent ordering of terms in the equation  (\ref{governingequationadditional}) according to the different integral powers of $\epsilon_A^{\frac{1}{2}}$ leads  to the following set of equations:
\begin{equation}
 O(\epsilon_A^{\frac{1}{2}}): \mathscr D_1 \mathbf u_1 = \mathbf 0,
\label{eqnforu1}
\end{equation}

\begin{equation}
 O(\epsilon_A):  \mathscr D_1 \mathbf u_2 = - \mathscr D_2 \mathbf u_1 + \mathscr N_2 = \mathscr {(RHS)}^{(1)},
\label{eqnforu2}
\end{equation}
 
\begin{equation}
 O(\epsilon_A^{\frac{3}{2}}):  \mathscr D_1 \mathbf u_3 = - \mathscr D_3 \mathbf u_1 - \mathscr D_2 \mathbf u_2 + \mathscr N_3 = \mathscr {(RHS)}^{(\frac{3}{2})},
\label{eqnforu3}
\end{equation}
where $(\mathscr D_1, \mathscr D_2, \mathscr D_3)$ are asymptotic expansions of $\mathscr D$, the matrix derivative operator, 
and $\mathscr N_2, \mathscr N_3$ asymptotic expansions of $\mathscr N$, the nonlinear term, in (\ref{doperatorswirl}). The right-hand sides, $(\mathscr {RHS})^{(1)}$ and $(\mathscr {RHS})^{(\frac{3}{2})}$ are column vectors.
For reasons that will soon be evident we will refer to the matrices $\mathscr D_2$ and $\mathscr D_3$ in (\ref{eqnforu2}, \ref{eqnforu3}) as derivatives 
w.r.t. the {\bf slow/long scale} variables of the problem. Precise forms for the column vectors $\mathscr N_2, \mathscr N_3$ for the flow example in question will be introduced 
later, in subsections of Sec. 5.

A hypothesis of pivotal significance in the present work will be seen to be the postulated relation between the departure of the flow parameters from their values on the
surface of neutral stability and the amplitude parameter $\epsilon_A$. The chain of thoughts underlying this hypothesis is essentially along the lines of Stewartson and Stuart. The classical solution, viewed from a
certain perspective, provides a logical foundation to its extension to handle the effect of the nonlinear terms neglected in the former, and hence may be regarded as the 
keystone for our present work. For this purpose, it suffices to note that the solution to the leading order, viz. $\mathbf u_1$, obeys the homogeneous equation
$\mathscr D \mathbf u = \mathbf 0$. The matrix differential operator $\mathscr D_1$ is formally the same as $\mathscr D$, with the elements given in {\bf Appendix 3}, both in {\it conventional variables} and in {\it Taylor variables}. The matrix differential operators $\mathscr D_2$ and $\mathscr D_3$ will be introduced later. The right-hand sides of these equations, $\mathscr {(RHS)}^{(1)}$ and
$\mathscr {(RHS)}^{(\frac{3}{2})}$, are column vectors wherein the superscripts set in paranthesis indicate the power to which the amplitude parameter $\epsilon_A$ is raised for
the immediately preceding expressions in (\ref{eqnforu2}) and (\ref{eqnforu3}) to hold. The components of the column vectors just introduced will be denoted by the subscripts 
$(OS, Sq, Co)$, chosen to indicate their source which are the {\bf Orr-Sommerfeld, Squire and Continuity} equations in that order. Later in the course of this work, in Sec. 5, 
the superscripts in paranthesis will be supplemented when necessary by symbols to indicate the periodicity w.r.t. the {\bf fast/short} scales.

Rewriting the equations in the extended set of variables for the particular flow problem in question, followed by substitution of the asymptotic expansions for
$\mathbf u$ and $\mathscr D$ according to (\ref{eqnforu1} - \ref{eqnforu3}) therein and subsequent ordering of the equations according to powers of $\epsilon_A^{\frac{1}{2}}$, leads to
the equations (\ref{eqnforu1}, \ref{eqnforu2}, \ref{eqnforu3}), which, although formally appear to be the same as in a conventional scheme, differ slightly in their meaning. Since this
difference in meaning is crucial for the further course of the present work, it is in order to draw attention to the same, so we briefly summarise these in words in the following:

\begin{itemize}
 \item The dependent variables, these are $(\mathbf u_1, \mathbf u_2, \mathbf u_3)$, should be regarded as functions of the {\bf extended set of independent variables} of the particular problem, 
although, on a cursory glance, this may seem to be unnecessary since the differential operator on the left-hand-side, $\mathscr D_1$, contains derivatives w.r.t. the original set of
independent variables only, which describe changes in the {\bf fast/short scale variables}. In particular, it is mandatory to adopt this viewpoint already for the leading term of the 
asymptotic expansion, $\mathbf u_1$. The task is then to derive the equation that describes the changes of $\mathbf u_1$ in the {\bf slow/long scale variables}. This, it will be seen, 
will call for fulfilment of {\bf solvability conditions} that assures consistency of the scheme of equations as a whole, (\ref{eqnforu1}, \ref{eqnforu2}, \ref{eqnforu3}), to be a {\bf rational approximation} in the sense of Van Dyke.

 \item For the flow example presently under consideration, a decisive step in reformulation of the problem along the guidelines of Stewartson and Stuart is relating the amplitude 
parameter $\epsilon_A$, which characterises the growth of a wave-packet of disturbances, to the departure of the flow parameters from their values on the surface of neutral stability. 
We postpone a discussion of this point to a later section (Sec. 5.2.2).
\end{itemize}
 
%


\subsection{The postulated relation between the amplitude parameter and the flow parametrers}
A point of importance hitherto left unaddressed concerns the relationship between the departure of the wavenumber and of the set of flow parameters, from their values at the chosen point on
the surface of neutral stability. This is a crucial step in the approach of Stewartson and Stuart since this sets the relation between the amplitude parameter in terms of which the
problem has been now reformulated to lead to a solvable scheme of rational approximation in the sense of Van Dyke, and the flow parameters defining the state of the flow. It is more or less understood that
attention has to be paid to {\bf consistency of the approximation} which may be viewed in terms of the amplitude parameter or the flow parameters. For the classical
channel-flow problem of their study, Stewartson and Stuart postulate a {\bf linear relationship} between the amplitude parameter characterizing the growth of disturbances and the
departure of the flow parameter from its critical value. This postulate, to be carried over conceptually to the present flow problems, requires an extension in respect of two questions
that do not arise in the bench-mark flows (the plane channel flow and the flow in the gap between a pair of concentric cylinders of which the inner is rotating),  but do in the class of flow problems presently addressed. Firstly it is regarding the vectorial nature of the wavenumber and, secondly the larger
number of flow parameters at present. It may be written in a form convenient to be represented by a matrix as follows:

\begin{eqnarray}
   \lambda_x - \lambda_{xN} = a_{xRe} (Re - Re_N) + a_{xS} (S - S_N)  \nonumber   \\
   n_{\varphi} - n_{\varphi N} = a_{\varphi Re} (Re - Re_N) + a_{\varphi S} (S -  S_N)
\label{defnmatrixaflow1}
\end{eqnarray}
In (\ref{defnmatrixaflow1}) the quantities $(a_{xRe}, a_{xS}, a_{\varphi Re}, a_{\varphi S})$ are all real numbers.
They may be expected to depend upon the location of the chosen point on the surface of neutral stability.

The relationship sought, which is between the amplitude parameter and the departure of the wavenumber vector and of the set of flow parameters from their values on the surface
of neutral stability, is obtained on combining
(\ref{defnmatrixaflow1}) with the expression to follow in  (\ref{epsilonaflow1}). It is:

\subsubsection*{Spiral Poiseuille Flow}
\begin{eqnarray}
   \epsilon_A = (R e - Re_N) \left \{ \left(\frac{\partial \omega}{\partial \lambda_x} \right)_{N,i} a_{xRe} +  \left(\frac{\partial \omega}{\partial Re} \right)_{N,i}\right \}   +    \nonumber   \\
( S -  S_N) \left \{ \left(\frac{\partial \omega}{\partial \lambda_x} \right)_{N,i} a_{xS} +  \left(\frac{\partial \omega}{\partial S} \right)_{N,i}\right \}
\label{relationepsilonaflow1}
\end{eqnarray}

The expression (\ref{relationepsilonaflow1}) shows that the {\bf amplitude parameter} $\epsilon_A$ may be regarded as a {\bf weighted sum} of the departures of the flow parameters
from their values on the surface of neutral stability. The expression (\ref{relationepsilonaflow1}) also shows that the {\bf sign of the amplitude parameter} is decided by a
{\bf suitably weighted algebraic sum of the magnitude and sign of the partial derivatives in the dispersion relationship} of the flow in question.

For setting up the equations to the orders $O(\epsilon_A)$ and $O(\epsilon_A^{\frac{3}{2}})$, i.e. (\ref{eqnforu2}, \ref{eqnforu3}), it is necessary to take note of the following points:
\begin{itemize}
 \item The left-hand sides of (\ref{eqnforu2}, \ref{eqnforu3}) are to be evaluated at {\bf the chosen point on the surface of neutral stabilty}, which we indicate by
adding the subscript $N$ to the operator $\mathscr D_1$, i.e. by writing the same as $\mathscr D_{1N}$. It may be recalled that, with this choice, $\mathscr D_{1N}$ is not
invertible.
 \item In the derivation of the right-hand sides it is necesssary to use, besides the asymptotic expansions for $\mathbf u$ and its derivatives, (\ref{asymptexpmathbfumathscrd}),
asymptotic expansions for the departure of the flow parameters from their values at the chosen point on the surface of neutral stability. A closer examination of these
expansions shows that they enter only the equation to the order $O(\epsilon_A^{\frac{3}{2}})$, not $O(\epsilon_A)$.
\end{itemize}
It will be necessary to specify order relations between $(\epsilon_{Re}, \epsilon_{S})$ and $\epsilon_A$. However, we wish to state here itself
that the specification of the order relations will have to be such that they enter only when formulating the equations at the order $O(\epsilon_A^{\frac{3}{2}})$ (Sec. 5.3), and not
in $O(\epsilon_A)$. For obtaining the asymptotic expansions for the {\bf dependent variables}, the {\bf derivatives} and the {\bf flow parameters} it is necessary to work with
order relations between the various parameters. In anticipation of the requirement for the present work these will be set as follows:

\begin{itemize}
 \item {\bf Spiral Poiseuille Flow}
  \begin{itemize}
   \item $\epsilon_{Re} = O(\epsilon_A): \epsilon_{Re} = d_{Re} \epsilon_A $
   \item $\epsilon_{S} = O(\epsilon_A): \epsilon_{S} = d_{S} \epsilon_A $
  \end{itemize}
\end{itemize}

It is appropriate at this juncture to draw attention to a salient difference between the bench-mark flows and the present work. The difference arises from the number of flow parameters influencing transition. Transition in the bench-mark flows is characterised by only one parameter, which is the Reynolds number, $Re$, for $S= 0$ and the Taylor number, $Ta$ for $S = \infty$. The extensive work that has been hitherto accumulated on the bench-mark flows has contributed to the present state of understanding of transition characteristics of these flows, one of which is on the shape of the neutrally stable curves on the wave-number vs. $Re$ or $Ta$ plot. The well-known {\bf U-shape}, of this plot  enables a minimum value, $Re_{min}$ or $Ta_{min}$ at which $\frac{d \omega_i}{d Re}$ or $\frac{d \omega_i}{d Ta}$ are $0$, to be ascertained. For studies of the effects of nonlinearity on transition in the bench-mark flows, the departure of the flow paramaters, $Re$ or $Ta$, from the values of $Re_{min}$ or $Ta_{min}$ respectively may be regarded as a suitable measure for the amplitude parameter. Carrying over such a concept to the present flow of interest, in which transition is influenced by two or more parameters instead of one, requires closer knowledge of the topography of the {\bf surface of neutral stability} than is presently available. A brief discussion of this point will follow in Section 6.

%
\section{The ansatz for solution of the problem to the different orders of the amplitude parameter $ \epsilon_A $ in (\ref{eqnforu1}, \ref{eqnforu2}, \ref{eqnforu3}). }
With the understanding that, after the governing equation for the flow problem $\mathscr D \mathbf u = \mathscr N$ has been recast in the {\bf relevant set of extended variables}, see (\ref{defntauxixswirl}), the next steps are:
\begin{itemize}
\item  defining the elements of the matrix differential operators in the terms of the asymptotic expansion,
\item laying out a suitable procedure for seeking a solution for $\mathbf u$ in the form of an asymptotic expansion (\ref{asymptexpmathbfumathscrd}) in terms in this set of extended variables, and
\item setting up the equations through definition of
the elements of the matrix differential operators occuring in (\ref{eqnforu1}, \ref{eqnforu2}, \ref{eqnforu3}).
\end{itemize}
Among these three, the problem of finding a solution to the
leading order is different in character from the others, making it appropriate to accord it a prerogative role in the present work. Summarising these differences in short, solving for $\mathbf u_1$ is an eigenvalue
problem that involves finding $\mathbf u_1$ and the (complex) eigenvalue $\omega$, whereas this is not so for $\mathbf u_2, \mathbf u_3$ and further terms in the set of  asymptotic expansion, if necessary. In other words,
obtaining a non-trivial
solution for $\mathbf u_1$ and the eigenvalue $\omega$ hinges upon the operator $\mathscr D_1$ being noninvertible. In contrast, with $\omega$ chosen as an eigenvalue and considered 
known, the equations for $\mathbf u_2, \mathbf u_3$ are {\bf inhomogeneous} with precisely the {\bf same noninvertible operator}, $\mathscr D_1$, appearing on the left-hand side! In other words, should one choose to convert the differential equations, (\ref{eqnforu1}, \ref{eqnforu2}, \ref{eqnforu3})  into their algebraic counterparts and solve the resulting matrix equations, the task would be one in which the leading equation would be, in Strang's notation ({\it vide} \cite{strang1, strang2, strang3}) of the type $\mathbf A \mathbf x = \lambda \mathbf x$ whereas the rest would belong to the type $ \mathbf A \mathbf x - \lambda \mathbf x = \mathbf b$. It is therefore
appropriate to deal with the leading order problem on its own separately, and the division into subsections reflects this viewpoint. In general, it may be stated that although the 
task of derivation of the operators to the different orders, viz. integral powers of $\epsilon_A^{\frac{1}{2}}$, is straightforward in principle, it involves rather heavy algebra.

\subsection{The ansatz for the solution to the leading order,  $O(\epsilon_A^{\frac{1}{2}})$ }
Up to this stage we have followed the same procedure for the entire range of the swirl parameter $S$, from $0$ to $\infty$. The closer study to follow unveils salient differences in the nature of the dependence of the {\bf Amplitude function} on the swirl parameter, and this has to be determined. So, from this stage onwards it is meaningful to make a sharper distinction between the cases of {\bf mild swirl}, $S \rightarrow 0 $, and {\bf strong swirl}, $S \rightarrow \infty$, in the basic flow. The common ground existing between the two is however strong enough to allow the two to be dealt with parallel to each other.

\subsubsection{The operator $\mathscr D_1$ to the order $O(\epsilon_A^{\frac{1}{2}})$ and the structural form of the solution}
To keep the treatment transparent, the elements of the operator $ \mathscr D_1$ which are the same as of $\mathscr D$, are given in {\bf Appendix 3} for the two cases of small and large $S$ separately; for the former in {\it conventional variables}, and for the latter in {\it Taylor variables}. The elements of the operator $\mathscr D_1$ are, cf. (\ref{governingequation}):
\begin{eqnarray}
\mathscr D_1 = \left(
\begin{array}{ccc}
 \mathscr D_{1OSr}   &  \mathscr D_{1OS \varphi}   &  \mathscr D_{1OSx}\\
 \mathscr D_{1Sqr}   &  \mathscr D_{1Sq \varphi}   &  \mathscr D_{1Sqx}\\
 \mathscr D_{1Cor}   &  \mathscr D_{1Co \varphi}   &  \mathscr D_{1Cox}
\end{array}
\right).
\label{operatormathscrd1}
\end{eqnarray} 
%
%
%

Since the equation to the leading order $O(\epsilon_A^{\frac{1}{2}})$, (\ref{eqnforu1}), is homogeneous, $\mathbf u_1$ admits nontrivial solutions only when the dispersion relation, $\omega = \omega(\lambda_x, n_{\varphi}; Re, S, \epsilon_R) $, is satisfied.

{\bf At this point we depart from conventional practice and set the solution, not just in the classical form of a
wave-like disturbance, (\ref{ansatzelementarywave}), but in an extended form in which the classical wave form is multiplied by a scalar function $B$ of the slow/long scale variables 
introduced for the problem}, (\ref{defntauxixswirl}) :    
\begin{equation}
 \mathbf u = B \mathbf a \exp (\imath \Theta) + B^\ast \mathbf a^\ast \exp (- \imath \Theta).
\label{ansatzextendedwave}
\end{equation}
It may be noted that this modification does not change the analytical structure of the ansatz w.r.t. the {\bf fast/short scale variables}. The definition of $\Theta$ remains the same as in (\ref{defnthetaproblm1}).

Fourier-superposition of waves of the type (\ref{ansatzextendedwave}) over the entire permissible range of wavenumbers leads to the following expressions for the solution of $\mathbf u_1$ : 
\begin{equation}
 \mathbf u_1 = \int_{\lambda_x = -\infty}^{\lambda_x = \infty} \sum_{n_{\varphi}= 0} ^{\infty} B\left(\tau, \xi_x; \lambda_x, n_{\varphi}, Re, S \right) \mathbf a (y; \lambda_x, n_{\varphi}, Re, S) \exp \left(\imath \Theta(\lambda_x, n_{\varphi}) \right) d \lambda_x + c.c.
\end{equation}

It is straightforward to verify that the expression (\ref{fouriersuperpswirl}) satisfies the equation for $\mathbf u_1$, (\ref{eqnforu1}), if, for each
wavenumber vector, $(\lambda_x, n_{\varphi})$, the (complex) frequency is given by the dispersion relation of the flow problem, (\ref{dispersionrelclassical}).
The dispersion relation, as stated earlier, follows from setting the operator $\mathscr D_1$ for the particular flow problem to be 
noninvertible. The expressions, (\ref{fouriersuperpswirl}), which are products of functions of the {\bf slow/long scale variables} 
and {\bf short/fast scale variables}, describe analytically the dependence of $\mathbf u_1$ on the {\bf fast/short scale variables} which is periodic in the space variable
exponential with a complex argument w.r.t. time. At this stage the dependence of $\mathbf u_1$ on the {\bf slow/long scale variables} is not specified. This, it will be seen, will be 
given by the {\bf amplitude evolution equation} that should satisfy the {\bf solvability condition}, for which closer examination of the equations to the higher order, (\ref{eqnforu2}, \ref{eqnforu3})
is necessary. 

For the further course of this work, it suffices to approximate the {\it Fourier} integrals for $\mathbf u_1$ in (\ref{fouriersuperpswirl}) through sums of two waves only, {\it viz.}
$\mathbf u_1^{(wave1)}$ and $\mathbf u_1^{(wave2)}$, of which we choose $\mathbf u_1^{(wave1)}$ to lie on the surface of neutral stability itself, and $\mathbf u_1^{(wave2)}$ only a short distance away in the parameter space of the dispersion relation, but on the growing side, the measure of the departure being the {\bf amplitude parameter}
$\epsilon_A$. We set $\mathbf u_1^{(wave1)}$ and $\mathbf u_1^{(wave2)}$ to be as follows: 
\begin{equation}
 \mathbf u_1^{(wave1)} = B_N  \mathbf a_N(y) \exp (\imath \Theta_N) + B_N^\ast \mathbf a_N^\ast (y) \exp (- \imath \Theta_N)    
\label{waveusub1supwave1} 
\end{equation}
\begin{eqnarray}
 \mathbf u_1^{(wave2)} = (B_N + \epsilon_A B_2)\left(\mathbf a_N + \epsilon_A \mathbf a_2 \right) \exp (\imath (\Theta_N + \epsilon_A \Theta_2)) +    \nonumber    \\ 
            (B_N + \epsilon_A B_2)^\ast \left(\mathbf a_N + \epsilon_A \mathbf a_2 \right)^\ast \exp (- \imath (\Theta_N + \epsilon_A \Theta_2 )^\ast) .
\label{waveusub1supwave2}
\end{eqnarray}
The difference in the phases of the two waves, $\epsilon_A \Theta_2$, is then given through the following expression:
\begin{eqnarray}
\epsilon_A \Theta_2 = (\lambda_x - \lambda_{xN}) x + (n_{\varphi} - n_{\varphi_N}) \varphi -   \nonumber    \\    
 \left \{\left(\frac{\partial \omega}{\partial \lambda_x} \right)_N (\lambda_x - \lambda_{xN})+  \left(\frac{\partial \omega}{\partial Re} \right)_N (Re - Re_N) + \left(\frac{\partial \omega}{\partial S} \right)_N (S - S_N) \right \}t
\label{epsilonatheta2flow1}
\end{eqnarray}

 An inspection of (\ref{epsilonatheta2flow1}) shows the relation between the amplitude parameter $\epsilon_A$, which is the inverse time-scale for growth of 
disturbances, see (\ref{defnepsilona}), and the departure of the flow parameters from their values on the surface of neutral stability. For the flow example presently under 
consideration, this is:
\begin{equation}
\epsilon_A =
\left \{\left(\frac{\partial \omega}{\partial \lambda_x} \right)_N (\lambda_x - \lambda_{xN})+ \left(\frac{\partial \omega}{\partial Re} \right)_N (Re - Re_N) + \left(\frac{\partial \omega}{\partial S} \right)_N (S - S_N) \right \}_i
\label{epsilonaflow1}
\end{equation}

The expression in (\ref{epsilonaflow1}) also shows that the amplitude parameter $\epsilon_A$ may be identified with the imaginary part of a weighted sum of the partial derivatives of 
the (complex) frequency $\omega$ w.r.t. axial wavenumber $ \lambda_x$ and the flow parameters, $(Re, S)$. The weights themselves, which are the departures of the
quantities from their values on the surface of neutral stability, are all real.

A comparison of the expressions in (\ref{fouriersuperpswirl}, \ref{waveusub1supwave1}, \ref{waveusub1supwave2}) with (\ref{usub1supwave1plususub1supwave2}) brings out the difference between 
Fourier-superposition in the 
classical approach and its extension to the problems presently on hand. We conclude this subsection by keeping ready for use in the sections to follow the expressions for the sum 
of the two waves and the tensor product corresponding to (\ref{usub1supwave1otimesusub1supwave2}), {\it viz.} $\mathbf u_1^{(wave1)} \otimes \mathbf u_1^{(wave2)} $, when $\mathbf u_1^{(wave1)}$ and
$ \mathbf u_1^{(wave2)}$ are given by the extended forms (\ref{waveusub1supwave1}, \ref{waveusub1supwave2}) respectively instead of the classical form (\ref{fouriersuperpswirl}).
\begin{eqnarray} 
 \mathbf u_1 = \mathbf u_1^{(wave1)} + \mathbf u_1^{(wave2)} \simeq 2 B_N \mathbf a_N \exp (\imath  \Theta_N) + 2 B_N^\ast \mathbf a_N^\ast \exp (- \imath  \Theta_N) +  \nonumber   \\ 
  \epsilon_A \left \{ B_N \mathbf a_2 + B_2 \mathbf a_N + B_N \mathbf a_N (\imath \Theta_2) \right \}  \exp (\imath \Theta_N) +  \nonumber  \\ 
  \epsilon_A \left \{ B_N^\ast \mathbf a_2^\ast + B_2^\ast \mathbf a_N^\ast + B_N^\ast \mathbf a_N^\ast (-\imath \Theta_2^\ast) \right \}  \exp (-\imath \Theta_N)
\label{usub1supwave1plususub1supwave2} 
\end{eqnarray}
\begin{eqnarray}
\mathbf u_1^{(wave 1)} \otimes \mathbf u_1^{(wave2)} \simeq B_N^2 \mathbf a_N \otimes \mathbf a_N \exp (2 \imath \Theta_N) + B_N^{\ast 2} \mathbf a_N^\ast \otimes \mathbf a_N^\ast  \exp (-2 \imath \Theta_N) +   \nonumber    \\  
B_N B_N^\ast \mathbf a_N \otimes \mathbf a_N^\ast +  B_N^\ast B_N \mathbf a_N^\ast \otimes \mathbf a_N +     \nonumber   \\   
\epsilon_A \left[ \left\{B_N B_2 \mathbf a_N \otimes \mathbf a_N + B_N B_N \mathbf a_N \otimes \mathbf a_2 + B_N B_N \mathbf a_N \otimes\mathbf a_N (\imath \Theta_2) \right \} \exp (2 \imath \Theta_N) \right] +  \nonumber      \\
\epsilon_A \left[ \left\{B_N^\ast B_2^\ast \mathbf a_N^\ast \otimes \mathbf a_N^\ast + B_N^\ast B_N^\ast \mathbf a_N^\ast \otimes \mathbf a_2^\ast + B_N^\ast B_N^\ast \mathbf a_N^\ast \otimes \mathbf a_N^\ast (-\imath \Theta_2^\ast) \right \} \exp (-2 \imath \Theta_N) \right] +  \nonumber      \\
\epsilon_A \left[ \left\{B_N^\ast B_2 \mathbf a_N^\ast \otimes \mathbf a_N + B_N^\ast B_N \mathbf a_N^\ast \otimes \mathbf a_2 + B_N^\ast B_N \mathbf a_N^\ast \otimes \mathbf a_N (\imath \Theta_2) \right \} \right] +  \nonumber      \\
\epsilon_A \left[ \left\{B_N B_2^\ast \mathbf a_N \otimes \mathbf a_N^\ast + B_N B_N^\ast \mathbf a_N \otimes \mathbf a_2^\ast + B_N B_N^\ast \mathbf a_N \otimes \mathbf a_N^\ast (-\imath \Theta_2^\ast) \right \} \right] 
\label{usub1supwave1otimesusub1supwave2}
\end{eqnarray}
Attention is drawn to the following salient points in the expression for $\mathbf u_1^{wave1} \otimes \mathbf u_1^{wave2} $ according to (\ref{usub1supwave1otimesusub1supwave2}):
\begin{itemize}
 \item To both the orders, $O(1)$ and $O(\epsilon_A)$, the expression is a sum of a nonperiodic term and terms of period $2 \Theta_N$.
 \item $\Theta_N$ is only real whereas $\Theta_2$ is complex, given by (\ref{epsilonatheta2flow1}) for the flow example of present interest.
 \item In this form, the product $\mathbf u_1^{(wave1)} \otimes \mathbf u_1^{(wave2)} $, (\ref{usub1supwave1otimesusub1supwave2}), exhibits, through the {\bf amplitude parameter}, $\epsilon_A$,  a parametric dependence on the departure of the wavenumber vector as well as of the flow parameters from their
values on the surface of neutral stability.  
\end{itemize}

%
\subsection{The form of the solution and the solvability condition of the equation to the order $ O(\epsilon_A)$, (\ref{eqnforu2})}
A cursory inspection of the equation to the order $O(\epsilon_A)$, (\ref{eqnforu2}), suggests that the structural form of the solution to $\mathbf u_2$ would be closely related to that of the 
right-hand side of this equation. It is therefore appropriate that gaining an insight into the structural form of the sum $- \mathscr D_2 \mathbf u_1 + \mathscr N_2$ precede the closer examination to follow in this subsection. It is also in order to recall in this context that the operator $\mathscr D_2$, vide (\ref{dervexpswirl}),
involves differentiation of $\mathbf u_1$ both w.r.t. the {\bf fast/short scale} variables and the {\bf slow/long scale variables} of the problem, and $\mathscr N_2$ only differentiation 
w.r.t. the {\bf fast/short scale} variables of the tensor product $\mathbf u_1 \otimes \mathbf u_1$, vide (\ref{usub1supwave1otimesusub1supwave2}).

 Since the tensor $\mathbf u_1 \otimes \mathbf u_1$, as (\ref{usub1supwave1otimesusub1supwave2}) shows, is a sum of tensors of period $2\Theta_N$ and a non-periodic tensor, the vector $\mathscr N_2$ ( see for definition and derivation (\ref{nonlinelementsswirl}) in {\bf Appendix 1}), may also be written as a sum of a contribution of period $2 \Theta_N$ and a non-periodic contribution as
follows: 
\begin{equation}
\mathscr N_2 = \mathscr N_2^{(2)}\exp \left \{2i(\Theta_N)\right\} + \mathscr N_2^{(2 *)}\exp \left\{-2i(\Theta_N)\right\} + \mathscr N_2^{(0)},  
\label{tensorsumn2}
\end{equation}
where the number enclosed in parantheses in the superscript indicates the periodicity w.r.t. the {\bf fast/short scale} variables $(x, n_\varphi, t)$ .

Since $\mathbf u_1$ according to (\ref{ansatzextendedwave}) has only one period $\Theta_N$ and the operator $\mathscr D_2$, as will be seen shortly, does not introduce any further period
in the {\bf fast/short scale} variables, the structural form of the solution of $\mathbf u_2$ w.r.t. the {\bf fast/short scale} variables may be set as follows: 
\begin{eqnarray}
 \mathbf u_2 = \mathbf u_2^{(0)} + \mathbf u_2^{(1)} \exp (i \Theta_N) + \mathbf u_2^{(1)\ast} \exp (- i \Theta_N)   \nonumber   \\ 
                                 + \mathbf u_2^{(2)} \exp (2 i \Theta_N) + \mathbf u_2^{(2)\ast} \exp (- 2 i \Theta_N) + \mathbf u_{2hom}.
\label{summenansatzforu2}
\end{eqnarray}
In the ansatz for $\mathbf u_2$ above, (\ref{summenansatzforu2}), $\mathbf u_{2hom}$ is a solution of the {\bf homogeneous} equation corresponding to (\ref{eqnforu2}) and again, as in (\ref{tensorsumn2}), the number enclosed in parantheses
in the superscript indicates the periodicity w.r.t. the {\bf fast/short scale} variables $(x, n_\varphi, t)$. The expression (\ref{summenansatzforu2})
sheds light on the analytical structure of $\mathbf u_2$ as a sum of a nonperiodic part $\mathbf u_2^{(0)}$, and of quantities of periodicity $\Theta_N$ and $2 \Theta_N$, and of a homogeneous part. 
%

\subsubsection{The equations to be satisfied by the contributions to $\mathbf u_2$ in (\ref{summenansatzforu2})}
It is in order at this juncture to redraw attention to the consequences of the salient difference between the ansatz for $\mathbf u_1$ in the present study, (\ref{ansatzextendedwave}), 
and in the conventional form for $ \mathscr L \mathbf a = \omega \mathscr M \mathbf a $ . The form of the solution for $\mathbf u_2$, (\ref{summenansatzforu2}), when substituted into  (\ref{eqnforu2}), gives rise to
equations for the coefficients of the different periods. In the conventional approach in which the ansatz for $\mathbf u_1$, (\ref{ansatzextendedwave}), contains no dependence on
{\bf slow/long scale variables}, these equations are ordinary differential equations in $y$ containing, in addition to the flow parameters of the problem, the wavenumber components 
also as parameters. In contrast, in the present study, $( \mathbf u_2^{(0)}, \mathbf u_2^{(1)}, \mathbf u_2^{(2)})$ should be regarded as functions of $y$ and the {\bf slow/long scale variables} 
of the problem, hence the equations for these quantities as partial differential equations. The nature of dependence of $( \mathbf u_2^{(0)}, \mathbf u_2^{(1)}, \mathbf u_2^{(2)})$ 
on $y$ and on the {\bf slow/long scale variables} will emerge on closer study of the right-hand sides of the respective equations. Here we merely state the finding in advance which is as follows:  
%
\begin{eqnarray}
 \mathbf u_2^{(0)} = B_N B_N^\ast \mathbf w_2^{(0)}        \nonumber     \\ 
 \mathbf u_2^{(1)} = \frac{\partial (2 B_N)}{\partial \xi_x} \mathbf w_2^{(1)};                                       \nonumber     \\  
 \mathbf u_2^{(2)} = 4 B_N^2 \mathbf w_2^{(2)} ;  \mathbf u_2^{(2)\ast} = 4 B_N^{^\ast2} \mathbf w_2^{(2)\ast}.  
\label{usub2sup012flowex1}
\end{eqnarray} 

Both the similarities and the differences in structure between the contributions of different period are worthy of special note. They are: 
\begin{itemize} 
 \item There is a structural similarity between the three insofar as that they are products of functions of {\bf slow/long scale variables} and of $y$. The precise form of this 
functional dependence follows from examination of the respective right-hand sides that is to follow. Particular attention is drawn to the  
derivatives of $B_N$ w.r.t. the {\bf slow/long scale variables} of the problem, viz. $\xi_x$ for {\bf Spiral Poiseuille Flow}.
These occur only in the contribution of period $\Theta_N$, which is $\mathbf u_2^{(1)}$.
In contrast to the contribution of period $\Theta_N$, those of period $0$ and $2 \Theta_N$ are nonlinear in $B_N$.       
 \item Among the three in (\ref{usub2sup012flowex1}), setting up the ordinary differential equation for $\mathbf w_2^{(2)}$, which is associated with with $\mathbf u_2^{(2)}$, is most
straightforward, needing only substitution of the appropriate form into the equation. It follows from substitution of (\ref{summenansatzforu2}) in (\ref{eqnforu2}) that 
$\mathbf u_2^{(2)}$ should satisfy the following equation: 
\begin{equation}
 \mathscr D_1 \left( \mathbf u_2^{(2)} \exp (2 \imath \Theta_N) \right) = \mathscr N_2^{(2)}\exp \left \{2i(\Theta_N)\right\} ,  
\label{eqnforusub2sup2}
\end{equation} 
which in turn leads to the analytical forms for $\mathbf u_2^{(2)}$ in (\ref{usub2sup012flowex1}).  The left-hand side of (\ref{eqnforusub2sup2}) is indeed invertible, the period being 
$2 \Theta_N$, not $\Theta_N$.
\item The form of $\mathbf u_2^{(1)}$ follows from substitution of (\ref{summenansatzforu2}) in (\ref{eqnforu2}). One obtains for $\mathbf u_2^{(1)}$ the equation: 
\begin{equation}
 \mathscr D_1 \left( \mathbf u_2^{(1)} \exp (\imath \Theta_N) \right) = - \mathscr D_2 \left( 2 B_N \mathbf a_N \exp (\imath \Theta_N) \right),  
\label{eqnforusub2sup1}
\end{equation} 
with the left-hand side being noninvertible due to the appearance of the period $\Theta_N$ in $\mathscr D_1 \left( \mathbf u_2^{(1)} \exp (\imath \Theta_N) \right) $. Solution of 
(\ref{eqnforusub2sup1}) requires closer knowledge of the structure of the operator $\mathscr D_2$ (Sec. 5.2.2 and 5.2.3) and setting up a solvability condition (Sec. 5.2.4), which 
together go to determine the slow/long-scale dependent coefficient/coefficients of $\mathbf u_2^{(1)}$ in (\ref{usub2sup012flowex1}), 
viz. $\frac{\partial (2 B_N)}{\partial \xi_x} $ for the {\bf Spiral Poiseuille Flow}.

\item Some caution is warranted when carrying over a similar reasoning for setting up the equation for $\mathbf u_2^{(0)}$. The reason is that although the period-free nonlinear contribution is 
obtainable from (\ref{usub2sup012flowex1}), the step of arriving at the elements of $\mathscr N_2^{(0)}$ from $\mathscr N$ according to (\ref{nonlinelementsswirl}) involves
differentiaion w.r.t. the {\bf short-scale variable}, $x$,  which leads to a trivial forcing term on the right-hand side. But, as will be seen in the Section dealing with
the order $O(\epsilon_A^{\frac{3}{2}})$, a non-trivial $\mathbf u_2^{(0)}$ does enter the result,  so $\mathbf u_2^{(0)}$ has to be obtained from the equations of motion themselves, 
prior to any differentiation of the same to eliminate pressure. The situation is not diffrent in content in the bench-mark flows for this problem, see  \cite{stuart1, stewartsonstuart}.

\end{itemize}
%
%
\subsubsection{Elements of the differential operator $\mathscr D_2$ in (\ref{asymptexpmathbfumathscrd}), $O(\epsilon_A ^{\frac{1}{2}})$}
Elements of $\mathscr D_2$ for {\bf Spiral Poiseuille Flow in both connventional and Taylor variables} are listed in {\bf Appendix 3}.
The elements of $\mathscr D_2$ on the right-hand side of (\ref{eqnforu2}) follow on substitution of the appropriate expansions, i.e.
(\ref{asymptexpmathbfumathscrd}) in the expressions for its element operators which are listed in {\bf Appendix 3}.  It may be recalled that
$\mathscr D_2$ operating on $\mathbf u_1$ gives rise to terms of the order $O(\epsilon_A)$. 

%
\subsubsection{The solvability condition for the equation to the order $O(\epsilon_A)$}
The object pursued in this subsection is to examine the solution of (\ref{eqnforu2}) when the right-hand side is only $- \mathscr D_2 \mathbf u_1$ with $\mathbf u_1$ given by 
$\mathbf u_1 = 2 B_N \mathbf a_N (y) \exp (\imath \Theta_N) + 2 B_N^\ast \mathbf a_N^\ast (y) \exp (- i \Theta_N)$, vide (\ref{ansatzextendedwave}, \ref{waveusub1supwave1}, \ref{waveusub1supwave2}) in
Sec. 5.1.2. Such a focus of attention is a reflection of the reasoning that the left-hand side of (\ref{eqnforu2}), being of period $\Theta _N$, is then noninvertible and solutions for
$\mathbf u_2$ are possible only when the right-hand side fulfils a {\bf solvability condition} which leads to the analytical form for $\mathbf u_2^{(1)}$ in (\ref{usub2sup012flowex1}). 

In order to derive this solvability condition we substitute on the left-hand side $\mathbf u_2 =  \mathbf u_2^{(1)} \exp (\imath \Theta_N) + \mathbf u_2^{(1)\ast} \exp (- \imath \Theta_N) $ 
and evaluate $- \mathscr D_2 \mathbf u_1$ as just described with $\mathscr D_2$, given in {\bf Appendix 3} through the expressions  (\ref{mathscrd2ossqcontconven},  \ref{mathscrd2ossqconttaylor}). It should be noted at this juncture that there is a difference between the  expressions for basic flows of mild and strong swirl, see (\ref{usub2sup012flowex1}). Here
$\mathbf u_2^{(1)} = \frac{\partial (2 B_N)}{\partial \xi_x} \mathbf w_2^{(1)} (y) $.  We denote the left-hand side of (\ref{eqnforu2}) with
this substitution as the column vector $\left( (LHS)_{OS}^{(1, \Theta_N)}, (LHS)_{Sq}^{(1, \Theta_N)}, (LHS)_{Co}^{(1, \Theta_N)} \right)^T$, 
where the subscripts $(OS, Sq, Co)$ in that order stand for the Orr-Sommerfeld, Squire and Continuity equations respectively, and the superscripts in paranthesis $(1, \Theta_N)$ 
denote the order to the power of $\epsilon_A$ and the period, viz. $\Theta_N$ respectively. 

 Substitution of the expression for $\mathbf u_2$ into $\mathscr D_1 \mathbf u_2$ for the flow example presently under
 consideration, followed by carrying out the differentiations w.r.t. the {\bf fast/short scale variables} which are $(t, x, \varphi, y)$, yields the expressions for the elements of the column  vector sought, {\it viz.}
  $\left( (LHS)_{OS}^{(1, \Theta_N)}, (LHS)_{Sq}^{(1, \Theta_N)}, (LHS)_{Co}^{(1, \Theta_N)} \right)^T$. It should be noted here that the analytical expressions for the element operators are not the same for the cases of {\bf mild} and {\bf strong swirl}.
\subsubsection*{The left-hand sides}

\subsubsection*{For the case of strong swirl, $S \rightarrow \infty$, in Taylor variables:}
The expressions on the left-hand side are obtainable from the analytical expressions for the terms in the equations (\ref{taylorvariabmathscrlosrstoinfty} thru \ref{taylorvariabmathscrmosrvarphixstoinfty}).

For the column vector
$\left( (LHS)_{OS}^{(1, \Theta_N)}, (LHS)_{Sq}^{(1, \Theta_N)}, (LHS)_{Co}^{(1, \Theta_N)} \right)^T$ we then have:
%
\begin{eqnarray}
   (LHS)_{OS}^{(1, \Theta_N)} = - \imath \omega_N\left( \left(\lambda_{xN}^2 - \frac{d^2}{dy^2} \right)  w_{2r}^{(1)} \right) - Ta_N(1-\hat y) \lambda_{xN}^2  w_{2 \varphi}^ {(1)}+ \left( \lambda_{xN}^4 + \frac{d^4}{d \hat y^4} - 2 \lambda_{xN}^2 \frac{d^2}{dy^2} \right) w_{2r}^{(1)} +   \nonumber  \\  -  \left( \frac{(1- \hat y)}{2} \right) \left( (\imath n_{\varphi N}) \frac{d^2}{dy^2}- \lambda_{xN}^2 (\imath n_{\varphi N})  \right) w_{2r}^{(1)}  +   \frac{\sqrt{Ta_N}}{S_N \sqrt {\epsilon_R}} \left( -2(\imath \lambda_{xN}) - (1- \hat y^2)(\imath \lambda_{xN})(\frac{d^2}{dy^2} - \lambda_{xN}^2) \right) w_{2r}^{(1)} ,   \nonumber   \\
(LHS)_{Sq}^{(1, \Theta_N)} = - i \omega_N\left( i \lambda_{xN}  w_{2 \varphi}^{(1)} \right) - \frac{1}{2}(\imath \lambda_{xN}) w_{2r}^{(1)} \left(\imath \lambda_{xN}^3 + (\imath \lambda_{xN}) \frac{d^2}{d \hat y^2} \right) +  \frac{\sqrt {Ta_N}}{S_N \sqrt {\epsilon_R}} \left( (1- \hat y^2) \lambda_{xN}^2 \right) ,  \nonumber   \\
(LHS)_{Co}^{(1, \Theta_N)} = 0   .
\label{lhssup1thetanswirl}
\end{eqnarray}
\subsubsection*{For the case of mild swirl, $S \rightarrow 0$ , in conventional variables:}
The analytical expressions in conventional variables are rather lengthy and are obtainable from the expressions in $\left( (LHS)_{OS}^{(1, \Theta_N)}, (LHS)_{Sq}^{(1, \Theta_N)}, (LHS)_{Co}^{(1, \Theta_N)} \right)^T$ in the equations  (\ref{mathscrlosrsto0} thru \ref{mathscrthemssto0}) in {\bf Appendix 3}. They should be evaluated on the surface of neutral stability at which the imaginary part of the frequency, $\omega_i$ is $=0$.
It may be noted that whereas for the case of mild swirl there are three parameters, $(Re, S, \epsilon_R )$, occuring on their own, for the case of strong swirl there are only two, viz. $(Ta, S \sqrt{\epsilon_R})$, with $S$ and $\epsilon_R$ occuring only in the combination $S \sqrt{\epsilon_R}$.

\subsubsection*{The right-hand sides}
The contribution of periodicity $\Theta_N$ on the right-hand side of (\ref{eqnforu2}), which is  $- \mathscr D_2 \mathbf u_1$, may be evaluated by substitution of 
$\mathbf u_1 = 2 B_N \mathbf a_N (y) \exp (\imath \Theta_N) $, in conventional variables in  (\ref{mathscrd2ossqcontconven}), and in Taylor variables in ( \ref{mathscrd2ossqconttaylor}). We
denote this contribution by the column vector $ \left( \mathbf {RHS}^{(1, \Theta_N)} \right)= \left( (RHS)_{OS}^{(1, \Theta_N)}, (RHS)_{Sq}^{(1, \Theta_N)}, (RHS)_{Co}^{(1, \Theta_N)} \right)^T$ where again
the subscripts $(OS, Sq, Co)$ stand in that order for the Orr-Sommerfeld, Squire and Continuity equations respectively, and the superscripts in paranthesis $(1, \Theta_N)$ 
denote the order to the power of $\epsilon_A$ and the period, {\it viz.} $\Theta_N$ respectively. There arises no contribution from $\mathscr N_2$ to terms of period $\Theta_N$ at the
level $O(\epsilon_A)$. We then have for the right-hand sides: 

\subsubsection*{For the case of strong swirl, $S \rightarrow \infty $, in Taylor variables}

\begin{eqnarray}
(RHS)_{OS}^{(1, \Theta_N)} =
 \left[2 (- \imath \omega_N) (\imath \lambda_{xN}) A_{Nr} \frac{\partial (2 B_N)}{\partial \xi_x} - \left(\frac{\partial \omega}{\partial \lambda_x} \right)_N \left((\imath \lambda_{xN})^2 A_{Nr} +  \frac{d^2 A_{Nr}}{dy^2}\right) \frac{\partial (2 B_N)}{\partial \xi_x}\right]    \nonumber   \\
- \left[4 (\imath \lambda_{xN})^3 A_{Nr} \frac{\partial (2 B_N)}{\partial \xi_x} + 2 (\imath \lambda_{xN}) \frac{\partial (2 B_N)}{\partial \xi_x} \frac{d^2 A_{Nr}}{d y^2}   \right]   \nonumber   \\ 
- \frac{\sqrt {Ta_N}}{ S_{N} \sqrt{\epsilon_R}} \left[ -2 A_{Nr} \frac{\partial (2 B_N)}{\partial \xi_x} - (1 - y^2) \left( \frac{\partial (2 B_N)}{\partial \xi_x} \frac{d^2 A_{Nr}}{d y^2} + 3 (\imath \lambda_{xN})^2 A_{Nr} \frac{\partial (2 B_N)}{\partial \xi_x}\right)  \right]     \nonumber   \\
- Ta_N (1-y) \left[2 (\imath \lambda_{xN}) A_{N \varphi} \frac{\partial (2 B_N)}{\partial \xi_x} \right]   ; \nonumber   \\  
(RHS)_{Sq}^{(1, \Theta_N)} =  \frac{1}{2} A_{Nr} \frac{\partial (2 B_N)}{\partial \xi_x} + \left[ (- \imath \omega_N) A_{N \varphi} \frac{\partial (2 B_N)}{\partial \xi_x} - \left(\frac{\partial \omega}{\partial \lambda_x} \right)_N (\imath \lambda_{xN})A_{N \varphi} \frac{\partial (2 B_N)}{\partial \xi_x} \right]  \nonumber   \\
+ \left[ 3 (\imath \lambda_{xN})^2 A_{N \varphi} \frac{\partial (2 B_N)}{\partial \xi_x} \right] + \frac{\sqrt {Ta_N}}{ S_{N} \sqrt{\epsilon_R}} \left[ (1-y^2) 2 (\imath \lambda_{xN}) A_{N \varphi} \frac{\partial (2 B_N)}{\partial \xi_x} \right] ;  \nonumber   \\
(RHS)_{Co}^{(1, \Theta_N)} = - A_{Nx} \frac{\partial (2 B_N)}{\partial \xi_x} .
\end{eqnarray}
We slightly rearrange the above expression as follows to render the product form involving multiplication with $\frac{\partial (2 B_N)}{\partial \xi_x}$ more transparent:

\begin{eqnarray}
(RHS)_{OS}^{(1, \Theta_N)} = + \left[2 (- \imath \omega_N) (\imath \lambda_{xN}) A_{Nr}  - \left(\frac{\partial \omega}{\partial \lambda_x} \right)_N \left((\imath \lambda_{xN})^2 A_{Nr} +  \frac{d^2 A_{Nr}}{dy^2}\right) \right] \frac{\partial (2 B_N)}{\partial \xi_x}   \nonumber   \\       
- \left[4 (\imath \lambda_{xN})^3 A_{Nr}  + 2 (\imath \lambda_{xN})  \frac{d^2 A_{Nr}}{d y^2}   \right]\frac{\partial (2 B_N)}{\partial \xi_x}   \nonumber   \\ 
- \frac{\sqrt {Ta_N}}{ S_{N}\sqrt{\epsilon_R}} \left[ -2 A_{Nr} - (1 - y^2) \left(  \frac{d^2 A_{Nr}}{d y^2} + 3 (\imath \lambda_{xN})^2 A_{Nr} \right)  \right] \frac{\partial (2 B_N)}{\partial \xi_x}    \nonumber   \\
- Ta_N (1-y) \left[2 (\imath \lambda_{xN}) A_{N \varphi}\right] \frac{\partial (2 B_N)}{\partial \xi_x}   ; \nonumber   \\  
(RHS)_{Sq}^{(1, \Theta_N)} = + \frac{1}{2} A_{Nr} \frac{\partial (2 B_N)}{\partial \xi_x} + \left[ (- \imath \omega_N) A_{N \varphi} - \left(\frac{\partial \omega}{\partial \lambda_x} \right)_N (\imath \lambda_{xN}) A_{N \varphi} \right] \frac{\partial (2 B_N)}{\partial \xi_x} \nonumber   \\
+ \left[ 3 (\imath \lambda_{xN})^2 A_{N \varphi} \right] \frac{\partial (2 B_N)}{\partial \xi_x} + \frac{\sqrt {Ta_N}}{ S_{N}\sqrt{\epsilon_R}} \left[ (1-y^2) 2 (\imath \lambda_{xN}) A_{N \varphi} \right] \frac{\partial (2 B_N)}{\partial \xi_x}  ;  \nonumber   \\
(RHS)_{Co}^{(1, \Theta_N)} = - A_{Nx} \frac{\partial (2 B_N)}{\partial \xi_x} .
\label{rhssup1thetanswirl}
\end{eqnarray}

\subsubsection*{For the case of mild swirl, $S \rightarrow 0 $, in conventional variables}
Substitution of $\mathbf u_1 = 2 B_N \mathbf a_N (y) \exp (\imath \Theta_N)$ in (\ref{mathscrd2ossqcontconven}) leads to the column vector \\
$\left( (RHS)_{OS}^{(1, \Theta_N)}, (RHS)_{Sq}^{(1, \Theta_N)}, (RHS)_{Co}^{(1, \Theta_N)} \right)^T$. This will not be written out explicitly here since, despite its length, writing this out does not involve complications. It has the same structure as the corresponding expression for strong swirl given above, except for the appearence of $(Re, S, \epsilon_R)$ as the set of parameters instead of $ (Ta,S \sqrt {\epsilon_R})$ .

The equation (\ref{rhssup1thetanswirl})  and its corresponding equation for mild swirl  bring out a feature in the analytical structure of the column vector
$\mathbf {(RHS)}^{(1, \Theta_N)}$. It is that $\mathbf {(RHS)}^{(1, \Theta_N)}$ with the elements $ \left( (RHS)_{OS}^{(1, \Theta_N)}, (RHS)_{Sq}^{(1, \Theta_N)}, (RHS)_{Co}^{(1, \Theta_N)} \right)^T$ assumes importance for the further course of the present work. It is its product form which is as follows:
\begin{equation}
 \mathbf {(RHS)}^{(1, \Theta_N)} = \frac{\partial (2 B_N)}{\partial \xi_x} \left( \mathscr F_N + \left(\frac{\partial \omega}{\partial \lambda_x} \right)_N \mathscr G_N \right),
\label{vectorrhssup1thetanswirl}
\end{equation}
where the column vectors $(\mathscr F_N, \mathscr G_N)$ with $\mathscr F_N = (\mathscr F_{NOS}, \mathscr F_{NSq}, \mathscr F_{NCo})^T $ and $ \mathscr G_N = (\mathscr G_{NOS}, \mathscr G_{NSq}, \mathscr G_{NCo})^T$ are defined for the case of strong swirl as follows:
\begin{eqnarray}
\mathscr F_{NOS} =  + \left[2 (- \imath \omega_N) (\imath \lambda_{xN}) A_{Nr}  \right]
- \left[4 (\imath \lambda_{xN})^3 A_{Nr}  + 2 (\imath \lambda_{xN})  \frac{d^2 A_{Nr}}{d y^2}   \right]  \nonumber   \\ 
- \frac{\sqrt {Ta_N}}{ S_{N}\sqrt{\epsilon_R}} \left[ -2 A_{Nr} - (1 - y^2) \left(  \frac{d^2 A_{Nr}}{d y^2} + 3 (\imath \lambda_{xN})^2 A_{Nr} \right)  \right]
- Ta_N (1-y) \left[2 (\imath \lambda_{xN}) A_{N \varphi}\right]   ; \nonumber   \\  
\mathscr F_{NSq} =  + \frac{1}{2} A_{Nr} + \left[ (- \imath \omega_N) A_{N \varphi} \right]
+ \left[ 3 (\imath \lambda_{xN})^2 A_{N \varphi} \right] + \frac{\sqrt {Ta_N}}{ S_{N}\sqrt{\epsilon_R}} \left[ (1-y^2) 2 (\imath \lambda_{xN}) A_{N \varphi} \right]  ;  \nonumber   \\                                                              \nonumber   \\
\mathscr F_{NCo} =  - A_{Nx}                                                          \nonumber  \\
\mathscr G_N =  \left( \mathscr G_{NOS}, \mathscr G_{NSq}, \mathscr G_{NCo} \right)^T =      \left( - \left( (\imath \lambda_{xN})^2 A_{Nr} + \frac{d^2 A_{Nr}}{dy^2} \right), - \left( (\imath \lambda_{xN}) A_{N \varphi}\right), 0 \right)^T
\label{defnmathscrfngnswirl}
\end{eqnarray}
The appearance of $\frac{\partial (2 B_N)}{\partial \xi_x}$ as a coefficient in (\ref{vectorrhssup1thetanswirl}) is a confirmation of the product form set for $\mathbf u_2^{(1)}$ in 
(\ref{usub2sup012flowex1}). We may also draw attention to a feature of the analytical structure of this expression. It is aproduct of two functions, one of the {\bf slow/long-scale variables} and the other of the {\bf fast/short-scale variables}. The latter is evaluable from the solution of the {\bf classical problem in the initial iteration step}. It has to be now considered known.

The steps from (\ref{vectorrhssup1thetanswirl}, \ref{defnmathscrfngnswirl}) to the formulation of the solvability conditions that is sought are simple and straightforward. They are as detailed below:
\begin{enumerate}
 \item  Scalar multiplication of $(\mathscr F_N, \mathscr G_N) $ with the adjoint vector $\mathbf a^+$, vide {\bf The Adjoint Problem} in {\bf Appendix 2},  ;
 \item  Integration of the scalar product between the limits $-1 \le y \le +1$; 
 \item  Setting the integrated product $= 0$, which is an expression of the requirement that for the inhomogenius equation (\ref{eqnforu2}) to be solvable the column vector \\ $\left( (RHS)_{OS}^{(1,\Theta_N)},  (RHS)_{Sq}^{(1,\Theta_N)}, (RHS)_{Co}^{(1, \Theta_N)}) \right)$ should be  aligned with the column vector on the left-hand side of (\ref{eqnforu2}), see  \cite{strang1, strang2, strang3}.
\end{enumerate} 
The steps outlined above yield an expression for $\left( \frac{\partial \omega}{\partial \lambda_x}  \right)_N$ in terms of a ratio of integrals of products of the elements of the 
column vector $\mathbf a_N$, the conjugate of its adjoint $\mathbf a_N^{+\ast}$, and their derivatives, the subscript $N$ behind $(\mathbf a_N, \mathbf a_N^{+\ast})$ being employed to denote
that the quantity is to be evaluated at the chosen point on the surface of neutral stability. The result can be written in a compact vector notation as follows:
\begin{equation}
 \left( \frac{\partial \omega}{\partial \lambda_x} \right)_N =  \frac{\int_{-1}^{1} (\mathscr F_N \cdot \mathbf a_N^{+ \ast}) dy }{\int_{-1}^{1} (\mathscr G_N \cdot \mathbf a_N^{+ \ast}) dy}
\label{groupvelexprswirl}
\end{equation}
The real part of $ \left(  \frac{\partial \omega}{\partial \lambda_x}\right)_N$ may be considered as representing the {\bf group velocity} of the wave-packet, which is at this stage evaluable from the solution of the classical eigenvalue problem obtained in the initial step, see (\ref{defnmathscrfngnswirl}, \ref{groupvelexprswirl}).The signed imaginary part of $ \left(  \frac{\partial \omega}{\partial \lambda_x}\right)_N$ may be regarded as a measure of the temporal growth/decay of the wave-packet.

The imaginary part of the (complex) {\bf group velocity}, $\left( \frac{\partial \omega}{\partial \lambda_x} \right)_N$ represents, in a system that is {\bf moving with the wave-packet}, the {\bf temporal growth or decay of the Amplitude Function $B$}. It may be noted that this quantity is expressed in the {\bf slow scale variable$ ~ \tau$}.

It is in order to recall at this stage that the (complex) {\bf group velocity} that is evaluated through (\ref{groupvelexprswirl}) depends, besides upon the set of parameters $(Re, S, \epsilon_R)$,  upon the chosen point on the surface of neutral stability. A meaningful criterion for the choice of this point for purposes of the present work is a subject of discussion that will be taken up in Sec. 6.
%
%

%
\subsubsection{The set of solvable equations for the inhomogeneous problem to the order $O(\epsilon_A)$ for terms of period $\Theta_N$ on the right-hand-side}
The step to follow at setting up of the solvability condition for the contribution to $\mathbf u_2$ from the right-hand side of periodicity $\Theta_N$, which is (\ref{groupvelexprswirl}) 
for the flow example of present interest is to obtain solutions for the set of equations for the vector $\mathbf w_2^{(1)}$ in (\ref{usub2sup012flowex1}). To this end it is necessary to evaluate
the column vector $\left( (RHS)_{OS}^{(1, \Theta_N)}, (RHS)_{Sq}^{(1, \Theta_N)}, (RHS)_{Co}^{(1, \Theta_N)} \right)_{\perp}^T $ that is perpendicular to
$\left( (RHS)_{OS}^{(1, \Theta_N)}, (RHS)_{Sq}^{(1, \Theta_N)}, (RHS)_{Co}^{(1, \Theta_N)} \right)^T$. Here, to facilitate reading, we merely summarise the equations for solution of 
the vector $\mathbf w_2^{(1)} = (w_{2r}^{(1)}, w_{2\varphi}^{(1)}, w_{2x}^{(1)})^{T} $:

\subsubsection*{Spiral Poiseuille Flow, equation determining solvablity condition}
\begin{eqnarray}
- i \omega_N\left( \left(\lambda_{xN}^2 - \frac{d^2}{dy^2} \right)  w_{2r}^{(1)} \right) + \left( \mathscr L_{OSrN}  w_{2r}^{(1)} + \mathscr L_{OS\varphi N}  w_{2\varphi}^{(1)} + \mathscr L_{OSxN}  w_{2x}^{(1)} \right) = (RHS)_{OS}^{(1, \Theta_N)}  ,  \nonumber   \\
- i \omega_N\left( i \lambda_{xN}  w_{2 \varphi}^{(1)} \right) + \left( \mathscr L_{SqrN}  w_{2r}^{(1)} + \mathscr L_{Sq\varphi N}  w_{2 \varphi}^{(1)} + \mathscr L_{SqxN}  w_{2x}^{(1)} \right) = (RHS)_{Sq}^{(1, \Theta_N)}  ,  \nonumber   \\
 \left( \mathscr L_{CorN}  w_{2r}^{(1)} + \mathscr L_{Co\varphi N}  w_{2 \varphi}^{(1)} + \mathscr L_{CoxN}  w_{2x}^{(1)} \right) = (RHS)_{Co}^{(1, \Theta_N)},    
\end{eqnarray}
where the column vector $\mathbf {(RHS)}^{(1, \Theta_N)}$ is given by (\ref{vectorrhssup1thetanswirl}).

%
%
%

\subsection{The form of the solution and the solvability condition of the equation to the order $O(\epsilon_A^{\frac{3}{2}})$, (\ref{eqnforu3})}
The structural form of the solution for $\mathbf u_3$, i.e. the form corresponding to (\ref{summenansatzforu2}) for $\mathbf u_2$, follows from that of the right-hand side of 
(\ref{eqnforu3}), together with $(\mathbf u_1, \mathbf u_2)$ according to (\ref{eqnforu2}, \ref{summenansatzforu2}). A closer inspection of the right-hand side of (\ref{eqnforu3})
shows that this is a sum of quantities of {\bf four} periods: $0, \Theta_N, 2 \Theta_N$, and $ 3 \Theta_N$. Therefore the ansatz for $\mathbf u_3$ corresponding to that for $\mathbf u_2$ according to (\ref{summenansatzforu2}) is:
\begin{eqnarray}
 \mathbf u_3 = \mathbf u_3^{(0)} +
 \mathbf u_3^{(1)} \exp(\imath \Theta_N) + c.c. +
 \mathbf u_3^{(2)} \exp(\imath 2 \Theta_N) + c.c +
 \mathbf u_3^{(3)} \exp(\imath 3 \Theta_N) + c.c
 \label{summenansatzforu3}
\end{eqnarray}

Of these four, it is only the contribution of period $\Theta_N$ that has an influence on the {\bf amplitude evolution} for the problem, so it suffices to restrict attention to this part only for establishing the solvability condition to the order $O(\epsilon_A^{\frac{3}{2}})$,  whose derivation is to follow.

In order to establish the solvability condition for (\ref{eqnforu3}), we note that the source of terms of period $\Theta_N$ in the solution of (\ref{eqnforu3}) for $\mathbf u_3$ are $ - \mathscr D_3 \mathbf u_1$, $ - \mathscr D_2 \mathbf u_2 $ and $  \mathscr N_3 $.
Of these, it is straightforward to obtain the contribution from $ \mathscr D_3 \mathbf u_1$ since $\mathbf u_1$ has only one period, see (\ref{usub1supwave1plususub1supwave2}), and $\mathscr D_3$ does
not introduce any further period in the {\bf fast/short scale} variables. 
For the contribution of period $\Theta_N$ arising from $  \mathscr D_2 \mathbf u_2 $, since $\mathbf u_2$ is a sum of contributions of different periods, see (\ref{summenansatzforu2}),
it is only necessary to attend to the relevant part of $\mathbf u_2$, which is $\mathbf u_2^{(1)}$. The contributions may be written as follows:
\begin{equation}
-\mathscr D_3 (\mathbf u_1) = - \mathscr D_3 \left( 2 B_N \mathbf a_N \exp (\imath  \Theta_N) + 2 B_N^\ast \mathbf a_N^\ast \exp (- \imath  \Theta_N) \right)
\end{equation}
\begin{equation}
- \mathscr D_2 (\mathbf u_2) = - \mathscr D_2 \left(\mathbf u_2^{(1)} \exp (i \Theta_N) + \mathbf u_2^{(1)\ast} \exp (- i \Theta_N)  \right)
\end{equation}
%
%

\subsubsection{Evaluation of $ \mathscr D_2 \mathbf u_2$ and $  \mathscr D_3 \mathbf u_1 $}
The elements of the matrix differential operators $\mathscr D_2$ and $\mathscr D_3$ on the right-hand side of the equation for $\mathbf u_3$, (\ref{eqnforu3}), follow on substitution of the appropriate 
expansions giving due regard to the asymptotic expansions of the difference in flow parameters from their values at the chosen point on the surface of neutral stability.
It may be recalled that $\mathscr D_2$ operating on $\mathbf u_2$ and $\mathscr D_3$
operating on $\mathbf u_1$ give rise to terms of the order 
$O(\epsilon_A^{\frac{3}{2}})$. The matrix differential operator $\mathscr D_2$ has already been given for the flow of present interest in {\bf Appendix 3}, there  (\ref{mathscrd2ossqcontconven}, \ref{mathscrd2ossqconttaylor}) . Elements of the matrix
differential operator $\mathscr D_3$ will be given in ths subsection following the evaluation of $\mathscr D_2 \mathbf u_2$.

\subsubsection*{a) Evaluation of $  \mathscr D_2 \mathbf u_2 $}
The matrix differential operator $\mathscr D_2$ for the flow example of present interest has been derived and kept ready in Section 5.2.3. These are equations (\ref{groupvelexprswirl})
wherein the group velocity on the surface of neutral stability, $\left(\frac{\partial \omega}{\partial \lambda_x}\right)_N$ is now to be regarded known and given by (\ref{groupvelexprswirl}).
In evaluating $\mathscr D_2 \mathbf u_2$, for present purposes it is only necessary to consider the part contributed by 
$(\mathbf u_2^{(1)} \exp (i \Theta_N) + \mathbf u_2^{(1)\ast} \exp (- i \Theta_N)) $ in the ansatz for $ \mathbf u_2$, (\ref{summenansatzforu2}), since the others are of a period different from $\Theta_N$. The
part $\mathscr D_2 (\mathbf u_2^{(1)} \exp (i \Theta_N))$ itself is to be evaluated for $\mathbf u_2^{(1)} $ given by (\ref{usub2sup012flowex1}). 
The components of the column vector $\mathscr D_2 (\mathbf u_2^{(1)} \exp (i \Theta_N))$ may then be written in the following form that illuminates their analytical structure:
\begin{eqnarray}
 \left( \mathscr D_2 (\mathbf u_2^{(1)} \exp (i \Theta_N)) \right)_{OS} =   \frac{\partial ^2}{\partial \xi_x^2} (2 B_N) \mathscr P_{OS} (y) \exp (\imath \Theta_N)  \nonumber   \\
 \left( \mathscr D_2 (\mathbf u_2^{(1)} \exp (i \Theta_N)) \right)_{Sq} =  \frac{\partial ^2}{\partial \xi_x^2} (2 B_N) \mathscr P_{Sq} (y) \exp (\imath \Theta_N)                 \nonumber   \\
 \left( \mathscr D_2 (\mathbf u_2^{(1)} \exp (i \Theta_N)) \right)_{Co} =   \frac{\partial ^2}{\partial \xi_x^2} (2 B_N) \mathscr P_{Co} (y) \exp (\imath \Theta_N).
\label{mathscrd2mathbfu2swirl}
\end{eqnarray}
In (\ref{mathscrd2mathbfu2swirl}) above, the functions, $(\mathscr P_{OS}, \mathscr P_{Sq}, \mathscr P_{Co})$, depend only upon $y$, and are, for the case of strong swirl, as follows:
\begin{eqnarray}
\mathscr P_{OS} (y) = - \left(2 (- \imath \omega_N) (\imath \lambda_{xN}) - \left(\frac{\partial \omega}{\partial \lambda_x} \right)_N \left((\imath \lambda_{xN})^2 + \frac{d}{dy} \right) \right) w_{2r}^{(1)} +
\left( 4 (\imath \lambda_{xN})^3 + 2 (\imath \lambda_{xN}) \frac{d^2}{d y^2} \right) w_{2r}^{(1)} +   \nonumber  \\  
\frac{\sqrt {Ta_N}}{S_{N} \sqrt{\epsilon_R}} \left(-2 - (1-y^2) \left(\frac{d^2}{dy^2} + 3 (\imath \lambda_{xN})^2 \frac{d}{dy} \right) \right) w_{2r}^{(1)} + Ta_N (1-y) \left( 2 (\imath \lambda_{xN}) \right) w_{2 \varphi}^{(1)} ;  \nonumber   \\
\mathscr P_{Sq} (y) = - \frac{1}{2} w_{2r}^{(1)} - \left( (- \imath \omega_N) - \left( \frac{\partial \omega}{\partial \lambda_x} \right)_N (\imath \lambda_{xN}) \right) w_{2 \varphi}^{(1)} - \left( 3 (\imath \lambda_{xN})^2 + \frac{d^2}{dy^2} \right) w_{2 \varphi}^{(1)} -   \nonumber   \\
 \frac{\sqrt {Ta_N}}{ S_{N} \sqrt {\epsilon_R}} \left( (1-y^2)2 (\imath \lambda_{xN}) \right) w_{2 \varphi}^{(1)} ;  \nonumber   \\
\mathscr P_{Co} (y) = w_{2x}^{(1)}
\label{swirlpospsqpco}
\end{eqnarray}

In the case of mild swirl the components of the column vector $ \mathscr D_2(\mathbf u_2^{(1)} \exp (\imath \Theta_N))$ retain the form as in (\ref{mathscrd2mathbfu2swirl}, \ref{swirlpospsqpco}), in particular regarding the dependence on group velocity. However,  the expressions for $ \mathscr P $, although differing in detail regarding the appearance of parameters ($Re, S, \epsilon_R$) instead of ($Ta, S \sqrt{\epsilon_R} $),  exhibit a similar product structure as in (\ref{swirlpospsqpco}).

It is convenient at this stage to introduce the matrix $\mathscr P$ and the column vector $\mathbf w_2^{(1)}$ defined through
\begin{eqnarray}
 \mathscr P = \left(
\begin{array}{ccc}
  \mathscr P_{OSr}  & \mathscr P_{OS \varphi}  & \mathscr P_{OSx} \\
  \mathscr P_{Sqr}  & \mathscr P_{Sq \varphi}  & \mathscr P_{Sqx} \\
  \mathscr P_{Cor}  & \mathscr P_{Co \varphi}  & \mathscr P_{Cox}
\end{array}
\right), 
\mathbf w_2^{(1)} = \left( 
\begin{array}{c}
 w_{2r}^{(1)}  \\ 
 w_{2 \varphi}^{(1)}  \\ 
 w_{2x}^{(1)}
\end{array}
\right). 
\label{defnswirlmatrixp}
\end{eqnarray}
The elements of the matrix $\mathscr P$ follow from a straightforward comparison of its definition, (\ref{defnswirlmatrixp}) with the expressions in (\ref{swirlpospsqpco}).
The expressions in (\ref{mathscrd2mathbfu2swirl}, \ref{swirlpospsqpco}) may then be combined and the contribution of period $\Theta_N$ to the column vector $\mathscr D_2 \bf u_2$ 
may then be written in a compact analytical form which is as follows:  
\begin{equation}
 \mathscr D_2 \mathbf u_2 =  \frac{\partial ^2 (2 B_N)}{\partial \xi_x^2} \mathscr P \mathbf w_{2}^{(1)}   \exp \left(\imath \Theta_N \right) + cc
\label{mathscrd2mathbfusub2swirl}
\end{equation}

%

%
\subsubsection*{b) Evaluation of $ \mathscr D_3 \mathbf u_1$ }
For evaluating $ \mathscr D_3 \mathbf u_1$ , it is only necessary to substitute the ansatz for $\mathbf u_1$ according to (\ref{usub1supwave1plususub1supwave2}) in the expressions for the elements of
$\mathscr D_3$ which are as below. 

\subsubsection*{Spiral Poiseuille Flow with strong swirl}

\begin{eqnarray}
 \mathscr D_{3OSr} = - \left \{ \frac{\partial^3}{\partial t \partial \xi_x^2} -  \left( \frac{\partial \omega}{\partial \lambda_x} \right)_N \cdot 2 \frac{\partial ^3}{\partial x \partial \xi_x^2} + \frac{\partial ^3 }{\partial \tau \partial x^2} \right \}  - \frac{\partial ^3}{\partial \tau \partial y^2} +  \nonumber  \\
\left(6 \frac{\partial ^4}{\partial x^2 \partial \xi_x^2} \right) +  2 \frac{\partial ^4}{\partial \xi_x^2 \partial y^2} - 2  \frac{\sqrt{Ta_N}}{ S_{N} \sqrt{\epsilon_R}} \left(\frac{1}{2} d_{Ta} - d_{S} \right) \frac{\partial}{\partial x}   \nonumber  \\
   - \frac{\sqrt{Ta_N}}{ S_{N}\sqrt{\epsilon_R}} (1-y^2) \left(\frac{1}{2} d_{Ta} - d_{S} \right) \frac{\partial ^3}{\partial x \partial y^2}  \nonumber  \\
   - \frac{\sqrt{Ta_N}}{ S_{N} \sqrt{\epsilon_R}} (1-y^2) \left(\left(\frac{1}{2} d_{Ta} - d_{S} \right) \frac{\partial ^3}{\partial x^3} + 3 \frac{\partial ^3}{\partial x \partial \xi_x^2}\right) ; \nonumber   \\
 \mathscr D_{3OS \varphi} =  Ta_N (1-y)\left(\left(\frac{1}{2} d_{Ta} - d_{S} \right) \frac{\partial ^2}{\partial x^2} + \frac{\partial ^2}{\partial \xi_x^2} \right) ;    \nonumber  \\
 \mathscr D_{3OSx} = 0  ;  \nonumber    \\
 \mathscr D_{3Sqr} = - \frac{\sqrt {Ta_N}}{ S_{N}\sqrt{\epsilon_R}} (1-y^2)  \left( (\frac{1}{2} d_{Ta} - d_{S} )\frac{\partial ^2 }{\partial x^2} + \frac{\partial ^2}{\partial \xi_x^2} \right)  ;   \nonumber    \\
 \mathscr D_{3Sq \varphi} = 0  ;  \nonumber    \\
 \mathscr D_{3Sqx} = 0         ;  \nonumber    \\
 \mathscr D_{3Cor} = 0         ;  \nonumber    \\
 \mathscr D_{3Co \varphi} = 0  ;  \nonumber    \\
 \mathscr D_{3Cox} = 0 .
\label{elementsd3swirl}
\end{eqnarray}

Using $\mathbf u_1 \simeq 2 B_N \mathbf a_N \exp (\imath  \Theta_N) + 2 B_N^\ast \mathbf a_N^\ast \exp (- \imath  \Theta_N) + O(\epsilon_A) $, cf. (\ref{usub1supwave1plususub1supwave2}), in
(\ref{elementsd3swirl}) leads to expressions from which the elements of the vector resulting from the matrix differential operator $\mathscr D_3$ in 
(\ref{elementsd3swirl}) operating on $\mathbf u_1$ in the flow example in consideration may readily be obtained. For this purpose it is convenient to rewrite the elements of  
the differential operator $\mathscr D_3$ in terms of another differential operator, $\mathcal D $, in a form that takes the {\bf fast/short scale} structure of $\mathbf u_1$ according to (\ref{usub1supwave1plususub1supwave2}) into account,
so that the structural dependence of the quantity on the {\bf slow/long scale} and the {\bf fast/short scale} variables is exhibited more explicitly. This is equivalent to writing the column 
vector $\mathscr D_3 \mathbf u_1$ in terms of a matrix operator $ \mathcal { D_3}$, operating on the column vector of the eigenfunctions, $(A_{Nr}, A_{N \varphi}, A_{Nx})^T$
regarded as a diagonal matrix $\mathbf {a_N}$ and the temporal function $exp (\imath \Theta_N)$. We write:
\begin{equation}
\mathscr D_3 \mathbf u_1 = \Sigma_{rows} \left( \mathcal {D_3} \mathbf {a_N} \right) \exp (\imath \Theta_N)
\end{equation}
For the specific example of $  \mathcal D_{3OSr} u_{1r}$, taken here for illustrative purposes, it is
\begin{equation}
 \mathcal D_{3OSr} u_{1r} = \left(  \mathcal D_{3OSr} A_{Nr} \exp (\imath \Theta_N) +   \mathcal D_{3OSr}^\ast A_{Nr} ^\ast \exp (-\imath \Theta_N)  \right)
\label{bardsub3osr}
\end{equation}
and similarly for operations with the other elements, distinguishing between the two using the same indices as for $\mathscr D_3$ but with another scriptform ($\mathcal D$). The elements, again as an example for the case of {\bf Spiral Poiseuille Flow with strong swirl} may then bewritten as follows where the arrow $\rightarrow$ denotes the correspondence:


\begin{eqnarray}
\mathscr D_{3OSr} \rightarrow \mathcal D_{3OSr} = \nonumber  \\ - \left \{ (- \imath \omega_N) \frac{\partial ^2 (2 B_N)}{\partial \xi_x^2} -  \left( \frac{\partial \omega}{\partial \lambda_{x}} \right)_N \cdot 2 (\imath \lambda_{xN})\frac{\partial ^2(2 B_N)}{ \partial \xi_x^2} + (\imath \lambda_{xN})^2 \frac{\partial (2 B_N) }{\partial \tau } \right \}  - \frac{\partial (2 B_N)}{\partial \tau} \frac{d ^2}{ d y^2} +  \nonumber  \\
\left(6 (\imath \lambda_{xN})^2 \frac{\partial ^2 (2 B_N)}{ \partial \xi_x^2} \right) +  2 \frac{\partial ^2 (2 B_N)}{\partial \xi_x^2}  \frac{d^2}{d y^2} - 2  \frac{\sqrt{Ta_N}}{ S_{N}\sqrt{\epsilon_R}} \left(\frac{1}{2} d_{Ta} - d_{S} \right) (\imath \lambda_{xN})(2 B_N)   \nonumber  \\
   - \frac{\sqrt{Ta_N}}{ S_{N}\sqrt{\epsilon_R}} (1-y^2) \left(\left(\frac{1}{2} d_{Ta} - d_{S} \right) (\imath \lambda_{xN})^3 (2 B_N) + 3 (\imath \lambda_{xN}) \frac{\partial ^2 (2 B_N)}{\partial \xi_x^2}\right) ; \nonumber   \\
\mathscr D_{3OS \varphi} \rightarrow \mathcal D_{3OS \varphi} =  Ta_N (1-y)\left(\left(\frac{1}{2} d_{Ta} - d_{S} \right) (\imath \lambda_{xN})^2 (2 B_N) + \frac{\partial ^2 (2 B_N)}{\partial \xi_x^2} \right) ;    \nonumber  \\
\mathscr D_{3OSx} \rightarrow \mathcal D_{3OSx} = 0  ;  \nonumber    \\
\mathscr D_{3Sqr} \rightarrow \mathcal D_{3Sqr} = - \frac{\sqrt {Ta_N}}{ S_{N}\sqrt{\epsilon_R}} (1-y^2)  \left( (\frac{1}{2} d_{Ta} - d_{S} )(\imath \lambda_{xN})^2 (2 B_N) + \frac{\partial ^2 (2 B_N)}{\partial \xi_x^2} \right)  ;   \nonumber    \\
\mathscr D_{3Sq \varphi} \rightarrow \mathcal D_{3Sq \varphi} = 0  ;  \nonumber    \\
\mathscr D_{3Sqx} \rightarrow \mathcal D_{3Sqx} = 0         ;  \nonumber    \\
\mathscr D_{3Cor} \rightarrow \mathcal D_{3Cor} = 0         ;  \nonumber    \\
\mathscr D_{3Co \varphi} \rightarrow \mathcal D_{3Co \varphi} = 0  ;  \nonumber    \\
\mathscr D_{3Cox} \rightarrow \mathcal D_{3Cox} = 0 .
\label{elemented3swirl}
\end{eqnarray}

When setting up the amplitude evolution equations for the present flow example, it is purposeful to render the nature of dependence of  $ \mathscr D_3 \mathbf u_1$ on 
the {\bf slow/long scale variables} more transparent. To this end we rewrite the equations (\ref{elemented3swirl}) in a form arranged according to the 
coefficients of $(2 B_N)$ and its derivatives that appear in the particular flow example. They are as follows: 
\subsubsection*{Spiral Poiseuille Flow with strong swirl}
\begin{eqnarray}
\mathcal D_{3OSr} = \frac{\partial (2 B_N)}{\partial \tau} \left [ +\lambda_{xN}^2 - \frac{d^2}{dy^2} \right ] +   \nonumber   \\
\frac{\partial ^2 (2 B_N)}{\partial \xi_x^2} \left[\imath \omega_N + \left(\frac{\partial \omega}{\partial \lambda_x} \right)_N \cdot 2 (\imath \lambda_{xN}) - 6 \lambda_{xN}^2 + 2 \frac{d^2}{dy^2} - \frac{\sqrt{Ta_N}}{ S_{N}\sqrt{\epsilon_R}} (1 - y^2) \cdot 3 (\imath \lambda_{xN}) \right] +   \nonumber   \\
(2 B_N) \frac{\sqrt {Ta_N}}{ S_{N}\sqrt{\epsilon_R}} \left [ -2 \left( \frac{1}{2} d_{Ta} - d_{S}\right) (\imath \lambda_{xN}) + (1-y^2) \left(\frac{1}{2} d_{Ta} - d_{S} \right) \imath \lambda_{xN}^3  \right ] ;  \nonumber   \\
\mathcal D_{3OS \varphi} =  \frac{\partial ^2 (2 B_N)}{\partial \xi_x^2} \left[ Ta_N (1-y)\right] + (2 B_N)  Ta_N (1-y)\left( -\left(\frac{1}{2} d_{Ta} - d_{S} \right) \lambda_{xN}^2  \right) ;    \nonumber  \\
\mathcal D_{3OSx} = 0  ;  \nonumber    \\
\mathcal D_{3Sqr} = \frac{\partial^2 (2B_N)}{\partial \xi_x^2} \left[ - \frac{\sqrt {Ta_N}}{ S_{N}\sqrt{\epsilon_R}} (1-y^2) \right] + (2 B_N) \left[ - \frac{\sqrt {Ta_N}}{ S_{N}\sqrt{\epsilon_R}} (1-y^2) \right]   \left(- (\frac{1}{2} d_{Ta} - d_{S} ) \lambda_{xN}^2  \right)  ;   \nonumber    \\
\mathcal D_{3Sq \varphi} = 0  ;  \nonumber    \\
\mathcal D_{3Sqx} = 0         ;  \nonumber    \\
\mathcal D_{3Cor} = 0         ;  \nonumber    \\
\mathcal D_{3Co \varphi} = 0  ;  \nonumber    \\
\mathcal D_{3Cox} = 0 .
\label{elemented3swirlrewritten}
\end{eqnarray}
A closer inspection of (\ref{elemented3swirlrewritten}) shows that all the elements of the matrix operator $ \mathcal D_3$ are sums of products of the amplitude function,$(2 B_N)$, or its
derivatives w.r.t. the {\bf slow/long scale} variables, $\frac{\partial (2 B_N)}{\partial \tau}$, $\frac{\partial ^2 (2 B_N)}{\partial \xi_x^2}$, with differential operators w.r.t. 
$y$. The matrix differential operator $ \mathcal D_3$ may be written in the form of a sum of products as follows to illuminate this structure.

\begin{eqnarray}
\mathcal D_3 = \left(
\begin{array}{ccc}
 \mathcal D_{3OSr} & \mathcal D_{3OS \varphi} & \mathcal D_{3OSx} \\
 \mathcal D_{3Sqr} & \mathcal D_{3Sq \varphi} & \mathcal D_{3Sqx} \\
 \mathcal D_{3Cor} & \mathcal D_{3Co \varphi} & \mathcal D_{3Cox} \\
\end{array} 
\right)  \nonumber   \\   
= \frac{\partial (2 B_N)}{\partial \tau}  \mathcal D_3^{(\tau)} + \frac{\partial ^2 (2 B_N)}{\partial \xi_x^2} \mathcal D_3^{(\xi_x \xi_x)} + (2 B_N) \mathcal D_3^{(0)}
\label{productformofd3flow1}
\end{eqnarray}
The definition of the matrices $ \mathcal D_3^{(\tau)}$, $ \mathcal D_3^{(\xi_x \xi_x)} $ and $ \mathcal D_3^{(0)} $ follow directly from a comparison of (\ref{productformofd3flow1})
with (\ref{elemented3swirlrewritten}), with the superscripts set in paranthesis, $(\tau)$, $(\xi_x \xi_x)$, $(0)$ indicating the {\bf slow/long scale} variables w.r.t. which derivatives of 
$(2 B_N)$ are the coefficients appearing in the product sum.

The superscript to $\mathcal D_3$ set in paranthesis, $(\tau)$, $(\xi_x \xi_x)$ and $(0)$ indicate in that order that the matrix differential
operators are coefficients of the derivatives of the amplitude function, $(2 B_N)$, w.r.t. to the {\bf slow/long scale variables} in the paranthesis. 

\subsubsection{Evaluation of $\mathscr N_3$}
For evaluating $\mathscr N_3$ it is useful to keep ready that part of the tensor product $(\mathbf u_1 \otimes \mathbf u_2 + \mathbf u_2 \otimes \mathbf u_1)$ that has a period
$\Theta_N$ and which we therefore denote as $\left(\mathbf u_1 \otimes \mathbf u_2 + \mathbf u_2 \otimes \mathbf u_1 \right)^{(\Theta_N)}  $. This is as follows where $\mathbf u_N$ is an abbreviation for $2 B_N \mathbf a_N$  in (\ref{waveusub1supwave1}), and $\mathbf u_2^{(0)}$ and $\mathbf u_2^{(2)}$ are related to $\mathbf u_2$ through
(\ref{summenansatzforu2}). We may write:
\begin{eqnarray}
 \left(\mathbf u_1 \otimes \mathbf u_2 + \mathbf u_2 \otimes \mathbf u_1 \right)^{(\Theta_N)} =            \nonumber   \\
\left(\mathbf u_N \otimes \mathbf u_2^{(0)} + \mathbf u_2^{(0)} \otimes \mathbf u_N  + \mathbf u_N^\ast \otimes \mathbf u_2^{(2 \Theta_N)} + \mathbf u_2^{(2 \Theta_N)} \otimes \mathbf u_N^\ast  \right) \exp (\imath \Theta_N)   +   \nonumber   \\
\left(\mathbf u_N^\ast \otimes \mathbf u_2^{(0)} + \mathbf u_2^{(0)} \otimes \mathbf u_N^\ast  + \mathbf u_N \otimes \mathbf u_2^{(2 \Theta_N)\ast} + \mathbf u_2^{(2 \Theta_N)\ast} \otimes \mathbf u_N  \right) \exp (- \imath \Theta_N)  ,
\label{u1otimesu2supthetan}
\end{eqnarray}
where the superscript set in paranthesis, $(\Theta_N)$, on the left-hand side of the above equation indicates that it is only the contribution of period $\Theta_N$ that has been written.

The expressions for $\mathbf u_1$, and $\mathbf u_2^{(0)}, \mathbf u_2^{(2)}$ according to (\ref{waveusub1supwave1} ) and (\ref{usub2sup012flowex1}) substituted
into (\ref{u1otimesu2supthetan}) leads to the following expression which clearly brings out the analytical structure of $\left(\mathbf u_1 \otimes \mathbf u_2 + \mathbf u_2 \otimes \mathbf u_1 \right)^{(\Theta_N)}$
as a product of functions of {\bf fast/short scale variables} and {\bf slow/long scale variables}:
\begin{eqnarray}
 \left(\mathbf u_1 \otimes \mathbf u_2 + \mathbf u_2 \otimes \mathbf u_1 \right)^{(\Theta_N)} =            \nonumber   \\
\left( 2 B_N (B_N B_N^\ast)(\mathbf a_N \otimes \mathbf w_2^{(0)} + \mathbf w_2^{(0)} \otimes \mathbf a_N)  + 2 B_N^\ast (4 B_N^2) (\mathbf a_N^\ast \otimes \mathbf w_2^{(2 )} + \mathbf w_2^{(2 )} \otimes \mathbf a_N^\ast)  \right) \exp (\imath \Theta_N)   +   \nonumber   \\
\left( 2 B_N^\ast (B_N^\ast B_N)(\mathbf a_N^\ast \otimes \mathbf w_2^{(0)} + \mathbf w_2^{(0)} \otimes \mathbf a_N^\ast)  + 2 B_N (4 B_N^{\ast 2})(\mathbf a_N \otimes \mathbf w_2^{(2 )\ast} + \mathbf w_2^{(2 )\ast} \otimes \mathbf a_N)  \right) \exp (- \imath \Theta_N)  =  \nonumber   \\
B_N (B_N B_N^\ast) \left( 2 (\mathbf a_N \otimes \mathbf w_2^{(0)} + \mathbf w_2^{(0)} \otimes \mathbf a_N)  + 8 (\mathbf a_N^\ast \otimes \mathbf w_2^{(2 )} + \mathbf w_2^{(2 )} \otimes \mathbf a_N^\ast)  \right) \exp (\imath \Theta_N)   +   \nonumber   \\
B_N^\ast (B_N B_N^\ast) \left( 2 (\mathbf a_N^\ast \otimes \mathbf w_2^{(0)} + \mathbf w_2^{(0)} \otimes \mathbf a_N^\ast)  + 8 (\mathbf a_N \otimes \mathbf w_2^{(2 )\ast} + \mathbf w_2^{(2 )\ast} \otimes \mathbf a_N)  \right) \exp (- \imath \Theta_N)  ,
\label{productformu1otimesu2supthetan}
\end{eqnarray}
from which the components of the column vector $ \mathscr N_3 = (\mathscr N_{3OS}, \mathscr N_{3Sq}, \mathscr N_{3Co})^T$ in (\ref{eqnforu3}) for the flow example in question
can be readily evaluated by appropriate differentiation w.r.t. the {\bf fast/short scale variables}, according to (\ref{nonlinelementsswirl}).

It may be noted that in (\ref{u1ru2xandu1xu2r}), $B_N^\ast B_N^2$ and $B_N B_N^{\ast2}$ may both written as $B_N (B_N B_N^\ast)$ and $B_N^\ast (B_N B_N^\ast)$
respectively, see {\bf Appendix 4}. It is useful for the further course of this work to introduce at this juncture a tensor $\mathscr T$ comprising only those $r-$ dependent coefficients of
$\exp (\imath \Theta_N)$ in the nonlinear terms to the order $\epsilon_A^{\frac{3}{2}}$ which enter the solvability condition to be formulated, which in turn leads to the amplitude
evolution equation sought. The scheme for derivation of the elements of $\mathscr T$ from the elements of $\mathscr H$ is clear on an examination of the illustrative
examples in {\bf Appendix 4}, see  (\ref{u1ru2xandu1xu2r}). The elements of $\mathscr T$ are as in (\ref{defnswirltensort}) below where the arrow
$\rightarrow$ has been used to indicate the correspondance between the elements of $\mathscr T$  and the source of the nonlinear term:
\subsubsection*{Spiral Poiseuille Flow}

\begin{eqnarray}
 u_x u_x \rightarrow  \mathscr T_{xx} =   2 \left( 2 ( A_{Nx}  w_{2x}^{(0)} +  w_{2x}^{(0)} A_{Nx})  + 8 ( A_{Nx}^\ast  w_{2x}^{(2 )} +  w_{2x}^{(2 )}  A_{Nx}^\ast)  \right)=  \nonumber   \\
                        \left( 8 A_{Nx} w_{2x}^{(0)} + 32 A_{Nx}^\ast w_{2x}^{(2)} \right) ;    \nonumber    \\
 u_r u_x = u_x u_r \rightarrow \mathscr T_{rx} = \mathscr T_{xr} =  \left( 2 (A_{Nr} w_{2x}^{(0)} + w_{2r}^{(0)} A_{Nx})  +    8 (A_{Nr}^\ast w_{2x}^{(2 )} +      w_{2r}^{(2 )} A_{Nx}^\ast)  \right) +   \nonumber   \\
 \left( 2 ( A_{Nx} w_{2r}^{(0)} +  w_{2x}^{(0)}  A_{Nr})  + 8 ( A_{Nx}^\ast w_{2r}^{(2 )} +  w_{2x}^{(2 )} A_{Nr}^\ast)  \right) ;  \nonumber   \\
 u_r u_r \rightarrow \mathscr T_{rr} = 2 \left( 2 (A_{Nr} w_{2r}^{(0)} + w_{2r}^{(0)} A_{Nr})  + 8 (A_{Nr}^\ast w_{2r}^{(2 )} + w_{2r}^{(2 )} A_{Nr}^\ast)  \right) =  \nonumber  \\
                       \left( 8 A_{Nr} w_{2r}^{(0)} + 32 A_{Nr}^\ast w_{2r}^{(2)} \right) ;    \nonumber    \\
 u_\varphi u_\varphi  \rightarrow \mathscr  T_{\varphi \varphi} =  2 \left( 2 ( A_{N \varphi}  w_{2 \varphi}^{(0)} +  w_{2 \varphi}^{(0)} A_{N \varphi})  + 8 (A_{N \varphi}^\ast w_{2 \varphi}^{(2 )} +  w_{2 \varphi}^{(2 )}  A_{N \varphi}^\ast)  \right)  =  \nonumber   \\
                                     \left( 8 A_{N \varphi} w_{2 \varphi}^{(0)} + 32 A_{N \varphi}^\ast w_{2 \varphi}^{(2)} \right) ;     \nonumber  \\
 u_x u_\varphi = u_\varphi u_x     \rightarrow \mathscr T_{x \varphi} = \mathscr T_{\varphi x} = \left( 2 ( A_{Nx}  w_{2 \varphi}^{(0)} + w_{2x}^{(0)} A_{N \varphi})  + 8 ( A_{Nx}^\ast  w_{2 \varphi}^{(2 )} +  w_{2x}^{(2 )} A_{N \varphi}^\ast)  \right) +   \nonumber   \\
  \left( 2 ( A_{N \varphi}  w_{2x}^{(0)} +  w_{2 \varphi}^{(0)} A_{Nx})  + 8 ( A_{N \varphi}^\ast  w_{2x}^{(2 )} +  w_{2 \varphi}^{(2 )} A_{Nx}^\ast)  \right) ;  \nonumber   \\
 u_r u_\varphi = u_\varphi u_r  \rightarrow \mathscr T_{r \varphi} = \mathscr T_{\varphi r} =   \left( 2 (A_{Nr} w_{2 \varphi}^{(0)} + w_{2r}^{(0)} A_{N \varphi})  + 8 (A_{Nr}^\ast w_{2 \varphi}^{(2 )} + w_{2r}^{(2 )} A_{N \varphi}^\ast)  \right)  +  \nonumber   \\
 \left( 2 ( A_{N \varphi}  w_{2r}^{(0)} +  w_{2 \varphi}^{(0)} A_{Nr})  + 8 ( A_{N \varphi}^\ast  w_{2r}^{(2 )} +  w_{2 \varphi}^{(2 )} A_{Nr}^\ast)  \right) .
\label{defnswirltensort}
\end{eqnarray}

From the expressions for the elements of the tensor $\mathscr T$ in (\ref{defnswirltensort}), the components of the column vector $\mathscr N_3$ in
(\ref{eqnforu3}) may readily be evaluated by carrying out the differentiations w.r.t. the {\bf short-scale variables}. $\mathscr N_3$ is seen to be of the form of a product as follows,
which exhibits the analytical nature of its structural dependence on both the {\bf fast/short scale variables} and the {\bf slow/long scale variables} explicitly:
\begin{equation}
\mathscr N_3 = B_N (B_N B_N^\ast) \left( \mathscr N_{3}^{(\Theta_N)}\right)^T \exp (\imath \Theta_N)  +  c.c.
\label{productformofmathscrn3}
\end{equation}
where $\mathscr N_3^{(\Theta_N)} (y) = (\mathscr N_{3OS}^{(\Theta_N)}, \mathscr N_{3Sq}^{(\Theta_N)}, \mathscr N_{3Co}^{(\Theta_N)})^T $  with the components as follows:
\subsubsection*{Spiral Poiseuille Flow}
\begin{eqnarray}
 \mathscr N_{3OS}^{(\Theta_N)} = - \frac{d}{d y} \left( (\imath \lambda_{xN})^2 \mathscr T_{xx} \right)  -  \frac{d^2}{d y^2} \left( (\imath \lambda_{xN}) \mathscr T_{rx} \right) +  (\imath \lambda_{xN})^3 \mathscr T_{xr} + \frac{d}{dy} \left((\imath \lambda_{xN})^2  \mathscr T_{rr}\right) - Ta (\imath \lambda_{xN})^2 \mathscr T_{\varphi \varphi} \nonumber   \\
                        \nonumber   \\
\mathscr N_{3Sq}^{(\Theta_N)} =  - (\imath \lambda_{xN})^2 \mathscr T_{x \varphi} + \frac{d}{dy} \left((\imath \lambda_{xN}) \mathscr T_{r \varphi} \right)                           \nonumber   \\
\mathscr N_{3Co}^{(\Theta_N)} =  0  ,
\label{productformofmathscrn3swirl}
\end{eqnarray}
with the elements of the tensor $\mathscr T$ given by (\ref{defnswirltensort}).

\subsubsection{The right-hand side of the set of equations }
After evaluation of $\mathscr N_3$ (Sec 5.3.1)  and of $ \mathscr D_2 \mathbf u_2$ and $  \mathscr D_3 \mathbf u_1 $ (Sec. 5.3.2), it is a straightforward matter to set up the 
column vector on the right-hand side of (\ref{eqnforu3}), {\it viz.} $\mathbf {(RHS)}^{(\frac{3}{2})}$.
The right-hand side is a sum of three terms, viz. $- \mathscr D_3 \mathbf u_1 - \mathscr D_2 \mathbf u_2 + \mathscr N_3$, of which the contributions $\mathscr D_3 \mathbf u_1$ and 
$\mathscr D_2 \mathbf u_2$ are only of periodicity $\Theta_N$, whereas $\mathscr N_3$ contains terms of other period as well. For setting up the condition for solvability of (\ref{eqnforu3}), we recall that 
it is only necessary to examine the contribution of period $\Theta_N$ on the right-hand side of this vector equation, i.e. $\mathbf {(RHS)}^{(\frac{3}{2}, \Theta_N)}$ .  It therefore 
suffices for the present purpose to extract from $\mathscr N_3$ the contribution of period $\Theta_N$ alone, which has been done in Sec. 5.3.1.. With $\mathscr D_3 \mathbf u_1$ and 
$\mathscr D_2 \mathbf u_2$ given by the appropriate expressions derived in {\bf Sec 5.3.2 a} and {\bf Sec 5.3.2 b} respectively, and with the contribution of period $\Theta_N$ to 
$\mathscr N_3$ by (\ref{productformofmathscrn3}), the vector equation (\ref{eqnforu3}) can be broken down into its components and the right-hand sides, after division by 
$\exp (\imath \Theta_N)$, assume a form detailed as under for the flow example under study.

{\bf For the current Flow Example}:

The right-hand sides of the Orr-Sommerfeld, Squire and Continuity equations, which are the components of the column vector $\mathbf {(RHS)}^{(\frac{3}{2}, \Theta_N)}$, are as follows: 
\begin{eqnarray} 
 - \frac{\partial ^2(2 B_N)}{\partial \xi_x^2} P_{OS} - \left( \mathcal D_{3OSr} A_{Nr} + \mathcal D_{3OS \varphi} A_{N \varphi} + \mathcal D_{3OSx} A_{Nx}  \right) +  B_N (B_N B_N^\ast) \mathscr N_{3OS}^{(\Theta_N)} ;  \nonumber  \\
 - \frac{\partial ^2(2 B_N)}{\partial \xi_x^2} P_{Sq} - \left( \mathcal D_{3Sqr} A_{Nr} + \mathcal D_{3Sq \varphi} A_{N \varphi} + \mathcal D_{3Sqx} A_{Nx}  \right) +  B_N (B_N B_N^\ast) \mathscr N_{3Sq}^{(\Theta_N)} ;  \nonumber  \\
 - \frac{\partial ^2(2 B_N)}{\partial \xi_x^2} P_{Co} - \left( \mathcal D_{3Cor} A_{Nr} + \mathcal D_{3Co \varphi} A_{N \varphi} + \mathcal D_{3Cox} A_{Nx}  \right) +  B_N (B_N B_N^\ast) \mathscr N_{3Co}^{(\Theta_N)} ,
\label{finalrhseqnforu3swirl}
\end{eqnarray}
where $(P_{OS}, P_{Sq}, P_{Co})$ are functions $y$ alone defined through (\ref{swirlpospsqpco}). In (\ref{finalrhseqnforu3swirl}) above, 
$( \mathcal D_{3OSr}, \mathcal D_{3OS \varphi}, \mathcal D_{3OSx})$, $(\mathcal D_{3Sqr}, \mathcal D_{3Sq \varphi}, \mathcal D_{3Sqx})$ and $( \mathcal D_{3Cor}, \mathcal D_{3Co \varphi}, \mathcal D_{3Cox})$
may be regarded as the rows of the matrix $ \mathcal D_3$, see (\ref{elemented3swirlrewritten}). The column vector of the nonlinear terms $\mathscr N_3$ is as in
(\ref{productformofmathscrn3swirl}) together with (\ref{productformofmathscrn3}). 

The column vector $\mathbf {(RHS)}^{(\frac{3}{2}, \Theta_N)}$ with components (\ref{finalrhseqnforu3swirl}) may then be written in a compact matrix-vector notation as follows: 

\begin{eqnarray}
\mathbf {(RHS)}^{(\frac{3}{2}, \Theta_N)} = 
 - \frac{\partial (2 B_N)}{\partial \tau} \mathcal D_{3 \tau} \mathbf a_N - \frac{\partial^2 (2 B_N)}{\partial \xi_x^2} \left( \mathcal D_{3 \xi_x \xi_x} \mathbf a_N + \mathscr P \mathbf {w_2^{(1)}} \right)  \nonumber   \\  - (2 B_N) \mathcal D_{30} \mathbf a_N  + B_N (B_N B_N^\ast) \mathscr N_3^{(\Theta_N)}
\label{mathbfrhssup3by2thetanswirl}
\end{eqnarray}

The {\bf solvability condition} for (\ref{eqnforu3}) that is sought may then be expressed in words as follows: 

 The inner product of the vector $\mathbf {(RHS)}^{(\frac{3}{2}, \Theta_N)}$ with the adjoint of $\mathbf {(LHS)}^{(\frac{3}{2}, \Theta_N)}$ which is $\mathbf a^{+\ast}$, integrated between 
the limits $ y = \pm 1$ is to be set $= 0$.

In the matrix notation introduced in the present work this may be written concisely as follows:
\begin{eqnarray}
  - \frac{\partial (2 B_N)}{\partial \tau} \int_{-1}^{1} \mathcal D_{3 \tau} \mathbf a_N \cdot \mathbf a_N^{+\ast} dy -
\frac{\partial^2 (2 B_N)}{\partial \xi_x^2} \int_{-1}^{1} \left( \mathcal D_{3 \xi_x \xi_x} \mathbf a_N + \mathscr P \mathbf {w_2^{(1)}} \right) \cdot \mathbf a_N^{+\ast} dy  \nonumber   \\
 - (2 B_N) \int_{-1}^{1} \mathcal D_{30} \mathbf a_N \cdot \mathbf a_N^{+\ast} dy  + B_N (B_N B_N^\ast) \int_{-1}^{1} \mathscr N_3^{(\Theta_N)} \cdot \mathbf a_N^{+\ast} dy = 0
\label{solvabcondswirl}
\end{eqnarray}

 It is worthy of note that the solvability condition just derived, (\ref{solvabcondswirl}) is an equation for the amplitude $(2B_N)$ in terms of the {\bf slow/long-
 scale} variables,
$(\tau, \xi_x)$, in which nonlinearity enters only through $\mathscr N_3^{(\Theta_N)}$ in the form $B_N (B_N B_N^\ast)$. All the other terms are linear.

To facilitate comparison of the solvability conditions derived for the cases of mild and strong swirl in {\bf Spiral Poiseuille Flow}, the effects of nonlinearity during transition may be written in a generalised common form of the equation (\ref{solvabcondswirl}), which involves operations of forming the inner product and integration. This would be as follows:
\begin{equation}
 \alpha_{\tau} \frac{\partial (2 B_N)}{\partial \tau} + \alpha_{\xi_x \xi_x} \frac{\partial^2 (2 B_N)}{\partial \xi_x^2} + \alpha_{0} (2 B_N) + \beta B_N (B_N B_N^\ast) = 0,  
\label{solvabcondswirlaux}
\end{equation}
where $(\alpha_{\tau}, \alpha_{\xi_x \xi_x}, \alpha_{0})$ and $\beta$ are numbers that can be evaluated from the eigensolutions of the classical problem for infinitesimally small disturbances. They depend upon whether the specific basic {\bf Spiral Poiseuille Flow} under study is one with mild swirl or stong swirl, and also on
the chosen point on the surface of neutral stability of basic flow.

%
\section{Discussion, Concluding Remarks and Outlook}
The effects of nonlinearity on the flow behaviour during transtion is brought explicitly  to light through the {\bf cubic term of the Amplitude Function}  in the solvability condition for Spiral Poiseuille Flow, {\it viz.} $ B_N (B_N  B_N^{\ast})$.
Since the method of derivation of (\ref{solvabcondswirlaux}) is based on first principles, the relation expressed through this equation may be expected to hold over the entire range of the swirl parameter, $S$, from $0$ to $\infty$. This inference should not be understood to mean that the numbers that are evaluable from the classical solution of the eigenvalue problem (for infinitesimally small disturbances), $ \alpha_{\tau}, \alpha_{\xi_x \xi_x}, \alpha_0$ and $\beta$, are ''universal'' for all Spiral Poiseuille flows. The numbers depend not only upon the swirl parameter of the basic flow, $S$, but also upon the chosen point on the surface of neutral stability with values $Re_N$ and $S_N$, or $Ta_N$ and $S_N \sqrt{\epsilon_{RN}}$, from which effects of small but finite magnitudes of the further disturbance are reckoned. These relate the values of the flow parameter to the Amplitude Parameter $\epsilon_A$, see Section 4.2.

An examination of the solvability condition derived also indicates that its linear part is already influenced by the Reynolds number, $Re$, the swirl parameter in the basic flow, $S$, and the transverse curvature parameter, $\epsilon_R$, through the eigenfunctions of the classical eigenvalue problem for infinitesimally small disturbances. Furthermore, regardless of whether the {\bf swirl} in the basic {\bf Spiral Poiseuille Flow} is mild or strong, the equation for the {\bf Amplitude function} bears a striking resemblace to that in the bench-mark flows derived by Stuart and Stewartson to capture the effect of nonlinearitiy during transition. This equation, derived from first principles, has a structure closely resembling that of the Ginzburg-Landau equation. The numbers that appear as coefficients in this equation however depend upon various details of the basic flow which include the swirl number, $S$, the topology of the surface of neutral stabilty and the point on the surface of neutral stability according to the classical theory (for infinitesimally small disturbances) for this problem. In short, within the framework of the theory proposed in this paper, nonlinearity affects only the magnitude  (amplitude) of the transition inducing disturbance wave, leaving its shape essentially unchanged.

Particularly noteworthy from the Author's viewpoint, is the pronounced anisotropy of the fluctuating motion, which depends not upon the swirl as such, but upon the transverse curvature parameter, $ \epsilon_R $. A direct evidence for this is the difference in the orders of magnitude of the azimuthal component of the velocity disturbance and of the axial and radial components of the same quantity. Nonlinearity does not modify this difference which is already present in the classical linearized problem,  see \cite{vvr}.

A further point worthy of note is the relation between the {\bf Amplitude Parameter}, $\epsilon_A$, and the kinetic energy in the wave of the transition inducing fluctuating motion. With the velocity disturbance to the leading order according to (\ref{ansatzextendedwave}) and the Amplidude Function $B$ obeying the solvability condition (\ref{solvabcondswirlaux}) we have, on integration over one period w.r.t. the fast/short scale variable, an expression from which the kinetic energy in the disturbance wave can be evaluated. This is for $\mathbf u$ in (\ref{ansatzextendedwave}) together with (\ref{solvabcondswirlaux}). We have, on an  \underline{average over a period w.r.t. the fast/short-scale variables}:
\begin{equation}
 \mathbf u \cdot \mathbf u = 2 B B^{\ast} \mathbf  a \cdot {\mathbf a^{\ast}}  ,
\end{equation}
with the above relation holding to the order $O(\epsilon_A)$.

\subsubsection*{Outlook}

We conclude this paper by drawing attention to an important point regarding the post-transition behaviour of the flow pattern that, for  multifarious reasons, had to be left unaddressed in this work. It concerns the computational route for following the amplified disturbance waves in the space of the dispersion relation on onset of transition. The problem posed hereby can be briefly summarised through the following words: \\
{\bf The keystone of the method proposed is knowledge on the topological features of the surface of neutral stability, conveyed through the dispersion relationship of the classical eigenvalue problem}. This is for propagation characteristics of infinitesimally small disturbances. Although analytically the dispersion relationship may formally be written as $\omega = \omega(Re, S, \epsilon_R, \epsilon_A)$, the parametrization of infinitesimally small disturbances through Fourier analysis makes it more appropriate to write the dispersion relationship as
$\omega = \omega(Re, S, \epsilon_R, \epsilon_A = 0; \lambda_x, n_{\varphi})$. Here $n_{\varphi} =0 $ represents toroidal modes of disturbance waves and $n_{\varphi} \ne 0 $ represent helical modes.

Given that, in the present state of development, one has to resort primarily to numerical methods to obtain the dispersion relation, determination of a quantity of interest such as the set of critical parameters for transition, involves computation of $\omega$ over a range of six parameters that include the parameters $S$ and $n_{\varphi}$ too. Recent computational work, e.g. of Brockmann, \cite{brockmannetal}, has indicated that the onset of transition may occur, depending upon the values of the set $(Re, S, \epsilon_R)$, for toroidal modes or helical modes of disturbance, distinguishable through the value of $n_{\varphi}$. This means that the pattern of post-transitional flow may change from toroidal to helical as the flow is taken through the parameter range of $S$ from $0$ to $\infty$. In particular, since for both the {\bf bench-mark flows}, $S=0$ and $S=\infty$, it is known from the wealth of previous work that the onset of transition occurs for toroidal disturbances, the onset of transition with $n_{\varphi} \ne 0$ within the region $0 > S < \infty$ implies a change followed by its reversal in the post-transitional flow-pattern. This qualitative feature in the behaviour of the post-transitional flow pattern makes this a step of possibly fundamentak significance in gaining insight into the phenomenon of transition in {\bf Spiral Poiseuille Flows}.

The above observation hints at the highly complex topological structure of the surface of neutral stability of {\bf Spiral Poiseuille Flows}.  The derivation in Sec. 5.2.3 suggests a computational procedure for following the growth of disturbance waves which would be as follows: in the space of $(Re, S, n_{\varphi})-$ one could take a computational route in the direction of the maximum growth rate of the imaginary part of $\left( \frac{\partial \omega}{\partial \lambda_x} \right)_N $. This quantity which is obtainable on the surface of neutral stability of infinitesimally small disturbances is  representative of the growth rate of a packet of wave disturbances at a chosen point on the surface of neutral stability.

The wealth of knowledge hitherto accumulated for the two {\bf bench-mark flows}, for which the dispersion relation assumes a relatively simpler form with only one parameter, {\it viz.} $Re$ or $Ta$, put together with the requirement of having to ascertain from computations the point where the derivative $\frac{\partial \lambda_x}{\partial Re}$  or $\frac{\partial \lambda_x}{\partial Ta}$ is $=0$, has enabled the acquired insight into the onset of transition in these flows to be gained through the study of toroidal disturbances alone, for which $n_{\varphi}= 0$.

In particular, the  computational work of Brockmann, \cite{brockmannetal}, indicates that disturbances with $n_{\varphi} \ne 0$, which are {\bf not toroidal} but {\bf helical}, may be primarily instrumental for transition in an {\bf intermediate parameter range} of $S$ between $0$ and $\infty$. Summarising these observations, the Author sees a necessity for a more thoroughgoing investigation of the propagation characteristics of infinitesimally small transition inducing wave disturbances in the intermediate parameter range of $S$ between $0$ and $\infty$ which covers both {\bf toroidal} and {\bf helical} disturbances. This is a computational task of proportions falling clearly outside the scope of the present paper.

With the  knowledge gained through a broader and deeper study of disturbance propagation based on fundamental principles, the Author sees the possibility of the present study leading to a deeper insight into further features of interest such as e.g. {\bf intermittency in Spiral Poiseuille Flows} during transition.

\subsubsection*{Acknowledgement}

The Author dedicates this paper to the memory of: \\ Prof. Roddam Narasimha, his mentor and teacher in spirit, and trusted friend. \\
He also acknowledges with gratitude the benefit of discussions with: Prof. J. T. Stuart, Prof. K. R. Sreenivasan, Dr. J. Schlitter, Dr.  J. F. Spittler, Prof. Jeanette Hussong, Dr. P. Brockmann.

%
\section*{Appendix 1: The generalised nonlinear Orr-Sommerfeld, Squire and Continuity equations}
In this Appendix the governing equations for the problem in consideration are given in three sets of variables. These are: the set of {\bf conventional}, {\bf strong swirl} and {\bf Taylor variables}
\subsection*{The governing equations in conventional variables}
The starting point for our work is the set of momentum equations governing fluid flow, and the continuity equation for an incompressible fluid, which are known from standard 
text-books, e.g. \cite{lamb}, \cite{batchelor}, \cite{schlichtinggersten}. From these equations pressure may be eliminated as an unknown to arrive at the 
{\bf generalised Orr-Sommerfeld, Squire} and {\bf Continuity equations}, (\ref{generalisednonlinos}, \ref{generalisednonlinsq}, \ref{continuityinuxuruvarphi}), in which the 
unknowns are only the three components of the velocity. For details of the procedure for elimination of the pressure disturbance as an unknown the reader is referred to 
\cite{vvr}. For purposes of enhancing transparency in the further course of this work, we rewrite the {\bf generalised non-linear Orr-Sommerfeld, Squire} and 
{\bf Continuity equations}, (\ref{generalisednonlinos}, \ref{generalisednonlinsq}, \ref{continuityinuxuruvarphi}), ordered according to powers of the geometrical parameter 
$\epsilon_R = \frac{(R_o - R_i)}{(R_o + R_i)} $. These are then as follows: \\ 

 {\bf The generalised nonlinear Orr-Sommerfeld equation}
\begin{eqnarray}
 - \frac{\partial}{\partial t} \left( \frac{\partial^2 u_r}{\partial x^2} + \frac{\partial^2 u_r}{\partial y^2} \right) - \frac{2}{\sqrt{(1+S^2)}} \frac{\partial u_r}{\partial x} - \frac{(1-y^2)}{\sqrt{(1+S^2)}} \left( \frac{\partial^3 u_r}{\partial x \partial y^2} + \frac{\partial^3 u_r}{\partial x^3}\right)   \nonumber  \\ 
+ \frac{1}{Re} \left( \frac{\partial^4 u_r}{\partial x^4} + \frac{\partial^4 u_r}{\partial y^4} + 2 \frac{\partial^4 u_r}{\partial x^2 \partial y^2} \right)   
+ \epsilon_R \left[- \frac{\partial}{\partial t} \left( \frac{\partial u_r}{\partial y} \right) + \frac{4y}{\sqrt{(1+S^2)}} \left(\frac{\partial u_r}{\partial x} \right)  \right]    \nonumber   \\ 
+ \epsilon_R \left[- \frac{(1-y^2)}{\sqrt{(1+S^2)}} \frac{\partial^2 u_r}{\partial y \partial x} + \frac{1}{\sqrt{(1+S^2)}} \frac{y(1-y^2)}{3} \left( \frac{\partial^3 u_r}{\partial x \partial y^2} + \frac{\partial^3 u_r}{\partial x^3} \right) \right]   \nonumber  \\  
+ \epsilon_R \frac{S}{\sqrt{(1+S^2)}} \left[ \frac{(1-y)}{2} \left( - \frac{\partial^3 u_r}{\partial y^2 \partial \varphi} - \frac{\partial ^3 u_r}{\partial x^2 \partial \varphi} + 2 \frac{\partial^2 u_{\varphi}}{\partial x^2} \right)  \right] + \frac{2 \epsilon_R}{Re} \left[ \left(\frac{\partial^3 u_r}{\partial x^2 \partial y} - \frac{\partial ^3  u_r}{\partial y^3} \right) \right]  \nonumber  \\ 
= 
- \frac{\partial ^3 (u_x u_x)}{\partial x^2 \partial y} - \frac{\partial ^3 (u_r u_x)}{\partial x \partial y^2} + \frac{\partial ^3 (u_x u_r)}{\partial x^3} + \frac{\partial ^3 (u_r u_r)}{\partial x^2 \partial y}   \nonumber  \\ 
- \epsilon_R \left[ \frac{\partial ^2 (u_r u_x)}{\partial x \partial y} + \frac{\partial ^3 (u_\varphi u_x)}{\partial x \partial y \partial \varphi}      
+  \frac{\partial ^3 (u_x u_\varphi)}{\partial x \partial y \partial \varphi} +  \frac{\partial ^3 (u_r u_\varphi)}{\partial y^2 \partial \varphi }       
- \frac{\partial ^2 (u_r u_r)}{\partial x^2} - \frac{\partial ^3 (u_\varphi u_r)}{\partial x^2 \partial \varphi} \right] - \epsilon_R \frac{\partial ^2 (u_\varphi u_\varphi)}{\partial x^2}  
\label{generalisednonlinos}
\end{eqnarray}
{\bf The generalised nonlinear Squire equation}: 
\begin{eqnarray}
- \frac{\partial}{\partial t} \left(\frac{\partial u_{\varphi}}{\partial x} \right) - \frac{1}{\sqrt{(1+S^2)}} \left( 1-y^2\right) \left(\frac{\partial^2 u_{\varphi}}{\partial x^2} \right) + \frac{S}{\sqrt{(1+S^2)}} \frac{1}{2} \frac{\partial u_r}{\partial x}   
+ \frac{1}{Re} \left( \frac{\partial^3 u_{\varphi}}{\partial x^3} + \frac{\partial^3 u_{\varphi}}{\partial x \partial y^2} \right)   \nonumber  \\  
+ \epsilon_R \left[ \frac{\partial}{\partial t}\left(\frac{\partial u_x}{\partial \varphi} \right) \right] + \frac{\epsilon_R}{\sqrt{(1+S^2)}} \left[ \left( \frac{y(1-y^2)}{3} \right) \left( \frac{\partial ^2 u_{\varphi}}{\partial x^2} \right) + \left( 1-y^2 \right) \frac{\partial ^2 u_x}{\partial x \partial \varphi} \right]  \nonumber  \\ 
+ \epsilon_R \frac{S}{\sqrt{(1+S^2)}} \left[\left(\frac{(1-y)}{2} \right) \left(- \frac{\partial^2 u_{\varphi}}{\partial \varphi \partial x} - \frac{\partial u_r}{\partial x} \right) - y \frac{\partial u_r}{\partial x} \right]   
- \frac{\epsilon_R}{Re} \left[ \frac{\partial^3 u_x}{\partial x^2 \partial \varphi} + \frac{\partial^3 u_x}{\partial y^2 \partial \varphi} - \frac{\partial^2 u_{\varphi}}{\partial x \partial y} \right]  \nonumber   \\  
=  
-\frac{\partial ^2 (u_x u_\varphi)}{\partial x^2} - \frac{\partial ^2 (u_r u_\varphi)}{\partial x \partial y} - \epsilon_R \left[ \frac{\partial (u_r u_\varphi)}{\partial x} + \frac{\partial ^2 (u_\varphi u_\varphi)}{\partial x \partial \varphi} + \frac{\partial (u_r u_\varphi)}{\partial x}     
-  \frac{\partial ^2 (u_x u_x)}{\partial \varphi \partial x} - \frac{\partial ^2 (u_r u_x)}{\partial \varphi \partial r} \right]
\label{generalisednonlinsq}
\end{eqnarray}
{\bf The Continuity Equation}
\begin{eqnarray}
\frac{\partial u_x}{\partial x} + \frac{\partial u_r}{\partial y} + \epsilon_R \left[ u_r + \frac{\partial u_{\varphi}}{\partial \varphi}  \right] = 0. 
\label{continuityinuxuruvarphi}
\end{eqnarray}
%

%

 {\bf Operators on the non-linear terms}
\begin{eqnarray}
\mathscr N_{OS} =           
- \frac{\partial ^3 (u_x u_x)}{\partial x^2 \partial y} - \frac{\partial ^3 (u_r u_x)}{\partial x \partial y^2}  + \frac{\partial^3 (u_x u_r)}{\partial x^3} + \frac{\partial^3 (u_x u_r)}{\partial x^2 \partial y}       \nonumber   \\ 
 - \epsilon_R \left( \frac{\partial ^2 (u_r u_x)}{\partial x \partial y} +  \frac{\partial ^3 (u_\varphi u_x)}{\partial x \partial y \partial \varphi}      
+  \frac{\partial ^3 (u_x u_\varphi)}{\partial x \partial y \partial \varphi} + \frac{\partial ^3 (u_r u_\varphi)}{\partial y^2 \partial \varphi }        
- \frac{\partial ^2 (u_r u_r)}{\partial x^2} -  \frac{\partial ^3 (u_\varphi u_r)}{\partial x^2 \partial \varphi} \right) - \epsilon_R \frac{\partial ^2 u_\varphi u_\varphi}{\partial x^2} . 
                                           \nonumber    \\ 
\mathscr N_{Sq} =     
-\frac{\partial ^2 (u_x u_\varphi)}{\partial x^2} - \frac{\partial ^2 (u_r u_\varphi)}{\partial x \partial y} - \epsilon_R \left( \frac{\partial (u_r u_\varphi)}{\partial x} + \frac{\partial ^2 (u_\varphi u_\varphi)}{\partial x \partial \varphi} + \frac{\partial (u_r u_\varphi)}{\partial x}     
-  \frac{\partial ^2 (u_x u_x)}{\partial \varphi \partial x} - \frac{\partial ^2 (u_r u_x)}{\partial \varphi \partial r} \right) .
\nonumber                                                            \nonumber    \\ 
\mathscr N_{Co} =  0 .
\label{nonlinelementsswirl}
\end{eqnarray} 
%
\subsection*{The governing equations in strong swirl variables}

The definition of the set of {\bf strong swirl variables}, denoted by a tilde over the quantity, $\tilde t, \tilde u_r, \tilde u_x, \tilde u_{\varphi}$, is as follows:
\begin{eqnarray}
\tilde t = t ;
\tilde x, \tilde y, \tilde \varphi  = x, y, \varphi \nonumber \\
\tilde u_r = \frac{u_r}{\epsilon_R} ;
\tilde u_x = \frac{u_x}{\epsilon_R} ;
\tilde u_{\varphi} = u_{\varphi} ;
\label{defntildevariables}
\end{eqnarray}
Attention is drawn in this definition, (\ref{defntildevariables}), to the difference in scaling between $(u_r, u_x)$ and $u_{\varphi}$ with respect to the transverse curvature parameter $\epsilon_R$.

Recasting the set of governing equations, (\ref{generalisednonlinos}, \ref{generalisednonlinsq}, \ref{continuityinuxuruvarphi}),
in terms of the unknowns $(\tilde u_r, \tilde u_x, \tilde u_\varphi)$ defined through
(\ref{defntildevariables}), followed by division through $\epsilon_R$, and noting that $\frac{1}{(1+S^2)^{\frac{1}{2}}} \simeq \frac{1}{S} \left(1- \frac{1}{2S^2} \right)$ for large $S$
leads, on rearrangement according to powers of $\epsilon_R$ to the following set of equations:

{\bf The Orr-Sommerfeld equation}:
\begin{eqnarray}
	- \frac{\partial}{\partial t} \left(  \frac{\partial^2 \tilde u_r}{\partial x^2} + \frac{\partial^2 \tilde u_r}{\partial y^2}   \right)
	- \frac{2}{S} \left(1 - \frac{1}{2 S^2} \right) ~\frac{\partial \tilde u_r}{\partial x}                                                              \nonumber \\
  -   \frac{1}{S} \left(1 - \frac{1}{2 S^2} \right)  (1-y^2)   \left(  \frac{\partial^3 \tilde u_r}{\partial x \partial y^2}   + \frac{\partial^3 \tilde u_r}{\partial x^3}    \right)    \nonumber   \\
    -  \left(1 - \frac{1}{2 S^2} \right)  ~\left( \frac{(1-y)}{2}  \right)~\left(
     \frac{\partial^3 \tilde u_r}{\partial y^2 \partial \varphi}
  + \frac{\partial^3 \tilde u_r}{\partial x^2 \partial \varphi}     \right)        \nonumber   \\
  + \left(1 - \frac{1}{ 2 S^2} \right)  ~\left( 1-y  \right)~\left(
     ~\frac{\partial^2 \tilde u_{\varphi}}{\partial x^2}   \right)                                                     \nonumber \\
  + \frac{1}{Re} \left(  \frac{\partial^4 \tilde u_r}{\partial x^4}   + \frac{\partial^4 \tilde u_r}{\partial y^4} + 2 \frac{\partial^4 \tilde u_r}{\partial y^2 \partial x^2} \right) - \epsilon_R \left[ \frac{\partial}{\partial t} \left( \frac{\partial \tilde u_r}{\partial y} \right) \right]  \nonumber \\
 - \epsilon_R \left[  \frac{1}{S} \left(1- \frac{1}{2 S^2} \right) \left((1-y^2)\frac{\partial ^2 \tilde u_r}{\partial x \partial y} + \frac{y(1-y^2)}{3} \left( \frac{\partial ^3 \tilde u_r}{\partial x \partial y^2} + \frac{\partial ^3 \tilde u_r}{\partial x^3} \right)\right)  \right]
   \nonumber   \\
   + \epsilon_R \left[ \left(1 - \frac{1}{2 S^2} \right)  ~\left(  \frac{(1-y^2)}{2} \right)~\left(  \frac{\partial^3 \tilde u_r}{\partial y^2 \partial \varphi}
  + \frac{\partial^3 \tilde u_r}{\partial x^2 \partial \varphi}     \right)  \right]       \nonumber   \\
  - \epsilon_R \left[ \left(1 - \frac{1}{2 S^2} \right)  ~\left(  1-y^2 \right)~\left(
     ~\frac{\partial^2 \tilde u_{\varphi}}{\partial x^2}   \right) \right]                                                     \nonumber \\
  + \epsilon_R \left[ \frac{2}{Re} \left(  ~\frac{\partial^3 \tilde u_r}{\partial y \partial x^2}   +  ~\frac{\partial^3 \tilde u_r}{\partial y^3  } \right)  \right]    \nonumber \\
 =
\epsilon_R \left[ 2 \frac{\partial^3(\tilde u_{\varphi} \tilde  u_{x})} {\partial \tilde x \partial \tilde y \partial \varphi} + \frac{\partial^3(\tilde u_r \tilde u_{\varphi})}{\partial y^2 \partial \varphi} - \frac{\partial^2 (\tilde u_{\varphi} \tilde u_{\varphi})}{\partial \tilde x^{2}} \right] + O(\epsilon_R^2) .
\label{rearrangednonlinearostildevariab}
\end{eqnarray}

{\bf The Squire equation}:
\begin{eqnarray}
	 - \frac{\partial}{\partial t} \left( \frac{\partial \tilde u_{\varphi}}{\partial x} \right)
	- \frac{1}{S} \left(1 - \frac{1}{2 S^2} \right)   (1-y^2)  \left( \frac{\partial^2 \tilde u_{\varphi}}{\partial x^2}  \right)
        + \frac{1}{Re} \left(  \frac{\partial^3 \tilde u_{\varphi}}{\partial x^3} + \frac{\partial^3 \tilde u_{\varphi}}{\partial x \partial y^2}  \right)    \nonumber \\
- \epsilon_R \left[ \frac{1}{S} \left(1 - \frac{1}{2 S^2} \right)  \left( \frac{y(1-y^2)}{3}\right) \left( \frac{\partial^2 \tilde u_{\varphi}}{\partial x^2} \right) -  \left(1 - \frac{1}{2 S^2} \right)  \left(\frac{1}{2} \frac{\partial \tilde u_r}{\partial x}\right) \right]    \nonumber   \\
- \epsilon_R \left[ \left(1 - \frac{1}{2 S^2} \right)  \left( \frac{(1-y)}{2}\right) \left( \frac{\partial ^2 \tilde u_{\varphi}}{\partial \varphi \partial x} \right) - \frac{1}{Re} \left( \frac{\partial ^2 \tilde u_{\varphi}}{\partial x \partial y} \right)  \right]
= - \epsilon_R \left[ \frac{\partial ^2 (\tilde u_x u_{\varphi})}{\partial \tilde x^2} + \frac{\partial ^2 (\tilde u_r \tilde u_{\varphi})}{\partial \tilde x \partial y} \right]  + O(\epsilon_R^2).
\label{rearrangednonlinearsqtildevariab}
\end{eqnarray}

{\bf The Continuity equation}:

\begin{eqnarray}
   \frac{\partial \tilde u_x}{\partial x}
+  \frac{\partial \tilde u_r}{\partial y} +  \epsilon_R \tilde u_r
+  \frac{\partial \tilde u_\varphi}{\partial \varphi}
&=& 0
\label{rearrangedkontitildevariab}
\end{eqnarray}
{\bf Operators on the non-linear terms}

\begin{eqnarray}
\mathscr N_{OS} =
\epsilon_R \left[ 2 \frac{\partial^3(\tilde u_{\varphi} \tilde  u_{x})} {\partial \tilde x \partial \tilde y \partial \varphi} + \frac{\partial^3(\tilde u_r \tilde u_{\varphi})}{\partial y^2 \partial \varphi} - \frac{\partial^2 (\tilde u_{\varphi} \tilde u_{\varphi})}{\partial \tilde x^{2}} \right] + O(\epsilon_R^2) .
\end{eqnarray}
\begin{eqnarray}
\mathscr N_{Sq}= - \epsilon_R \left[ \frac{\partial ^2 (\tilde u_x u_{\varphi})}{\partial \tilde x^2} + \frac{\partial ^2 (\tilde u_r \tilde u_{\varphi})}{\partial \tilde x \partial y} \right]  + O(\epsilon_R^2).
\end{eqnarray}
\begin{eqnarray}
\mathscr N_{Co} = 0.
\end{eqnarray}

\subsection*{The governing equations in Taylor variables}
When the flow undergoes transition primarily through the {\bf Taylor Mechanism}, effects of the other transition inducing mechanism, viz. the {\bf Tollmien-Schlichting Mechanism}, may be expected to recede in importance, making it meaningful to view the transition phenomenon through {\bf Taylor variables}. We prepare ground for this perspective in the following. For this we
chose as the reference velocity, instead of $U_{ref}$ as  defined earlier, {\it vide} Sec. 2.1, the characteristic velocity derivable from the rotation of the inner cylinder, viz. $\Omega_i R_i= U_{ref \varphi}$, designating the Reynoldsnumber $\frac{\Omega_i R_i  H}{\nu}$ as $Re_T$.

The definition of the set of {\bf Taylor variables}, designated by a hat on the quantity, ($ \hat  u_r, \hat u_x, \hat u_{\varphi}$), is as follows:
\begin{eqnarray}
\hat t = \frac{t}{Re_T};
\hat x, \hat y, \hat \varphi = x, y, \varphi \nonumber \\
\hat u_{r} =  \tilde u_{r} Re_T;
\hat u_{x} =  \tilde u_{x} Re_T ; \hat u_{\varphi} = \tilde u_{\varphi} = u_{\varphi}.
\label{defntaylorvariables}
\end{eqnarray}
Attention is drawn in this definition, (\ref{defntaylorvariables}), to the presence of the Reynoldsnumber, $Re_T$, in the reference quantity for time as well as for the radial and the axial components of the velocity disturbance. In dimensional terms the definition in (\ref{defntaylorvariables}) implies that time is referred to $\frac{H^2}{\nu}$, lengths to $H$, the component of the velocity disturbance in the azimuthal direction to $ \Omega R_i $, and in the radial and axial  directions to $ \frac{\nu}{H}$.

Rewriting the set of equations (\ref{generalisednonlinos}, \ref{generalisednonlinsq}, \ref{continuityinuxuruvarphi}), in the variables defined through (\ref{defntaylorvariables}), and carrying out the limiting process, $Re_T \rightarrow \infty$, $\epsilon_R \rightarrow 0$ such that the {\bf Taylor number, Ta} defined through  $ Ta = Re_T^2 \epsilon_R$ remains finite,
leads to the governing equations in {\bf Taylor variables} sought. They are as follows:

{\bf The nonlinear Orr-Sommerfeld equation in Taylor-Variables}
\begin{eqnarray}
 -\frac{\partial}{\partial \hat t} \left( \frac{\partial ^2 \hat u_{r}}{\partial \hat x^2} +  \frac{\partial ^2 \hat u_r}{\partial \hat y^2} \right)
+ Ta (1- \hat y) \frac{\partial^2 \hat u_{\varphi}}{\partial \hat x^2}+ \left( \frac{\partial ^4 \hat u_r}{\partial \hat x^4} +  \frac{\partial ^4 \hat u_r}{\partial \hat y^4}+ 2 \frac{\partial ^4 \hat u_r}{\partial \hat y^2 \partial \hat x^2}   \right)       \nonumber    \\
+ \frac{\sqrt{Ta}}{S \sqrt{\epsilon_R}} \left(-2 \frac{\partial \hat u_r}{\partial \hat x} - (1 - \hat y^2)\left(\frac{\partial ^3 \hat u_r}{\partial \hat x \partial \hat y^2}+ \frac{\partial ^3 \hat u_r}{\partial \hat x^3} \right) \right) =  \nonumber \\
- \left( \frac{\partial ^3 ( \hat u_x \hat u_x)}{\partial \hat x^2 \partial \hat y} + \frac{\partial ^3 ( \hat u_r \hat u_x)}{\partial \hat x \partial \hat y^2}  \right) +  \left( \frac{\partial ^3 (\hat u_x \hat u_r)}{\partial \hat x^3} + \frac{\partial ^3 ( \hat u_r \hat u_r)}{\partial \hat x^2 \partial \hat y} +  \right) - Ta \frac{\partial ^2 (\hat u_\varphi \hat u_\varphi )}{\partial \hat x^2}.
\label{erwnlos4}
\end{eqnarray}
{\bf The nonlinear Squire equation in Taylor-Variables}
\begin{eqnarray}
- \frac{\partial^2 \hat u_{\varphi}}{\partial \hat t \partial \hat x} - \frac{1}{2} \frac{\partial \hat u_{r}}{\partial \hat x} - \left( \frac{\partial ^3 \hat u_{\varphi}}{\partial \hat x^3} + \frac{\partial ^3 \hat u_{\varphi}}{\partial \hat x \partial \hat y^2} \right) - \frac{\sqrt{Ta}}{S \sqrt{\epsilon_R}}  \left((1 - \hat y ^2) \frac{\partial ^2 \hat u_{ \varphi}}{\partial \hat x^2} \right)  =                                  \nonumber \\
- \frac{\partial ^2 (\hat u_x \hat u_\varphi)}{\partial \hat x^2} - \frac{\partial ^2 (\hat u_r \hat u_\varphi)}{\partial \hat x \partial \hat y}  .
\label{erwnlsq4}
\end{eqnarray}

{\bf The Continuity equation in Taylor-Variables}, (\ref{continuityinuxuruvarphi}), reduces to:
\begin{eqnarray}
\frac{\partial \hat u_{x}}{\partial \hat x} + \frac{\partial \hat u_{r}}{\partial \hat y} = 0.
\label{sskontigl4}
\end{eqnarray}

{\bf Operators on the non-linear terms}
\begin{eqnarray}
\mathscr N_{OS}=
- \left( \frac{\partial ^3 ( \hat u_x \hat u_x)}{\partial \hat x^2 \partial \hat y} + \frac{\partial ^3 ( \hat u_r \hat u_x)}{\partial \hat x \partial \hat y^2}  \right) +  \left( \frac{\partial ^3 (\hat u_x \hat u_r)}{\partial \hat x^3} + \frac{\partial ^3 ( \hat u_r \hat u_r)}{\partial \hat x^2 \partial \hat y}   \right) - Ta \frac{\partial ^2 (\hat u_\varphi \hat u_\varphi )}{\partial \hat x^2}.
\end{eqnarray}
\begin{eqnarray}
\mathscr N_{Sq}=
- \frac{\partial ^2 (\hat u_x \hat u_\varphi)}{\partial \hat x^2} - \frac{\partial ^2 (\hat u_r \hat u_\varphi)}{\partial \hat x \partial \hat y}  .
\end{eqnarray}
\begin{eqnarray}
\mathscr N_{Co} = 0.
\end{eqnarray}

The following points are worthy of note on a comparison of the sets of the governing equations written in the three sets of variables:
\begin{itemize}
\item \underline{ The number of parameters} influencing the solutions of the governing equations is not the same!
\item They are $3$, $(Re, S, \epsilon_R)$ in the sets of conventional and strong swirl variables, whereas they are only $2$, $(Ta, S \sqrt{\epsilon_R})$ in the set of Taylor variables.
\item It may be noted that in the set of equations (\ref{erwnlos4}, \ref{erwnlsq4}, \ref{sskontigl4}) the effect of the axial pressure gradient is manifested only in the combination $S \sqrt  \epsilon_R$. The transverse curvature parameter, $\epsilon_R$ does not appear as a parameter by itself.
\end{itemize}
%
\section*{Appendix 2: The matrix differential  operators in the classical eigenvalue problem for propagation of infinitesimally small disturbances in Spiral Poiseuille flow }
Substitution of the ansatz for an elementary wave according to (\ref{ansatzelementarywave}) in the governing equation (\ref{governingequation}) with the right-hand side $\mathscr N = 0 $ leads to the set of ordinary differential equations governing the propagation of intinitesimally smaff disturbances. This may be written in a compact matrix form as follows:
\begin{equation}
 \mathscr L {\bf a} =  \omega \mathscr M \mathscr {\bf  a},
 \nonumber
\end{equation}
where $\bf a$ is a column vector and $\mathscr L$ and $\mathscr M$ are matrix operators as follows:
\begin{eqnarray}
\mathscr L = \left(
\begin{array}{ccc}
 \mathscr L_{OSr}   &  \mathscr L_{OS \varphi}   &  \mathscr L_{OSx}\\
 \mathscr L_{Sqr}   &  \mathscr L_{Sq \varphi}   &  \mathscr L_{Sqx}\\
 \mathscr L_{Cor}   &  \mathscr L_{Co \varphi}   &  \mathscr L_{Cox}
\end{array}
\right),
\mathscr M = \left(
\begin{array}{ccc}
 \mathscr M_{OSr}   &  \mathscr M_{OS \varphi}   &  \mathscr M_{OSx}\\
 \mathscr M_{Sqr}   &  \mathscr M_{Sq \varphi}   &  \mathscr M_{Sqx}\\
 \mathscr M_{Cor}   &  \mathscr M_{Co \varphi}   &  \mathscr M_{Cox}
\end{array}
\right).
\label{
}
\end{eqnarray}
The elements of the column vector of eigenfunctions $\bf a$ are:
\begin{equation}
{\bf  a} = ( A_r, A_{\varphi}, A_x)^T.
\end{equation}

Analytical expressions for the element operators, $\mathscr L$ and $\mathscr M$, are dependent upon whether the swirl parameter, $ S $, is mild or strong. Since both are needed in the course of this work they are listed below.

{\bf For the case of mild swirl, $S \rightarrow 0 $,  the element operators are} :
\begin{eqnarray}
\mathscr L_{OSr} =
+ \frac{1}{\sqrt{(1+S^2)}} (1-y^2) \left[-(\imath \lambda_x) \frac{d^2 }{dy^2} - (\imath \lambda_x)^3   \right]  \nonumber   \\
+ \frac{\epsilon_R}{\sqrt{(1 + S^2)}} \left[ (1-y^2) \left((\imath n_{\varphi}) \frac{d^2 }{dy^2} - (\imath \lambda_x) \frac{d}{dy} - \right) - \frac{y (1- y^2)}{3} \left(-(\imath \lambda_x) \frac{d^2 }{dy^2} \right)  - (\imath \lambda_x)^3 \right] \nonumber  \\
+ \frac{\epsilon_R}{\sqrt{(1+S^2)}} \left( -2y - \epsilon_R \left(\frac{1}{3} - y^2 \right) \right) \left[  (\imath \lambda_x)  +  + O(r^{-2}) \right] + \frac{1}{\sqrt{(1+S^2)}} (-2 + 2 \epsilon_R y )     \nonumber   \\
+ \frac{\epsilon_R S}{\sqrt{(1+S^2)}} \left(\frac{(1-y)}{2} \right) \left[-(\imath n_{\varphi}) \frac{d^2 }{dy^2} - (\imath \lambda_x)^2 (\imath n_{\varphi})  +  + O(r^{-2}) \right]   \nonumber   \\
+ \frac{\epsilon_R S}{\sqrt{(1+S^2)}}\left( - \frac{1}{2} \right) \left[ (\imath n_{\varphi}) \frac{d }{dy} + + O(r^{-2}) \right]    \nonumber   \\
+ \frac{1}{Re} \left[(\imath \lambda_x)^4  + \frac{d^4 }{dy^4} + 2 (\imath \lambda_x)^2 \frac{d^2 }{d y^2} + \epsilon_R \left( 2 (\imath \lambda_x)^2 \frac{d }{dy} +  \right) \right]
    \nonumber   \\
\label{mathscrlosrsto0}
\end{eqnarray}
\begin{eqnarray}
\mathscr L_{OS \varphi} =
+ \frac{\epsilon_R}{\sqrt{(1 + S^2)}} \left[ (1-y^2) \left( - (\imath n_{\varphi} ) \frac{d }{dy} \right) \right]   \nonumber  \\
 + \frac{\epsilon_R}{\sqrt{(1+S^2)}} \left( -2y - \epsilon_R \left(\frac{1}{3} - y^2 \right) \right) \left[  + (\imath n_{\varphi}) (\imath \lambda_x)  + O(r^{-2}) \right]    \nonumber   \\
 + \frac{\epsilon_R S}{\sqrt{(1+S^2)}} \left(\frac{(1-y)}{2} \right) \left[ + 2 (\imath \lambda_x)^2  + O(r^{-2}) \right]   \nonumber   \\
%
+ \frac{1}{Re} \left[+ \epsilon_R \left(  2 \frac{d^3  }{dy^3} \right) \right]
    \nonumber   \\
\label{mathscrlosvarphisto0}
\end{eqnarray}
\begin{eqnarray}
\mathscr L_{OSx} =
+ \frac{\epsilon_R S}{\sqrt{(1+S^2)}} \left( - \frac{1}{2} \right) \left[ + (\imath \lambda_x) (\imath n_{\varphi})  + O(r^{-2}) \right]    \nonumber   \\
%
\label{mathscrlosxsto0}
\end{eqnarray}
\begin{eqnarray}
\mathscr L_{Sqr} =
\frac{1}{\sqrt{(1+S^2)}} \left( (1-y^2) - \epsilon_R \frac{y(1-y^2)}{3} + O(\epsilon_R^2) \right) \left[+ \epsilon_R \frac{d }{dy}  \right] \nonumber \\
+\frac{1}{\sqrt{(1+S^2)}} \left(-2y - \epsilon_R \left(\frac{1}{3} - y^2 \right) \right) \left[(\imath \lambda_x)  \right]   \nonumber   \\
+ \frac{S}{\sqrt{(1+S^2)}} \left(\frac{(1-y)}{2} - \epsilon_R \frac{(1-y^2)}{2} \right) \left[ + \epsilon_R (\imath \lambda_x)   \right] \nonumber \\
\label{mathscrlsqrsto0}
\end{eqnarray}
\begin{eqnarray}
\mathscr L_{Sq \varphi} =
\frac{1}{\sqrt{(1+S^2)}} \left( (1-y^2) - \epsilon_R \frac{y(1-y^2)}{3} + O(\epsilon_R^2) \right) \left[(\imath \lambda_x)^2  \right] \nonumber \\
+ \frac{S}{\sqrt{(1+S^2)}} \left(\frac{(1-y)}{2} - \epsilon_R \frac{(1-y^2)}{2} \right) \left[ \epsilon_R (\imath n_{\varphi}) (\imath \lambda_x)   +  \right] \nonumber \\
+ \frac{1}{Re} \left[(\imath \lambda_x)^3 + (\imath \lambda_x) \frac{d^2 }{ dy^2} + \epsilon_R \left( (\imath \lambda_x) \frac{d }{ d y} \right) \right]  \nonumber  \\
\label{mathscrlsqvarphisto0}
\end{eqnarray}
\begin{eqnarray}
\mathscr L_{Sqx} =
+ \frac{1}{Re} \left[ + \epsilon_R \left( - (\imath \lambda_x)^2 (\imath n_{\varphi})  - (\imath n_{\varphi}) \frac{d^2 }{dy^2} \right) \right]. \nonumber  \\
\label{mathscrlsqxsto0}
\end{eqnarray}
\begin{eqnarray}
\mathscr L_{Cor} = \frac{d }{dy} + \epsilon_R       \nonumber    \\
\mathscr L_{Co \varphi}  = \epsilon_R (\imath n_{\varphi})   \nonumber    \\
\mathscr L_{Cox} = (\imath \lambda_x)
\label{mathscrlcorvarphixsto0}
\end{eqnarray}
\begin{eqnarray}
\mathscr M_{OSr} = - \frac{d^2}{dy^2} + {\lambda_x}^2 - \epsilon_R \frac{d}{dy};
\mathscr M_{OS \varphi} = 0 ;
\mathscr M_{OSx} = 0;   \nonumber   \\
\mathscr M_{Sqr} = 0;
\mathscr M_{Sq \varphi} = \imath \lambda_x;
\mathscr M_{Sqx} = - \epsilon_R n_{\varphi}    \nonumber     \\
\mathscr M_{Cor} = \frac{d}{dy} + \epsilon_R;
\mathscr M_{Co \varphi}  = \imath n_{\varphi};
\mathscr M_{Cox} = \imath \lambda_x.
\label{mathscrthemssto0}
\end{eqnarray}
{\bf For strong swirl, $S \rightarrow \infty, \epsilon_R \rightarrow 0 $, in Taylor variables, the element operators}  are:

The element operators, $\mathscr L$ and $\mathscr M$, written in Taylor variables, exhibit a dependence on the parameters $S$ and $\epsilon_R$ only through the product $S \sqrt{\epsilon_ R}$, not on their own. Besides, instead of the Reynolds number, $Re$, only the Taylor number, $Ta$, appears as the flow parameter.
\begin{eqnarray}
\mathscr L_{OSr} =
    -  \left(1 - \frac{1}{2 S^2} \right)  ~\left( \frac{(1-y)}{2}  \right)~\left(
     (\imath n_{\varphi}) \frac{d^2 }{d y^2 }
  + (\imath \lambda_x)^2 (\imath n_{\varphi})       \right)
  \nonumber   \\
+  \frac{Ta}{S \sqrt{\epsilon_R}} \left( - 2(\imath n_{\varphi}) - (1-y^2)  (\imath \lambda_x)^4    + \frac{d^4 }{d y^4} + 2 (\imath \lambda_x)^2 \frac{d^2 }{d y^2 } \right)    \nonumber   \\
   \nonumber   \\
 \label{taylorvariabmathscrlosrstoinfty}
\end{eqnarray}
\begin{eqnarray}
\mathscr L_{OS \varphi} = Ta (1-y) (\imath \lambda_x)^2
 \nonumber   \\
\mathscr L_{OSx} = 0 .
\label{taylorvariabmathscrlosvarphixstoinfty}
\end{eqnarray}

\begin{eqnarray}
 \mathscr L_{Sqr} = - \frac{1}{2} (\imath \lambda_x)  \nonumber   \\
 \mathscr L_{Sq \varphi} = \left( (\imath \lambda_x)^3 + (\imath \lambda_x) \frac{d^2}{dy^2} \right) - \frac{Ta}{S \sqrt{ \epsilon_R}} \left( (1-y^2)(\imath \lambda_x)^2 \right)
 .  \nonumber   \\
 \mathscr L_{Sqx} = 0.
\label{taylorvariabmathscrlsqrvarphixstoinfty}
\end{eqnarray}

\begin{eqnarray}
\mathscr M_{OSr} = - \left( \frac{d^2}{dy^2} + (\imath \lambda_x)^2 \right) - \epsilon_R \left( \frac{d}{dy}  \right);
\mathscr M_{OS \varphi} = 0;
\mathscr M_{OSx} = 0;   \nonumber  \\
\mathscr M_{Sqr} = 0;
\mathscr M_{Sq \varphi} = - \imath \lambda_x;
\mathscr M_{Sqx} = 0;  \nonumber   \\
\mathscr M_{Cor} = 0;
\mathscr M_{Co \varphi}  = 0;
\mathscr M_{Cox} = 0.
\label{taylorvariabmathscrmosrvarphixstoinfty}
\end{eqnarray}
{\bf THE ADJOINT FUNCTION and OPERATOR}  \\Since, in the course of the present work use will have to be made of the solution of the {\bf adjoint problem} too, a concise note on this aspect is included in this {\bf Appendix 2} to facilitate reading. The relationship between the pair of the posed problem, $(\mathbf A, \mathscr L)$, and its adjoint $( \mathbf A^{+\ast}, \mathscr L^{+\ast})$ is by definition as follows, {\it vide} e.g. \cite{schmidhenningson}, Chapter 3.3, page 85:
\begin{equation}
\int_{y=-1}^{y=+1} {\bf A^{+*} \left(\mathscr L \bf A \right)}dy = \int_{y=-1}^{y=+1} {\bf A \left(\mathscr L^+ \bf A^+ \right)^*} dy = \int_{y=-1}^{y=+1} {\bf A \left(\mathscr L^{+\ast} \bf A^{+\ast} \right)} dy
\label{defnadjoint}
\end{equation}
where $\ast $ denotes the complex conjugate. The basis of the relationship between the two is {\bf mutual  orthogonality}.

Analytical expressions for the element operators between the cases of {\bf mild} and {\bf strong} swirl are not identical, therefore both are listed separately here:  \\

{\bf For the case of mild swirl}, $S \rightarrow 0 $   \\
\begin{eqnarray}
 \mathscr L_{OSr}^{+ \ast} =  \frac{1}{\sqrt{(1+S^2)}} \left[ -(\imath \lambda_x) \left( (1-y^2) \frac{d^2}{dy^2} - 4y \frac{d}{dy} -2 \right) + (\imath \lambda_x^3) (1-y^2)  \right]      \nonumber  \\
                            + \frac{\epsilon_R}{\sqrt{(1+S^2)}} \left[ (\imath n_{\varphi})  \left( (1-y^2) \frac{d^2}{dy^2} - 4y \frac{d}{dy} -2 \right) - (\imath \lambda_x) \left( (1-y^2) \frac{d}{dy} + 2y \right) \right]    \nonumber   \\
                            + \frac{\epsilon_R}{\sqrt{(1+S^2)}} \left[ (\imath \lambda_x) \left( \frac{y(1.y^2)}{3} \frac{d^2}{dy^2} - \frac{2}{3} \left( (1-3y^2) \frac{d}{dy} - 6y \right) + (\imath \lambda_x^3) \right)  \right]   \nonumber   \\
                            - \frac{\epsilon_R}{\sqrt{(1+S^2)}} \left( 2y (\imath \lambda_x)  \right) + \frac{1}{\sqrt{(1+S^2)}} (-2 + 2 \epsilon_R y)    \nonumber   \\
                            + \frac{\epsilon_R S}{\sqrt{(1+S^2)}} \left[ - (\imath n_{\varphi}) \left( \frac{(1-y)}{2} \frac{d^2}{dy^2} \frac{d}{dy} \right) + \lambda_x^2(\imath n_{\varphi}) \frac{(1-y)}{2} + \frac{1}{2} (\imath n_{\varphi}) \frac{d}{dy} \right]    \nonumber   \\
                            + \frac{1}{Re} \left[ \lambda_x^4 + \frac{d^4}{dy^4} - 2 \lambda_x^2 \frac{d^2}{dy^2} + \epsilon_R (2 \lambda_x^2 \frac{d}{dy})  \right]
\label{mathscrlosradjsto0}
\end{eqnarray}
\begin{eqnarray}
 \mathscr L_{OS \varphi}^{+ \ast} =  \frac{\epsilon_R}{\sqrt{(1+S^2)}} \left[ (\imath n_{\varphi}) \left( (1-y^2) \frac{d}{dy} + 2y \right) + n_{\varphi} \lambda_x \cdot 2y  \right]    \nonumber   \\
                                    - \frac{\epsilon_R S}{\sqrt{(1+S^2)}} \left[ \lambda_x^2 (1-y) \right] - \frac{2}{Re} \frac{d^3}{dy^3}
\label{mathscrlosvarphiadjsto0}
\end{eqnarray}
\begin{eqnarray}
 \mathscr L_{OSx}^{+ \ast} = \frac{\epsilon_R S}{\sqrt{(1+S^2)}} \cdot \frac{1}{2} \left[ \lambda_x n_{\varphi} \right]
\label{mathscrlosxadjsto0}
\end{eqnarray}
\begin{eqnarray}
 \mathscr L_{Sqr}^{+ \ast} = \frac{\epsilon_R}{\sqrt{(1+S^2)}} \left[ (1-y^2) \frac{d}{dy} + 2y \right] + \frac{1}{\sqrt{(1+S^2)}} \left[(-2y)(\imath \lambda_x) \right] + \frac{\epsilon_R S}{\sqrt{(1+S^2)}} \left[ (\imath \lambda_x) \frac{(1-y)}{2} \right]
\label{mathscrlsqradjsto0}
\end{eqnarray}
\begin{eqnarray}
 \mathscr L_{Sq \varphi}^{+ \ast} = \frac{1}{\sqrt{(1+ S^2)}} \left[ - (\lambda_x^2)(1-y^2) - \lambda_x^2 \cdot \epsilon_R \frac{y(1-y^2)}{3} \right] - \frac{\epsilon_R S}{\sqrt{(1+S^2)}} \left[ (n_{\varphi} \lambda_x) \frac{(1-y)}{2} \right]  \nonumber   \\
                                    + \frac{1}{Re} \left[ - (\imath \lambda_x^3) + (\imath \lambda_x) \frac{d^2}{dy^2} - \epsilon_R (\imath \lambda_x) \frac{d}{dy}  \right]
\label{mathscrlsqvarphiadjsto0}
\end{eqnarray}
\begin{eqnarray}
 \mathscr L_{Sqx}^{+ \ast} = \frac{1}{Re} \left[ \epsilon_R \left( \lambda_x^2 (\imath n_{\varphi}) \right) - (\imath n_{\varphi}) \frac{d^2}{dy^2} \right]
\label{mathscrlsqxadjsto0}
\end{eqnarray}
\begin{eqnarray}
 \mathscr L_{Co \varphi}^{+ \ast} = \epsilon_R (\imath n_{\varphi}) ; ~
 \mathscr L_{Cor}^{+ \ast} = - \frac{d}{dy} + \epsilon_R ; ~
\mathscr L_{Cox}^{+ \ast} = (\imath \lambda_x)
\label{mathscrlcorvarphixadjsto0}
\end{eqnarray}

{\bf For the case of strong swirl}, $S \rightarrow \infty $  \\
\begin{eqnarray}
 \mathscr L_{OSr}^{+ \ast} = - \frac{2}{S} \left(1 - \frac{1}{2 S^2} \right) (\imath \lambda_x) - \frac{1}{S} \left( 1 - \frac{1}{2S^2} \right) \left[ (\imath \lambda_x) \left( (1-y^2) \frac{d^2}{dy^2} - 4y \frac{d}{dy} - 2 \right) - (\imath \lambda_x^3) (1-y^2) \right]    \nonumber   \\
                             - \left( 1- \frac{1}{2S^2}\right) \left[ \frac{(\imath n_{\varphi})}{2} \left( (1-y) \frac{d^2}{dy^2} - 2 \frac{d}{dy} \right) - \lambda_x^2 (\imath n_{\varphi}) (1-y) \right]
                             + \frac{1}{Re} \left[ \lambda_x^4 + \frac{d^4}{dy^4} - 2 \lambda_x^2 \frac{d^2}{dy^2} \right]   \nonumber   \\
                             - \epsilon_R \left[ \frac{1}{S}\left(1- \frac{1}{2S^2} \right) \left((\imath \lambda_x)\left( (1-y^2) \frac{d}{dy} + 2y \right) + (\imath \lambda_x) \left( \frac{y(1-y^2)}{3} \frac{d^2}{dy^2} - 2(1-3y) \frac{d}{dy} -6y \right) \right) \right]   \nonumber   \\
 + \epsilon_R \left[ \frac{1}{S} \left(1-\frac{1}{2S^2} \right) \left( (\imath \lambda_x^3)\left(\frac{y(1-y^2)}{3} \right) \right) \right]   \nonumber   \\
 + \epsilon_R \left[ \left( 1- \frac{1}{2S^2} \right) \left( (\imath n_{\varphi}) \left( \frac{(1-y^2)}{2} \frac{d^2}{dy^2} - 2y \frac{d}{dy} -1 \right) - (\imath n_{\varphi}) \lambda_x^2 \frac{(1-y^2)}{2}  \right)  \right]
 + \epsilon_R \left[ \frac{2}{Re} \left( \lambda_x^2 \frac{d}{dy} - \frac{d^3}{dy^3} \right) \right]
\label{mathscrlosradjstoinfty}
\end{eqnarray}
\begin{eqnarray}
 \mathscr L_{OS \varphi}^{+ \ast} = \left( 1 - \frac{1}{2S^2} \right) \left( -\imath \lambda_x^2 (1-y) \right) + \epsilon_R \left[ \left( 1-\frac{1}{2S^2} (\lambda_x^2) (1-y^2) \right) \right]
\label{mathscrlosvarphiadjstoinfty}
\end{eqnarray}
\begin{eqnarray}
 \mathscr L_{OSx}^{+ \ast} = 0
\label{mathscrlosxadjstoinfty}
\end{eqnarray}
\begin{eqnarray}
 \mathscr L_{Sqr}^{+ \ast} = 0
\end{eqnarray}
\begin{eqnarray}
 \mathscr L_{Sq \varphi}^{+ \ast} = \frac{1}{S} \left(1- \frac{1}{2S^2} \right) \lambda_x^2 (1-y^2) + \frac{1}{Re} \left( -(\imath \lambda_x^3) + (\imath \lambda_x) \frac{d^2}{dy^2} \right)
                                  + \epsilon_R \frac{1}{S} \left(1- \frac{1}{2S^2} \right) \lambda_x^2 \left( \frac{y(1-y^2)}{3} \right)   \nonumber   \\
                                  - \epsilon_R \left[ \left(1- \frac{1}{2S^2}  \right) (- n_{\varphi} \lambda_x) \left(\frac{(1-y)}{2} \right) - \frac{1}{Re} \left(- (\imath \lambda_x) \frac{d}{dy}  \right)   \right]
\end{eqnarray}
\begin{eqnarray}
 \mathscr L_{Sqx}^{+ \ast} = 0
\end{eqnarray}
\begin{eqnarray}
 \mathscr L_{Co \varphi}^{+ \ast} = \epsilon_R (\imath n_{\varphi}) ; ~
 \mathscr L_{Cor}^{+ \ast} = - \frac{d}{dy} + \epsilon_R ; ~
\mathscr L_{Cox}^{+ \ast} = (\imath \lambda_x)
\label{mathscrlcorvarphixadjstoinfty}
\end{eqnarray}

%
%


\section*{Appendix 3: The Element Operators in $\mathscr D$  to the orders $O(1)$, $O(\epsilon_A^{\frac{1}{2}})$ and $O(\epsilon_A)$, equations (\ref{eqnforu1}), (\ref{eqnforu2}) and (\ref{eqnforu3}). }
The element operators to the leading order in its matrix form, denoted $\mathscr D_1$, is formally the same as $\mathscr D$, {\it vide} Appendix 1, there (\ref{generalisednonlinos}, \ref{generalisednonlinsq},  \ref{continuityinuxuruvarphi}, \ref{rearrangednonlinearostildevariab},  \ref{rearrangednonlinearsqtildevariab},  \ref{rearrangedkontitildevariab},  \ref{erwnlos4}, \ref{erwnlsq4},  \ref{sskontigl4}) . However it should be noted that $\mathscr D_1$ assumes different forms when written in {\bf conventional} and {\bf Taylor} variables.

For ease of reading, the asymptotic expansions for $\mathbf u$, $\mathscr D$ and $\mathscr N$ from (\ref{asymptexpmathbfumathscrd}) are repeated here.
\begin{eqnarray}
 \mathbf  u \simeq \epsilon_A^{\frac{1}{2}} \mathbf u_1 + \epsilon_A \mathbf u_2 + \epsilon_A^{\frac{3}{2}} \mathbf u_3 + O(\epsilon_A^2)    \nonumber   \\
 \mathscr D \simeq \mathscr D_1 + \epsilon_A^{\frac{1}{2}} \mathscr D_2 + \epsilon_A \mathscr D_3 + O(\epsilon_A^{\frac{3}{2}})    \nonumber   \\
 \mathscr N \simeq \epsilon_A^{\frac{1}{2}} \mathscr N_1 + \epsilon_A \mathscr N_2 + \epsilon_A^{\frac{3}{2}} + O(\epsilon_A^2)      \nonumber
\end{eqnarray}

\subsection*{The Element Operators to the order $O(1)$ in conventional variables:}

 {\bf  In the Orr-Sommerfeld equation:}
\begin{eqnarray}
\mathscr D_{1OSr} =  - \frac{\partial}{\partial t} \left( \frac{\partial^2 }{\partial x^2} + \frac{\partial^2 }{\partial y^2} \right) - \frac{2}{\sqrt{(1+S^2)}} \frac{\partial }{\partial x} - \frac{(1-y^2)}{\sqrt{(1+S^2)}} \left( \frac{\partial^3 }{\partial x \partial y^2} + \frac{\partial^3 }{\partial x^3}\right)   \nonumber  \\
+ \frac{1}{Re} \left( \frac{\partial^4 }{\partial x^4} + \frac{\partial^4 }{\partial y^4} + 2 \frac{\partial^4 }{\partial x^2 \partial y^2} \right)
+ \epsilon_R \left[- \frac{\partial}{\partial t} \left( \frac{\partial }{\partial y} \right) + \frac{4y}{\sqrt{(1+S^2)}} \left(\frac{\partial }{\partial x} \right)  \right]    \nonumber   \\
+ \epsilon_R \left[- \frac{(1-y^2)}{\sqrt{(1+S^2)}} \frac{\partial^2 }{\partial y \partial x} + \frac{1}{\sqrt{(1+S^2)}} \frac{y(1-y^2)}{3} \left( \frac{\partial^3 }{\partial x \partial y^2} + \frac{\partial^3 }{\partial x^3} \right) \right]   \nonumber  \\
+ \epsilon_R \frac{S}{\sqrt{(1+S^2)}} \left[ \frac{(1-y)}{2} \left( - \frac{\partial^3 }{\partial y^2 \partial \varphi} - \frac{\partial ^3 }{\partial x^2 \partial \varphi}   \right)  \right] + \frac{2 \epsilon_R}{Re} \left[ \left(\frac{\partial^3 }{\partial x^2 \partial y} - \frac{\partial ^3  }{\partial y^3} \right) \right]  \nonumber  \\
\mathscr D_{1OS \varphi} =  \epsilon_R \frac{S}{\sqrt{(1+S^2)}}  2  \frac{\partial ^2}{\partial x^2}       \nonumber   \\
\mathscr D_{1OS x} = 0
\label{mathscrd1osconventional}
\end{eqnarray}
{\bf In the Squire equation:}:
\begin{eqnarray}
 \mathscr D_{1Sqr} =  \frac{S}{\sqrt{(1+S^2)}} \frac{1}{2} \frac{\partial}{ \partial x} +  \epsilon_R \frac{S}{\sqrt{(1+S^2)}} \left[ \frac{(1-y)}{2} \left( - \frac{\partial}{\partial x}\right) -y \frac{\partial}{\partial x} \right]       \nonumber    \\                                     \nonumber   \\
 \mathscr D_{1Sq \varphi} = -  \frac{\partial}{\partial t} \left(\frac{\partial }{\partial x} \right) - \frac{1}{\sqrt{(1+S^2)}} \left( 1-y^2\right) \left(\frac{\partial^2 }{\partial x^2} \right) +  \frac{1}{Re} \left( \frac{\partial^3 }{\partial x^3} + \frac{\partial^3 }{\partial x \partial y^2} \right)   \nonumber  \\
+ \frac{\epsilon_R}{\sqrt{(1+S^2)}} \left[ \left( \frac{y(1-y^2)}{3} \right) \left( \frac{\partial ^2 }{\partial x^2} \right)  \right]  \nonumber  \\
+ \epsilon_R \frac{S}{\sqrt{(1+S^2)}} \left[\left(\frac{(1-y)}{2} \right) \left(- \frac{\partial^2 }{\partial \varphi \partial x}  \right)  \right]
- \frac{\epsilon_R}{Re} \left[ - \frac{\partial^2 }{\partial x \partial y} \right]  \nonumber   \\
\mathscr D_{1Sqx}= - \frac{\epsilon_R}{Re} \left[ \frac{\partial ^3}{\partial x^2 \partial \varphi} + \frac{\partial ^3}{\partial y^2 \partial \varphi} \right]
\label{mathscrd1sqconventional}
\end{eqnarray}
{\bf In the Continuity Equation:}
\begin{eqnarray}
\mathscr D_{1Cor} = \frac{\partial}{\partial y} + \epsilon_R     \nonumber  \\
\mathscr D_{1Co \varphi} = \epsilon_R \frac{\partial}{\partial \varphi}     \nonumber   \\
\mathscr D_{1Cox} = \frac{\partial }{\partial x} .
\label{mathscrd1coconventional}
\end{eqnarray}
\subsection*{The Element Operators to the order $O(1)$ in Taylor Variables}
{\bf In the Orr-Sommerfeld equation: }
\begin{eqnarray}
\mathscr D_{1OSr} =  -\frac{\partial}{\partial \hat t} \left( \frac{\partial ^2 }{\partial \hat x^2} +  \frac{\partial ^2 }{\partial \hat y^2} \right)
+  \left( \frac{\partial ^4 }{\partial \hat x^4} +  \frac{\partial ^4 }{\partial \hat y^4}+ 2 \frac{\partial ^4 }{\partial \hat y^2 \partial \hat x^2}   \right)       \nonumber    \\
+ \frac{\sqrt{Ta}}{S \sqrt{\epsilon_R}} \left(-2 \frac{\partial }{\partial \hat x} - (1 - \hat y^2)\left(\frac{\partial ^3 }{\partial \hat x \partial \hat y^2}+ \frac{\partial ^3 }{\partial \hat x^3} \right) \right)        \nonumber \\
\mathscr D_{1OS \varphi} = Ta(1-\hat y) \frac{\partial^2}{\partial \hat x^2}             \nonumber    \\
\mathscr D_{1OSx} = 0
\label{mathscrd1ostaylor}
\end{eqnarray}
{\bf In the Squire equation: }
\begin{eqnarray}
\mathscr D_{1Sqr} = - \frac{1}{2} \frac{\partial}{\partial \hat x}      \nonumber   \\
\mathscr D_{1Sq \varphi} = - \frac{\partial^2 }{\partial \hat t \partial \hat x} - \left( \frac{\partial ^3 }{\partial \hat x^3} + \frac{\partial ^3 }{\partial \hat x \partial \hat y^2} \right) - \frac{\sqrt{Ta}}{S \sqrt{\epsilon_R}}  \left((1 - \hat y ^2) \frac{\partial ^2 }{\partial \hat x^2} \right)              \nonumber \\
\mathscr D_{1Sqx} = 0.
\label{mathscrd1sqtaylor}
\end{eqnarray}
{\bf In the Continuity equation: }
\begin{eqnarray}
\mathscr D_{1Cor} =  \frac{\partial }{\partial \hat y}            \nonumber  \\
\mathscr D_{1Co \varphi} = \frac{\partial}{\partial \hat x}    \nonumber  \\
\mathscr D_{1Cox} = 0
\label{mathscrd1cotaylor}
\end{eqnarray}
\subsection*{The Element Operators to the order $O(\epsilon_A^{\frac{1}{2}})$ in conventional variables}
{\bf In the Orr-Sommerfeld, Squire and Continuity equations}
\begin{eqnarray}
\mathscr D_{2OSr} = - \left[ \left( \frac{\partial \omega}{\partial \lambda_x} \right)_{N,r} \frac{\partial^3}{\partial \xi_x \partial x^2} + 2 \left( \frac{\partial^3}{\partial t \partial x \partial \xi_x}\right) \right]
- \left[ \left( \frac{\partial \omega}{\partial \lambda_x}\right)_{N,r} \frac{\partial^3}{\partial \xi_x \partial y^2} +  \right]  - \frac{2}{\sqrt{(1+S^2)}} \frac{\partial}{\partial \xi_x}     \nonumber   \\
- \frac{(1-y^2)}{\sqrt{(1+S^2)}} \left(\frac{\partial^ 3}{\partial \xi_x \partial y^2} +3 \frac{\partial^3}{\partial \xi_x \partial x^2} \right) +
\frac{1}{Re} \left( 4 \frac{\partial^4}{\partial x^3 \partial \xi_x} + 4 \frac{\partial^4}{ \partial x \partial \xi_x \partial y^2}\right) \nonumber   \\
+ \epsilon_R \left[ \left( \frac{\partial \omega}{\partial \lambda_x} \right)_{N,r} \frac{\partial^2}{\partial \xi_x \partial y} +   \right] + \frac{\epsilon_R}{\sqrt{(1+S^2)}} \left( 4y \frac{\partial}{\partial \xi_x} \right)  \nonumber    \\
+ \frac{\epsilon_R}{\sqrt{(1+S^2)}} \left[ -(1-y^2) \frac{\partial^2}{\partial \xi_x \partial y} +  \frac{y(1-y^2)}{3} \left( \frac{\partial^3}{\partial \xi_x \partial y^2} + 3 \frac{\partial^3}{\partial \xi_x \partial x^2}\right) \right]  \nonumber   \\
- \epsilon_R \frac{S}{\sqrt{(1+S^2)}} \frac{(1-y)}{2}\left( \frac{\partial^3}{\partial x \partial \varphi \partial \xi_x}\right) + \frac{2 \epsilon_R}{Re} \left( 2 \frac{\partial^3}{\partial x \partial \xi_x \partial y} \right)
\nonumber   \\
\mathscr D_{2OS \varphi} =  \epsilon_R \frac{S}{\sqrt{(1+S^2)}} \left( 4 \frac{\partial^2}{\partial x \partial \xi_x} \right)  \nonumber   \\
 \mathscr D_{2OSx}  =  0  \nonumber   \\
 \mathscr D_{2Sqr}   =  \frac{S}{\sqrt{(1+S^2)}} \frac{1}{2} \frac{\partial}{\partial \xi_x} + \epsilon_R \frac{S}{\sqrt{(1+S^2)}} \left[ - \frac{(1-y)}{2} \frac{\partial}{\partial \xi_x} - y \frac{\partial}{\partial \xi_x} \right]  \nonumber   \\
 \mathscr D_{2Sq \varphi}  =  - \left[ - \left( \frac{\partial \omega}{\partial \lambda_x} \right)_{N,r} \frac{\partial^2}{\partial \xi_x \partial x}  \right] - \frac{\partial^2}{\partial \xi_x \partial t}  - \frac{1}{\sqrt{(1+S^2)}}(1-y^2) \cdot 2 \frac{\partial^2}{\partial \xi_x \partial x} +      \nonumber   \\
\frac{S}{\sqrt{(1+S^2)}} \cdot \frac{1}{2} \frac{\partial}{\partial \xi_x} + \frac{1}{Re} \left( 3 \frac{\partial^3}{\partial \xi_x \partial x^2} + \frac{\partial^3}{\partial \xi_x \partial y^2} \right) + \frac{\epsilon_R}{\sqrt{(1+S^2)}}\left[ \left( \frac{y(1-y^2 )}{3} \right) \cdot 2 \frac{\partial^2}{\partial \xi_x \partial x} + (1-y^2) \frac{\partial^2}{\partial \xi_x \partial \varphi}  \right]  +    \nonumber   \\
\epsilon_R \frac{S}{\sqrt{(1+S^2)}} \left[ \left( \frac{(1-y)}{2} \right) \left(- \frac{\partial^2}{\partial \varphi \partial \xi_x} \right)  \right] + \frac{\epsilon_R}{Re} \left[  \frac{\partial^2}{\partial \xi_x \partial y} \right]    \nonumber    \\
 \mathscr D_{2Sqx}  =  -\frac{\epsilon_R}{Re} \left[ \frac{\partial^3}{\partial x^2 \partial \varphi} \right]  -  \epsilon_R \left[  \left(\frac{\partial \omega}{\partial \lambda_x} \right)_{N,r} \frac{\partial^2}{\partial \xi_x \partial \varphi}   \right]  \nonumber   \\
 \mathscr D_{2cor}   = 0  \nonumber   \\
 \mathscr D_{2co \varphi}  = 0  \nonumber   \\
 \mathscr D_{2cox}   = \frac{\partial}{\partial \xi_x}
 \label{mathscrd2ossqcontconven}
\end{eqnarray}

\subsection*{The Element Operators to the order $O(\epsilon_A^{\frac{1}{2}})$ in Taylor Variables}
{\bf In the Orr-Sommerfeld, Squire and Continuity equations}
\begin{eqnarray}
\mathscr D_{2OSr} = -\left[ \left( \frac{\partial \omega}{\partial \lambda_x} \right)_{N,r} \frac{\partial^3}{\partial \xi_x \partial \hat x^2}  +  \left(2 \frac{\partial^3}{\partial \hat t \partial \hat x \partial \xi_x} \right) \right]
- \left[ \left( \frac{\partial \omega}{\partial \lambda_x} \right)_{N,r} \frac{\partial^3}{\partial \xi_x \partial y^2} +  \right] +
\left( 4 \frac{\partial^4}{\partial \xi_x \partial \hat x^3} +2\frac{\partial^4}{\partial \hat x \partial \xi_x \partial y^2}\right) +  \nonumber   \\
\frac{\sqrt{Ta}}{S \sqrt{\epsilon_R}} \left(-2 \frac{\partial}{\partial \xi_x} - (1-y^2)\left( \frac{\partial^3}{\partial \xi_x \partial y^2} + 3 \frac{\partial^3}{\partial \xi_x \partial \hat x^2} \right) \right)   \nonumber   \\
\mathscr D_{2OS \varphi} = Ta(1-\hat y) \left(2 \frac{\partial^2}{\partial \hat x \partial \xi_x} \right)            \nonumber    \\
\mathscr D_{2OSx} = 0   \nonumber   \\
\mathscr D_{2Sqr}  =  - \frac{1}{2} \frac{\partial}{\partial \xi_x}    \nonumber   \\
\mathscr D_{2Sq \varphi} =  \left[  \left( \frac{\partial \omega}{\partial \lambda_x} \right)_{N,r} \frac{\partial^2}{\partial \xi_x \partial \hat x}  \right] - \left( 3 \frac{\partial^3}{\partial \xi_x \partial \hat x^2} + \frac{\partial^3}{\partial \xi_x \partial \hat y^2} \right) - \frac{Ta}{S \sqrt{\epsilon_R}} \left( 2(1-\hat y^2 )  \frac{\partial^2}{\partial \xi_x \partial \hat x} \right) \nonumber   \\
\mathscr D_{2Sqx} = 0     \nonumber   \\
\mathscr D_{2cor}   = 0    \nonumber   \\
\mathscr D_{2co \varphi}  = 0   \nonumber   \\
\mathscr D_{2cox}  = \frac{\partial}{\partial \xi_x}
\label{mathscrd2ossqconttaylor}
\end{eqnarray}

\subsection*{The Element Operators to the order $O(\epsilon_A)$ in conventional variables}
{\bf In the Orr-Sommerfeld, Squire and Contonuity equations}
\begin{eqnarray}
\mathscr D_{3OSr}  = - \left[ -2\left( \frac{\partial \omega}{\partial \lambda_x} \right)_{N,r} \frac{\partial^3}{\partial x \partial \xi_x^2}  + \frac{\partial^3}{\partial t \partial \xi_x^2} + \frac{\partial^3}{\partial \tau \partial x^2} \right] - \frac{\partial^3}{\partial \tau \partial y^2} -  \nonumber    \\
\frac{2}{\sqrt{(1+S^2)}} \frac{\partial}{\partial \xi_x} -  \frac{(1-y^2)}{\sqrt{(1+S^2)}} \left( \frac{\partial^3}{\partial \xi_x \partial y^2}+ 3\frac{\partial^3}{\partial x \partial \xi_x^2} \right) + \frac{1}{Re} \left( 6 \frac{\partial^4}{\partial x^2 \partial \xi_x^2} + 4 \frac{\partial ^4}{\partial x^2 \partial y^2} \right) +             \nonumber   \\  \frac{
\epsilon_R}{\sqrt{(1+S^2)}} \left[ - \frac{\partial^2}{\partial \tau \partial y} - {(1-y ^2)} \frac{\partial^2}{\partial \xi_x} + \frac{y(1-y^2)}{3}\left( 3 \frac{\partial^3}{\partial x \partial \xi_x^2} \right) \right] - \epsilon_R \frac{S}{\sqrt{(1+S^2)}} \left[ \frac{(1-y)}{2} \left( - \frac{\partial^3}{\partial \xi_x^2 \partial y}\right) \right] +      \nonumber   \\
\frac{2 \epsilon_R}{Re} \left[ \frac{\partial^3}{\partial \xi_x^2 \partial y} \right]                                \nonumber   \\
\mathscr D_{3OS \varphi}  =  \epsilon_R \frac{S}{\sqrt{(1+S^2)}} 2 \frac{\partial^2}{\partial \xi_x^2}   \nonumber   \\
\mathscr D_{3OSx}  = 0   \nonumber   \\
\mathscr D_{3Sqr}   =  0  \nonumber   \\
\mathscr D_{3Sq \varphi}  =  - \left[ -\left( \frac{\partial \omega}{\partial \lambda_x} \right)_{N,r} \frac{\partial}{\partial \xi_x}  \right]  - \frac{1}{\sqrt{(1+S^2)}} (1-y^2)\left( \frac{\partial^2}{\partial \xi_x^2} \right) + \frac{1}{Re} \left(3 \frac{\partial^3}{\partial x \partial \xi_x^2}  \right) +   \nonumber   \\
\frac{\epsilon_R}{\sqrt{(1+S^2)}} \left[ \left( \frac{y(1-y^2)}{3} \right)\frac{\partial^2}{\partial \xi_x^2} \right] \nonumber   \\
\mathscr D_{3Sqx}   =  \frac{\epsilon_R}{Re}\left[ \frac{\partial^3}{\partial \xi_x^2 \partial \varphi} \right]    \nonumber   \\
\mathscr D_{3cor}   =   0     \nonumber   \\
\mathscr D_{3co \varphi}   =   0      \nonumber   \\
\mathscr D_{3cox} =   0
\label{mathscrd3ossqcontconven}
\end{eqnarray}

\subsection*{The Element Operators to the order $O(\epsilon_A)$ in Taylor variables}
{\bf In the Orr-Sommerfeld, Squire and Continuity equations}
\begin{eqnarray}
 \mathscr D_{3OSr}   = - \left[-2\left( \frac{\partial \omega}{\partial \lambda_x} \right)_{N,r} \frac{\partial^ 3}{\partial x  \partial \xi_x^ 2}  + \frac{\partial^3}{\partial t \partial \xi_x^2} + \frac{\partial^3}{\partial x^2}  \right]  - \frac{\partial^3}{\partial \tau \partial y^2}  +    \nonumber   \\
 \left( 6 \frac{\partial^4}{\partial x^2 \partial \xi_x^2} + 2 \frac{\partial^ 4}{\partial \xi_x^2 \partial y^2}  \right) + \frac{Ta}{S \sqrt{\epsilon_R}}\left[- (1-y^2) \left( 3 \frac{\partial^3}{\partial x \partial \xi_x^2} \right) \right]        \nonumber   \\
 \mathscr D_{3OS \varphi} = \epsilon_R \frac{S}{\sqrt{(1+S^2)}} \cdot 2 \frac{\partial^2}{\partial \xi_x^2} \nonumber   \\
 \mathscr D_{3OSx}  =  0      \nonumber    \\
 \mathscr D_{3Sqr}   = \frac{S}{\sqrt{(1+S^2)}} \cdot 0 + \epsilon_R \frac{S}{\sqrt{(1+S^2)}} \cdot 0 =  0      \nonumber    \\
 \mathscr D_{3Sq \varphi}  =  \left[ - \left(\frac{\partial \omega}{\partial \lambda_x} \right)_{N,r} \frac{\partial^2}{\partial \xi_x^2} + \frac{\partial^ 2}{\partial \tau \partial x}  \right] - \left( 3 \frac{\partial^3}{\partial x \partial \xi_x^2}\right)  \nonumber   \\
 \mathscr D_{3Sqx}   =   - \frac{\epsilon_R}{Re} \left[ \frac{\partial^3}{\partial \xi_x^2 \partial \varphi} \right] \nonumber   \\
 \mathscr D_{3cor}  =  0   \nonumber   \\
 \mathscr D_{3co \varphi}   =  0   \nonumber   \\
 \mathscr D_{3cox}   =  0
 \label{mathscrd3ossqconttaylor}
\end{eqnarray}


%
%
%

\section*{Appendix 4: Influence of nonlinearity visible in the Ginzburg-Landau equation for Amplitude Evolution during transition in Spiral Poiseuille Flow.}

For the particular flow example
under consideration here the typical tensor of nonlinear terms in (\ref{productformu1otimesu2supthetan}) may be written in the following form:
\begin{equation}
 \left(\mathbf u_1 \otimes \mathbf u_2  \right)^{(\Theta_N)} = B_N (B_N B_N^\ast) \mathscr H (y) \exp (\imath \Theta_N) + B_N^\ast (B_N B_N^\ast) \mathscr H^\ast (y) \exp (- \imath \Theta_N),
\label{formofn3}
\end{equation}
where $\mathscr H$ and its conjugate complex $\mathscr H^\ast$ are symmetric tensors that depend only upon $y$. They are evaluable from the set of eigenfunctions of the classical
eigenvalue problem for small-amplitude disturbances at the chosen point on the surface of neutral stability. It is also clear that the tensor $\mathscr H$ differs from flow to
flow, and even in a certain flow, on the chosen point on the surface of neutral stability.
For the flow example presently in consideration, the tensor $\mathscr H $ may be written in terms of its elements as follows:
\begin{eqnarray}
\mathscr H = \left(
\begin{array}{ccc}
  \mathscr H_{OSr}   &  \mathscr H_{OS \varphi}   &   \mathscr H_{OSx}\\
  \mathscr H_{Sqr}   &  \mathscr H_{Sq \varphi}   &   \mathscr H_{Sqx}\\
  \mathscr H_{Cor}   &  \mathscr H_{Co \varphi}   &   \mathscr H_{Cox}
\end{array}
\right).
\label{defntensorh}
\end{eqnarray}

Comparison of (\ref{defntensorh}) with (\ref{productformu1otimesu2supthetan}) shows the elements of $\mathscr H$ to be as follows:

\subsubsection*{Spiral Poiseuille Flow}
\begin{eqnarray}
 \mathscr H_{OSr} = \left( 2 (\mathbf a_N \otimes \mathbf w_2^{(0)} + \mathbf w_2^{(0)} \otimes \mathbf a_N)  + 8 (\mathbf a_N^\ast \otimes \mathbf w_2^{(2 )} + \mathbf w_2^{(2 )} \otimes \mathbf a_N^\ast)  \right)_{rr} =    \nonumber   \\
\left( 2 (A_{Nr} w_{2r}^{(0)} + w_{2r}^{(0)} A_{Nr})  + 8 (A_{Nr}^\ast w_{2r}^{(2 )} + w_{2r}^{(2 )} A_{Nr}^\ast)  \right)   ;                                  \nonumber   \\
 \mathscr H_{OS \varphi} = \left( 2 (\mathbf a_N \otimes \mathbf w_2^{(0)} + \mathbf w_2^{(0)} \otimes \mathbf a_N)  + 8 (\mathbf a_N^\ast \otimes \mathbf w_2^{(2 )} + \mathbf w_2^{(2 )} \otimes \mathbf a_N^\ast)  \right)_{r \varphi} =    \nonumber   \\
 \left( 2 (A_{Nr} w_{2 \varphi}^{(0)} + w_{2r}^{(0)} A_{N \varphi})  + 8 (A_{Nr}^\ast w_{2 \varphi}^{(2 )} + w_{2r}^{(2 )} A_{N \varphi}^\ast)  \right)  ;    \nonumber   \\
 \mathscr H_{OSx} =  \left( 2 (\mathbf a_N \otimes \mathbf w_2^{(0)} + \mathbf w_2^{(0)} \otimes \mathbf a_N)  + 8 (\mathbf a_N^\ast \otimes \mathbf w_2^{(2 )} + \mathbf w_2^{(2 )} \otimes \mathbf a_N^\ast)  \right)_{rx} =    \nonumber   \\
\left( 2 (A_{Nr} w_{2x}^{(0)} + w_{2r}^{(0)} A_{Nx})  + 8 (A_{Nr}^\ast w_{2x}^{(2 )} +  w_{2r}^{(2 )} A_{Nx}^\ast)  \right) ;    \nonumber   \\                                  \nonumber   \\
\mathscr H_{Sqr} =   \left( 2 (\mathbf a_N \otimes \mathbf w_2^{(0)} + \mathbf w_2^{(0)} \otimes \mathbf a_N)  + 8 (\mathbf a_N^\ast \otimes \mathbf w_2^{(2 )} + \mathbf w_2^{(2 )} \otimes \mathbf a_N^\ast)  \right)_{\varphi r} =    \nonumber   \\
 \left( 2 ( A_{N \varphi}  w_{2r}^{(0)} +  w_{2 \varphi}^{(0)} A_{Nr})  + 8 ( A_{N \varphi}^\ast  w_{2r}^{(2 )} +  w_{2 \varphi}^{(2 )} A_{Nr}^\ast)  \right) ;                             \nonumber   \\
 \mathscr H_{Sq \varphi} =    \left( 2 (\mathbf a_N \otimes \mathbf w_2^{(0)} + \mathbf w_2^{(0)} \otimes \mathbf a_N)  + 8 (\mathbf a_N^\ast \otimes \mathbf w_2^{(2 )} + \mathbf w_2^{(2 )} \otimes \mathbf a_N^\ast)  \right)_{\varphi \varphi} =    \nonumber   \\
  \left( 2 ( A_{N \varphi}  w_{2 \varphi}^{(0)} +  w_{2 \varphi}^{(0)} A_{N \varphi})  + 8 (A_{N \varphi}^\ast w_{2 \varphi}^{(2 )} +  w_{2 \varphi}^{(2 )}  A_{N \varphi}^\ast)  \right)  ;                         \nonumber   \\
 \mathscr H_{Sqx} =   \left( 2 (\mathbf a_N \otimes \mathbf w_2^{(0)} + \mathbf w_2^{(0)} \otimes \mathbf a_N)  + 8 (\mathbf a_N^\ast \otimes \mathbf w_2^{(2 )} + \mathbf w_2^{(2 )} \otimes \mathbf a_N^\ast)  \right)_{\varphi x} =    \nonumber   \\
  \left( 2 ( A_{N \varphi}  w_{2x}^{(0)} +  w_{2 \varphi}^{(0)} A_{Nx})  + 8 ( A_{N \varphi}^\ast  w_{2x}^{(2 )} +  w_{2 \varphi}^{(2 )} A_{Nx}^\ast)  \right) ;                              \nonumber   \\
\mathscr H_{Cor} =    \left( 2 (\mathbf a_N \otimes \mathbf w_2^{(0)} + \mathbf w_2^{(0)} \otimes \mathbf a_N)  + 8 (\mathbf a_N^\ast \otimes \mathbf w_2^{(2 )} + \mathbf w_2^{(2 )} \otimes \mathbf a_N^\ast)  \right)_{xr} =    \nonumber   \\
    \left( 2 ( A_{Nx} w_{2r}^{(0)} +  w_{2x}^{(0)}  A_{Nr})  + 8 ( A_{Nx}^\ast w_{2r}^{(2 )} +  w_{2x}^{(2 )} A_{Nr}^\ast)  \right)  ;                          \nonumber   \\
 \mathscr H_{Co \varphi} =    \left( 2 (\mathbf a_N \otimes \mathbf w_2^{(0)} + \mathbf w_2^{(0)} \otimes \mathbf a_N)  + 8 (\mathbf a_N^\ast \otimes \mathbf w_2^{(2 )} + \mathbf w_2^{(2 )} \otimes \mathbf a_N^\ast)  \right)_{x \varphi} =    \nonumber   \\
       \left( 2 ( A_{Nx}  w_{2 \varphi}^{(0)} + w_{2x}^{(0)} A_{N \varphi})  + 8 ( A_{Nx}^\ast  w_{2 \varphi}^{(2 )} +  w_{2x}^{(2 )} A_{N \varphi}^\ast)  \right)  ;                \nonumber   \\
 \mathscr H_{Cox} =   \left( 2 (\mathbf a_N \otimes \mathbf w_2^{(0)} + \mathbf w_2^{(0)} \otimes \mathbf a_N)  + 8 (\mathbf a_N^\ast \otimes \mathbf w_2^{(2 )} + \mathbf w_2^{(2 )} \otimes \mathbf a_N^\ast)  \right)_{xx} =    \nonumber   \\
    \left( 2 ( A_{Nx} w_{2x}^{(0)} +  w_{2x}^{(0)} A_{Nx})  + 8 ( A_{Nx}^\ast w_{2x}^{(2 )} +  w_{2x}^{(2 )} A_{Nx}^\ast)  \right) .
\label{elementstensorhswirl}
\end{eqnarray}
The elements of $\mathscr H^\ast$ are obtainable from taking the complex conjugates of the elements of $\mathscr H$ from  (\ref{elementstensorhswirl}).

The structure of the elements of the tensor $\mathscr H$ in the flow example is worthy of note. Each element is a sum of two products, wherein the first product is a
multiplication of the shape of a velocity component on the surface of neutral stability with the shape of a component of the correction to the basic flow due to Reynolds
stresses. The former is an eigenfunction of the equations to the leading order and the latter is non-periodic. The second product is a multiplication of the conjugate complex of the
velocity component on the surface of neutral stability with the component of the correction of twice the period, $2 \Theta_N$, to the original velocity disturbance.

\subsubsection*{An illustrative example for the nonlinear term }
To facilitate transparency during the operation of using (\ref{elementstensorhswirl}) to evaluate $\mathscr N_3$ in (\ref{eqnforu3})  we present here
a few intermediate steps in the evaluation of a typical nonlinear term, for which we choose $u_r u_x$. To the order
in question here, viz.  $O(\epsilon_A^{\frac{3}{2}})$, and period $\Theta_N$, using (\ref{elementstensorhswirl}) we may write
$u_r u_x = u_{1r} u_{2x} + u_{2r} u_{1x}$ which, together with (\ref{usub1supwave1otimesusub1supwave2}), (\ref{summenansatzforu2}) and, (\ref{usub2sup012flowex1}),  may be written as follows:
\subsubsection*{Spiral Poiseille Flow}
\begin{eqnarray}
 u_{1r} u_{2x} = B_N (B_N B_N^\ast) \cdot 2 A_{Nr} w_{2x}^{(0)} \exp (\imath \Theta_N) + B_N^\ast (B_N B_N^\ast) \cdot 2 A_{Nr}^\ast w_{2x}^{(0)} \exp (-\imath \Theta_N)   \nonumber  \\
                + B_N^\ast B_N^2 \cdot 2 A_{Nr}^\ast \cdot 4 u_{2x}^{(2)} \exp (\imath \Theta_N) + B_N B_N^{\ast2} \cdot 2 A_{Nr} \cdot 4 u_{2x}^{(2) \ast} \exp (-\imath \Theta_N)   \nonumber    \\
u_{1x} u_{2r} = B_N (B_N B_N^\ast) \cdot 2 A_{Nx} w_{2r}^{(0)} \exp (\imath \Theta_N) + B_N^\ast (B_N B_N^\ast) \cdot 2 A_{Nx}^\ast w_{2r}^{(0)} \exp (-\imath \Theta_N)   \nonumber  \\
                + B_N^\ast B_N^2 \cdot 2 A_{Nx}^\ast \cdot 4 u_{2r}^{(2)} \exp (\imath \Theta_N) + B_N B_N^{\ast2} \cdot 2 A_{Nx} \cdot 4 u_{2r}^{(2) \ast} \exp (-\imath \Theta_N)
\label{u1ru2xandu1xu2r}
\end{eqnarray}
The correspondance between (\ref{elementstensorhswirl}) and (\ref{u1ru2xandu1xu2r}) is clear on a close inspection, as would be expected. The $y$-dependent coefficients of
$\exp (\imath \Theta_N)$ in the expressions for $u_{1r} u_{2x}$ and $u_{1x} u_{2r}$ may be verified to be identical in this order with $\mathscr H_{OSx}$ and $\mathscr H_{Cor}$ respectively.

\end{document}